\documentclass[letterpaper,11pt,fleqn]{article}
\pdfoutput=1
\usepackage{jheppub}

\usepackage{epstopdf}
\usepackage[T1]{fontenc}
\usepackage{extarrows}
\usepackage{bm,amsmath,amssymb}
\usepackage[mathscr]{eucal}
\usepackage{graphicx}
\usepackage{cancel}
\usepackage{subcaption}

\newcommand{\lan}{\langle}
\newcommand{\ran}{\rangle}
\long\def\comment#1{ }

\newcommand{\rnlo}{{\rm \scriptscriptstyle rNLO}}

\newcommand{\lo}{{\rm \scriptscriptstyle LO}}

\newcommand{\eqn}[1]{Eq.~\eqref{#1}}
\newcommand{\beq}{\begin{equation}}
\newcommand{\eeq}{\end{equation}}

\newcommand{\rmK}{{\rm K}}

\newcommand{\rmT}{{\rm T}}

\newcommand{\order}[1]{\mathcal{O}{(#1)}}
\newcommand{\abar}{\bar{\alpha}_s}
\newcommand{\nn}{\nonumber\\}
\newcommand{\rmd}{{\rm d}}
\newcommand{\rme}{{\rm e}}
\newcommand{\bk}{\bm{k}}
\newcommand{\bq}{\bm{q}}
\newcommand{\bp}{\bm{p}}
\newcommand{\bx}{\bm{x}}
\newcommand{\by}{\bm{y}}

\newcommand{\bz}{\bm{z}}

\title{Dihadron production in DIS at NLO: the real corrections}

\author[a]{Edmond Iancu}
\author[b]{and Yair Mulian}
\affiliation[a]{Institut de Physique Th\'eorique, Universite Paris Saclay, CNRS, CEA, F-91191, Gif-sur-Yvette, France}
\emailAdd{edmond.iancu@ipht.fr, yair25m@gmail.com}

\affiliation[b]{Departamento de Fisica de Particulas and IGFAE, Universidade de Santiago de Compostela,
15782 Santiago de Compostela, Galicia, Spain}

\abstract{By using the formalism of the light-cone wave function along with the colour glass condensate effective
theory, we consider next-to-leading order (NLO) corrections to the production of a pair of hadrons 
in electron-proton, or electron-nucleus, collisions at small Bjorken $x$. To the order of interest, the process involves 
the fluctuation of a virtual photon into a quark-antiquark pair, followed by the emission of a gluon 
from either the quark, or the antiquark. For the case of a virtual photon with transverse polarization, we compute
the real NLO corrections, where the emitted gluon is present in the final state. We first compute the tree-level
cross-section for the production of the quark-antiquark-gluon system and then deduce the real NLO corrections
to dihadron production by integrating out the kinematics of the gluon. We verify in detail that, in the limit where the 
gluon is soft, our calculation reproduces the (real piece of the) B-JIMWLK evolution of the leading-order cross-section 
for quark-antiquark production. Similarly, in the limit where the gluon is collinear with its emitter, 
we recover the real terms in the DGLAP evolution of the fragmentation function.
 The virtual NLO corrections to dihadron production will be presented by one of us in a subsequent publication.}

\begin{document}
\maketitle

\flushbottom

\section{Introduction}
One of the main discoveries of the HERA experimental program on deeply inelastic electron-proton scattering 
is the fact that the gluon density inside a hadron sharply rises
as one explores smaller and smaller values of Bjorken $x$ at a fixed virtuality $Q^{2}$ \cite{Aaron:2009aa,H1:2015ubc}.
While conceptually anticipated on the basis of the perturbative evolution in QCD with increasing energy, 
as encoded in the celebrated BFKL equation \cite{Lipatov:1976zz,Kuraev:1977fs,Balitsky:1978ic},
this rapid rise challenges our understanding of QCD scattering in the vicinity of the unitarity limit.
The natural solution to this problem, as also emerging from perturbative QCD, is the phenomenon of
{\it gluon saturation}, which is a consequence of the non-linear dynamics of the highly occupied gluon modes.
Predicted already before the advent of the HERA data \cite{Gribov:1984tu,Mueller:1985wy}, this phenomenon
finds its modern, pQCD-based, formulation in the effective theory for the {\it color glass condensate} (CGC)
\cite{Iancu:2002xk,Iancu:2003xm,Gelis:2010nm,Kovchegov:2012mbw,Blaizot:2016qgz}. The saturated gluon matter dubbed as the
CGC \cite{Iancu:2000hn} is believed to be a universal form of hadronic matter which controls the QCD scattering 
amplitudes for sufficiently high energies. The properties of this matter and the high-energy amplitudes
can be reliably computed within pQCD as soon as the characteristic transverse-momentum scale, the
{\it saturation momentum squared} $Q_s^2$ (a measure of the gluon density per unit transverse area),
is sufficiently hard: $Q_s^2\gg \Lambda_{_{\rm QCD}}^2$. This scale rises, roughly, as a power of $1/x$ and also 
(for a large nucleus) as a power of the nuclear mass number $A$: $Q_s^2(x,A)\propto A^{1/3}(1/x^\lambda)$ with 
$\lambda\simeq 0.2$. Hence pQCD should be a good tool for studies of gluon saturation at sufficiently small values
of $x$ and/or large values of $A$. 

This motivated the use of perturbative techniques for computing the
high-energy evolution of the gluon correlation functions --- a non-linear generalisation of the BFKL equation
known as the B-JIMWLK evolution\footnote{This is truly a functional renormalisation-group equation 
\cite{JalilianMarian:1997jx,JalilianMarian:1997gr,Kovner:2000pt,Iancu:2000hn,Iancu:2001ad,Ferreiro:2001qy}, 
equivalent to an infinite hierarchy of coupled equations \cite{Balitsky:1995ub},
and applies to gauge-invariant correlators built with products of Wilson lines.
In the limit of a large number of colors ($N_c\gg 1$), the first equation in this hierarchy reduces to a closed,
non-linear, equation for the dipole amplitude, known as the  Balitsky-Kovchegov equation  \cite{Balitsky:1995ub,Kovchegov:1999yj}.}
\cite{Balitsky:1995ub,JalilianMarian:1997jx,JalilianMarian:1997gr,Kovner:2000pt,Iancu:2000hn,Iancu:2001ad,Ferreiro:2001qy}
 --- as well as the hard impact factors which enter the CGC calculation of hadronic cross-sections  
 for ``dilute-dense'' collisions, like electron-nucleus ($eA$) or proton-nucleus ($pA$). 
 These ingredients have been originally computed to leading order (LO) in the
 QCD running coupling $\alpha_s$. However, one needs (at least) next-to-leading order (NLO) estimates in order to
 reach a perturbative accuracy of about 10\%, as required for realistic predictions for the phenomenology.
 And indeed, over the last years one assists at strenuous efforts aiming at promoting the CGC effective theory to NLO
 accuracy.  
 
 For the high-energy evolution, this program resulted not only in the NLO versions of the BK
 \cite{Balitsky:2008zza} and B-JIMWLK \cite{Balitsky:2013fea,Kovner:2013ona,Kovner:2014lca,Lublinsky:2016meo,Lappi:2020srm,Dai:2022imf}
 equations, but also in collinearly-improved
 equations \cite{Beuf:2014uia,Iancu:2015vea,Iancu:2015joa,Albacete:2015xza,Hatta:2016ujq,Ducloue:2019ezk,Ducloue:2019jmy,
Beuf:2020dxl} which resum to all orders the radiative corrections enhanced by large transverse logarithms (thus curing an instability
 of the strict NLO approximation \cite{Avsar:2011ds,Lappi:2015fma,Lappi:2016fmu} and partially including the 
 effects of the DGLAP evolution \cite{Gribov:1972ri,Altarelli:1977zs,Dokshitzer:1977sg}).
 Concerning the  ``'impact factors'' (the scattering matrix elements void of evolution), the NLO corrections have
 been computed for a few ``dilute-dense'' collisions. In the context of $pA$ collisions, this includes the
single inclusive hadron (or jet) production in $pA$ collisions
\cite{Chirilli:2011km,Chirilli:2012jd,Stasto:2013cha,Stasto:2014sea,Xiao:2014uba,Altinoluk:2014eka,Watanabe:2015tja,Ducloue:2016shw,Iancu:2016vyg,Ducloue:2017mpb,Ducloue:2017dit,Shi:2021hwx,Liu:2022ijp},
photon production at central rapidities \cite{Benic:2016uku},
and the ``real'' NLO corrections to dihadron production \cite{Iancu:2018hwa,Iancu:2020mos}
--- i.e. the NLO effects associated with a 3-parton final state, out of which only 2 are measured.
For electron-hadron ($ep$ or $eA$) deep  inelastic scattering  (DIS) at small Bjorken  $x$, one knows by now the
NLO corrections to the inclusive structure functions (including massive quarks) \cite{Balitsky:2010ze,Balitsky:2012bs,Beuf:2016wdz,Beuf:2017bpd,Beuf:2021qqa,Beuf:2021srj,Beuf:2022ndu},
to the exclusive (diffractive) production of (light or heavy) vector mesons and dijets \cite{Boussarie:2014lxa,Boussarie:2016ogo,Boussarie:2016bkq,Mantysaari:2021ryb,Mantysaari:2022bsp,Mantysaari:2022kdm}, to the inclusive photon + dijet production in  $eA$ collisions at small $x$ \cite{Roy:2019hwr},  to the inclusive
dijet (or dihadron) production \cite{Caucal:2021ent,Taels:2022tza,Bergabo:2022tcu,Caucal:2022ulg}
 (see also \cite{Ayala:2016lhd,Ayala:2017rmh,Altinoluk:2020qet} for an earlier 
calculation of polarized trijet production), and to semi-inclusive DIS (single inclusive hadron production) 
\cite{Bergabo:2022zhe}.
One has recently computed the real NLO corrections to the diffractive structure function \cite{Beuf:2022kyp}, 
as well as the diffractive production of 3 jets \cite{Iancu:2021rup,Hatta:2022lzj,Iancu:2022lcw}.

Among the processes above listed, the inclusive production of a pair of jets or hadrons is particularly interesting, as this is sensitive
to gluon saturation via the azimuthal correlations among the two measured particles
\cite{Marquet:2007vb,Albacete:2010pg,Dominguez:2011wm,Stasto:2011ru,Iancu:2013dta,Lappi:2012nh,Kotko:2015ura,Marquet:2016cgx,vanHameren:2016ftb,Albacete:2018ruq,Metz:2011wb,Dumitru:2015gaa,Altinoluk:2015dpi,Dumitru:2018kuw,Mantysaari:2019hkq}.
Multiple scattering should lead a broadening in their azimuthal distribution around the back-to-back peak.
Such a broadening has been observed in d+Au collisions
at RHIC \cite{Braidot:2011zj,Adare:2011sc}, although at this level it looks difficult to distinguish
the effects of saturation from those of the final state radiation  (the ``Sudakov factor'' \cite{Mueller:2013wwa}).
The kinematical conditions for studying this effect are expected to be better at the Electron-Ion Collider (EIC) 
 \cite{Zheng:2014vka}. This explains the interest in having accurate theoretical predictions for this process in $ep$, or $eA$ collisions.

It is our purpose here to present a calculation of the real NLO corrections to dihadron production at forward rapidities
in electron-hadron collisions in the CGC formalism. 
(The corresponding virtual corrections will be presented by one of us (Y.M.) in a separate paper.)
At LO, the respective cross-section involves the production of a quark-antiquark pair, which is initiated by the decay of the virtual photon
and put on mass-shell by the scattering off the hadronic target \cite{Dominguez:2011wm}. 
The real NLO corrections involve the emission of an additional
gluon, by either the quark or the antiquark, which is produced too in the final state, but it is not measured\footnote{Given our
results, it would be straightforward to deduce the NLO corrections for the case where the unmeasured parton is the
quark. or the antiquark: we will
present explicit results for three-parton ($q\bar q g$) production and the NLO contributions to di-hadron production
can be deduced by integrating out the kinematics of the unmeasured parton.}.
As aforementioned, there are already several  independent calculations of these NLO corrections in the literature \cite{Caucal:2021ent,Taels:2022tza,Bergabo:2022tcu,Ayala:2016lhd,Ayala:2017rmh,Altinoluk:2020qet}, 
with results which appear to agree with each other. So, one may wonder why there is need for yet another
calculation. In our opinion, there are in fact several reasons for that.

First, these are  complex and tedious calculations, which are technically involved and conceptually subtle. So, it is indeed useful to have several independent calculations, in order to cross check the previous results.
 Second, the previous calculations generally use different techniques and/or consider different final states: we use the light-cone wave function (LCWF) formalism together with light-cone perturbation theory  (LCPT)
 and look at dihadron final states. The same formalism has been used in Ref.~\cite{Taels:2022tza},
which however considered  dijet final states. Ref.~\cite{Caucal:2021ent} uses momentum-space (Feynman-rule) perturbation theory together with dijet final states, while Ref.~\cite{Bergabo:2022tcu} focuses on dihadrons (like ourselves), but uses
helicity techniques when computing Feynman graphs. Moreover, our results and those in Ref.~\cite{Bergabo:2022tcu} 
are complementary to each other, in the sense that they refer to different polarisation states for the virtual photon
--- transverse polarisation in our analysis and, respectively, longitudinal polarisation in \cite{Bergabo:2022tcu}.
Together, our respective papers cover all cases (for real NLO corrections at least).

Despite such formal differences, we have checked that our results for the real NLO corrections fully agree 
with the corresponding results in the literature, whenever direct comparisons were possible.
This is the case, e.g., for the LCWF of the virtual photon to the order of interest and 
for the tree-level cross-section for producing a system of three partons (a quark, an antiquark and a gluon)
--- the main ingredients for constructing the real NLO corrections to the production of both dihadrons and dijets.

But even when the results appear to agree with each other, we still believe that our analysis may offer
new insights, not only because of the use of different methods, but also because of the different emphasis
that we have put on various aspects of the calculation and the associated physical discussions. 
Notably, we feel that a strong point of our analysis
is the particularly  transparent discussion of the two special limits, which represent important checks
of the NLO results: the soft gluon limit, in which we shall recover the B-JIMWLK evolution of the LO cross-section,
and the limit where the gluon becomes collinear with its emitter, where we shall recover the
DGLAP evolution of the fragmentation function for the final quark, or antiquark. 

As compared to previous studies in the
literature, we shall analyse the soft gluon limit diagram by diagram and we shall thus demonstrate that only a particular
class of Feynman graphs contribute in this limit --- those where the gluon is emitted sufficiently close to
the scattering with the nuclear target. There are 16 such graphs and their respective contributions are found to be
in an one-to-one correspondence with the (real) terms of the B-JIMWLK equation for the quadrupole  
\cite{Dominguez:2011gc,Iancu:2011ns,Iancu:2011nj}.

Furthermore, we shall use an original strategy (that we introduced in the context of proton-nucleus collisions
\cite{Iancu:2020mos}) to verify the emergence of the DGLAP evolution in the collinear limit. While all the other
approaches in the literature rely on the transverse {\it momentum} representation for that purpose (see e.g. 
\cite{Bergabo:2022tcu} for a discussion in  a similar context), 
we have reformulated the collinear limit directly in term of transverse {\it coordinates}
(the natural representation for constructing the LCWF in the presence of multiple scattering).
While the two representations should be eventually equivalent, we feel that our method is more efficient
in practice, as it allows one to easily recognise the diagrams which contribute to the collinear limit and to extract
 the DGLAP evolution already before the final Fourier transform to momentum space.

This paper is structured as follows. In Sect.~\ref{sec:LCWF} we schematically describe the results of the LCPT for the 
virtual photon light-cone wavefunction to the order of interest --- that is, for the quark-antiquark
($q\bar q$) and the quark-antiquark-gluon ($q\bar q g$) Fock space components.  Notice that by ``LCWF''
we truly mean the outgoing state that would be probed by the detector and which also includes the effects
of the collision with the hadronic target. The general
but formal  results from Sect.~\ref{sec:LCWF} are subsequently used to derive fully explicit expressions for the respective
wavefunctions and scattering amplitudes. Specifically, in Sect.~\ref{sec:LO} we deduce
the amplitude and the cross-section for the leading order (LO) process, that is, the exclusive production
of a $q\bar q$ pair. Then, in Sect.~\ref{sec:qqgLCWF}
 we present the results for the quark-antiquark-gluon Fock component, 
where the gluon can be emitted from either the quark or anti-quark. In Sect.~\ref{sec:NLOreal} we deduce the real NLO
corrections to the dihadron cross-section by integrating out the kinematics of one (anyone) of the three final
partons. In particular, we explain the simplifications which occur in the colour structure (i.e. in the partonic 
$S$-matrices) due to the fact that one of the partons is not measured.  In Sect.~\ref{sec:soft} we show that in the limit where the unmeasured parton is a soft gluon, our NLO results reproduce, as expected, the (real part of the) JIMWLK evolution for the LO dijet cross-section. Finally, in Sect.~\ref{sec:realDGLAP}, we consider the limit where the unmeasured gluon is produced by a collinear splitting. We show that, in this limit, our NLO results develop collinear singularities, that we isolate to leading logarithmic accuracy and verify that they can be interpreted as one-step in the DGLAP evolution of the fragmentation function.

\section{The virtual photon light-cone wavefunction}
\label{sec:LCWF}

We consider high-energy electron-proton or electron-nucleus scattering 
 in a Lorentz frame where the  virtual photon $\gamma^*$ is a relativistic right-mover, with 4-momentum
$q^\mu= (q^+, -Q^2/2q^+,\bm{q} )$ in light-cone notations,  whereas the nucleus is an 
ultrarelativistic left-mover, which for the present purposes can be simply treated as a Lorentz-contracted
shockwave. 

We describe this collision using the light-cone wavefunction (LCWF) formalism, which aims at constructing
the wavefunction of the virtual photon as a superposition of Fock states built with ``bare 
quanta'' (the eigenstates of the free piece of the Hamiltonian). This superposition is evolving with time,
due to the radiation of new quanta, and is constructed via time-dependent perturbation theory. 
The bare quanta are assumed to be on their mass shall: a bare quark or gluon carries a
4-momentum $k^\mu= (k^+, k^-,\bm{k} )$, where $k^+>0$ and
the minus component is the LC energy: $k^-=E_k\equiv
k_\perp^2/2k^+$, with $k_\perp=|\bk|$. A bare single-particle state is specified by the particle 3-momentum 
$(k^+, \bk)$ and the corresponding quantum numbers, like spin, polarisation, and colour. E.g., a bare
quark state is written as 
$\left|q_{\lambda}^{\alpha}(k^{+},\,\bm{k})\right\rangle$, where $\lambda=\pm 1/2$ denotes the helicity and $\alpha=1,2,\dots,N_c$ is a colour index in the fundamental representation. Whenever no confusion is possible,
we shall denote bare single-particle states simply as $|q\rangle$, $|g\rangle$, and $|\gamma\ran$
for a quark, a gluon, and the original photon, respectively.

We would like to construct the final (``outgoing'') state at time $x^+\to\infty$ for the case where the incoming state at
$x^+\to -\infty$ is a bare, space-like, photon. We work in the projectile light-cone (LC) gauge $A^+_a=0$ and in the interaction representation. As already mentioned, the hadronic target is a shockwave localised
near $x^+=0$, so we can factorise the scattering from the ``initial-state'' and the ``final-state'' evolutions --- the
quantum evolutions occurring before ($x^+<0$) and  respectively after 
($x^+>0$) the collision. We can then write (in compact notations)
\begin{align}\label{outgamma}
\left| \gamma\right\rangle_{out}=\,U(\infty,\,0) \,\hat S \, U(0,\,-\infty)\,|\gamma\ran 
\end{align}
where $|\gamma\ran $ is the bare photon state, $\hat S$ is the scattering operator (or ``$S$--matrix'') describing
the collision with the shockwave in the eikonal approximation,
and the evolution operator is built with the interaction piece $H_{\rm int}=H_{_{\rm QCD}}^{\rm int} +H_{_{\rm QED}}^{\rm int}$ 
of the full (QCD plus QED) Hamiltonian :
\beq\label{Udef}
U(x^+,\,x_0^+) = {\rm T}\exp\left\{-i\int_{x_0^+}^{x^+}\rmd y^+ H_{\rm int}(y^+)\right\}.
\eeq
(The time-dependence in  $H_{\rm int}(y^+)$ is governed by the 
the free Hamiltonian $H_0=H_{_{\rm QCD}}^{0} +H_{_{\rm QED}}^{0}$, as standard in the  interaction representation.)
The QED Hamiltonian $H_{_{\rm QED}}^{\rm int}$ generates the splitting of the virtual photon into a quark-antiquark pair
in a colour singlet state (a ``colour dipole''), whereas the QCD Hamiltonian $H_{_{\rm QCD}}^{\rm int}$
governs the evolution of this $q\bar q$ pair via gluon emissions. 

The quark and the antiquark can be put on-shell by their scattering off the nuclear
shockwave and thus emerge as two ``jets'' in the final state. Our ultimate goal is to compute the cross-section
for dijet production to lowest order in the QED coupling, i.e to $\order{\alpha_{\rm em}}$,
and to next-to-leading order in the QCD coupling, i.e. to $\order{\alpha_s}$
(as usual, we have $\alpha_{\rm em}=e^2/4\pi$ and $\alpha_s=g^2/4\pi$). To that aim, it is sufficient to compute
single gluon emissions\footnote{Two real emissions would contribute to $\order{g^2}$ for the outgoing state, but only
to $\order{g^4}$ for the dijet cross-section.}, which can be either {\it real} --- the gluon exists in the final state, albeit it is not measured ---,
or {\it virtual} --- the gluon is emitted and reabsorbed at the level of the amplitude,  meaning that the final state is just
a (bare) $q\bar q$ state.

To the accuracy of interest, one  can expand the evolution operators in \eqn{outgamma} to linear order in\footnote{From now on
we omit the upper script ``int''  on the Hamiltonians, to simplify notations.} 
$H_{_{\rm QED}}^{\rm int}\equiv H_{_{\rm QED}}$, for instance
\begin{align}
U(0,\,-\infty)&\,=\,1-i\int_{-\infty}^0\rmd x^+ {\rm T}\left[\rme^{-i \int_{-\infty}^0\rmd y^+ H_{_{\rm QCD}}(y^+)}\,
H_{_{\rm QED}}(x^+)\right]
\nonumber \\*[0.2cm]
&\,\to\,1-i\int_{-\infty}^0\rmd x^+ U_{_{\rm QCD}}(0, x^+)H_{_{\rm QED}}(x^+),
\end{align}
with $U_{_{\rm QCD}}$ given by \eqn{Udef} with $H_{\rm int}(y^+)\to H_{_{\rm QCD}}(y^+)$.
The simpler expression in the second line is sufficient when acting on the photon state $|\gamma\ran $.  Using
this last expression for $U(0,\,-\infty)$ together with the corresponding one for $U_{I}(\infty,\,0) $, one finds that
\eqn{outgamma}  reduces to
\begin{align}\label{outgammaqq0}
\left| \gamma\right\rangle_{out}=\,\left| \gamma\right\rangle&\,-i U_{_{\rm QCD}}(\infty, \,0)
\,\hat S \int_{-\infty}^0\rmd x^+ U_{_{\rm QCD}}(0, x^+)H_{_{\rm QED}}(x^+)\,|\gamma\ran
\nonumber \\*[0.2cm]
 &\,-i\int_0^{\infty}\rmd x^+ U_{_{\rm QCD}}(\infty, x^+)H_{_{\rm QED}}(x^+)\,|\gamma\ran.
\end{align}

The physical interpretation of the above decomposition is quite transparent. The first corrective term, where  
$x^+$ takes negative values, describes a process which starts with the photon decay 
$\gamma\to q\bar q$ at some time $x^+ <0$, then the $q\bar q$ state evolves according to QCD 
until it hits the nuclear target at time $x^+=0$;  the scattering between the dressed $q\bar q$ state
and the shockwave is quasi-instantaneous (at least, within the eikonal approximation that we employ here) 
and is described by the $S$-matrix $ \hat S$; finally, 
the partonic state emerging from the scattering evolves up to the time of measurement $x^+\to\infty$.

The second corrective term in \eqn{outgammaqq0} describes the situation where the photon decays {\it after} crossing the
shockwave, at some time $x^+ >0$. In this case there is no scattering, but merely the evolution of the
$q\bar q$ state from $x^+$ up to infinity. It is convenient for what follows  
to observe that the effect of this second term  
is to subtract the no-scattering limit ($\hat S \to 1$) of the first term; that is, \eqn{outgammaqq0} is equivalent to
\begin{align}\label{outgammaqq}
\left| \gamma\right\rangle_{out}=\,\left| \gamma\right\rangle&\,-i U_{_{\rm QCD}}(\infty, \,0)
\,(\hat S -1)\int_{-\infty}^0\rmd x^+ U_{_{\rm QCD}}(0, x^+)H_{_{\rm QED}}(x^+)\,|\gamma\ran.
\end{align}
Indeed, in the absence of scattering, the virtual photon must eventually return to its initial state,
since a space-like photon cannot decay into a system of on-shell partons.
This means that $\left| \gamma\right\rangle_{out}\to \left| \gamma\right\rangle$ as $\hat S \to 1$, or
 $U(\infty,\,-\infty)| \gamma\ran=| \gamma\ran$, which 
to linear order in $H_{_{\rm QED}}$ implies the result in \eqn{outgammaqq}.

The perturbative expansion of the outgoing LCWF to the order of interest can now be obtained by expanding the evolution
operators in \eqn{outgammaqq} to second order in $H_{_{\rm QCD}}$:
\beq\label{pertU}
U_{_{\rm QCD}}(x^+,\,x_0^+) =1 - i\int_{x_0^+}^{x^+}\rmd y^+H_{_{\rm QCD}}(y^+)-\int_{x_0^+}^{x^+}\rmd y^+_2\int_{x_0^+}^{y^+_2}\rmd y_1^+ H_{_{\rm QCD}}(y^+_2) H_{_{\rm QCD}}(y^+_1)+\,\cdots.
\eeq

To evaluate the action of the interaction Hamiltonian, we shall work in the basis of bare multi-partonic Fock states, i.e. 
energy-momentum eigenstates of the free Hamiltonian with fixed numbers of free, on-shell, partons (quarks, antiquarks, and gluons). 
This not only matches the time-dependence of $H_{_{\rm QCD}}(y^+)$, but is also convenient for computing
particle production in the final state: indeed, for asymptotically large times, where the interactions are adiabatically switched off,
the bare Fock states are also eigenstates of the full Hamiltonian.
If $\left|i\right\rangle$ and $\left|j\right\rangle$  are two generic Fock states, then 
\beq
 \lan j | H_{_{\rm QCD}}(y^+) | i\ran =\rme^{i(E_j-E_i)y^+-\epsilon|y^+|} \lan j | H_{_{\rm QCD}} |i\ran,
 \eeq 
 with $E_i$ and $E_j$ the LC energies of the two states 
 (the sum of the LC energies of the constituent partons). The adiabatic factor $\rme^{-\epsilon|y^+|}$ was introduced in order
 to turn off the interactions at  large (positive or negative) times. This factor is useful in the intermediate calculations
 --- e.g. it ensures the convergence of time integrations like that in  \eqn{outgammaqq} ---, but the final, physical, results
 must have a finite limit when $\epsilon\to 0$.
Similarly,
\begin{align}\label{HQED}
H_{_{\rm QED}}(x^+)\,|\gamma\ran\,=\,\rme^{i(E_{q\overline{q}}-E_{\gamma})x^+ -\epsilon|x^+|}\,|q\bar q\ran \lan q\bar q| 
H_{_{\rm QED}}|\gamma\ran,\end{align}
where $|q\bar q\ran $ is a generic quark-antiquark bare state and the sum over all such states, i.e. over the momenta and the 
quantum numbers of the two bare fermions, is implicitly understood. Of course, this sum is constrained by the
conservation properties encoded in the matrix element of $H_{_{\rm QED}}$; in particular, the $q\bar q$
state created by the decay of the photon must be a colour-singlet (or ``colour dipole'').
In  what follows, we shall systematically use this convention, that repeated indices must be summed over.

Before we turn to a study of the NLO corrections, let us derive the leading-order (LO) result for the $q\bar q$
Fock space component, as obtained by replacing both QCD evolution operators by unity:
\begin{align}\label{gamma0qq}
\left| \gamma\right\rangle_{out}^{(0),\,q\bar q}&\,=-i(\hat S -1)\int_{-\infty}^0\rmd x^+ 
\rme^{i(E_{q\overline{q}}-E_{\gamma})x^+ +\epsilon x^+}\,|q\bar q\ran \lan q\bar q| 
H_{_{\rm QED}}|\gamma\ran,\nonumber \\*[0.2cm]
&\,=-|q_1\bar q_1\ran
\lan q_1\bar q_1|\hat S -1 |q\bar q\ran 
\,\frac{\lan q\bar q| H_{_{\rm QED}}\,|\gamma\ran}{E_{q\overline{q}}-E_{\gamma}-i\epsilon}\,.
 \end{align}
In the final result, one can neglect the $i\epsilon$ prescription in the denominator, since the space-like
photon cannot be degenerate with the on-shell quark-antiquark pair.

\begin{figure}
  \centering
    \includegraphics[width=0.9\textwidth]{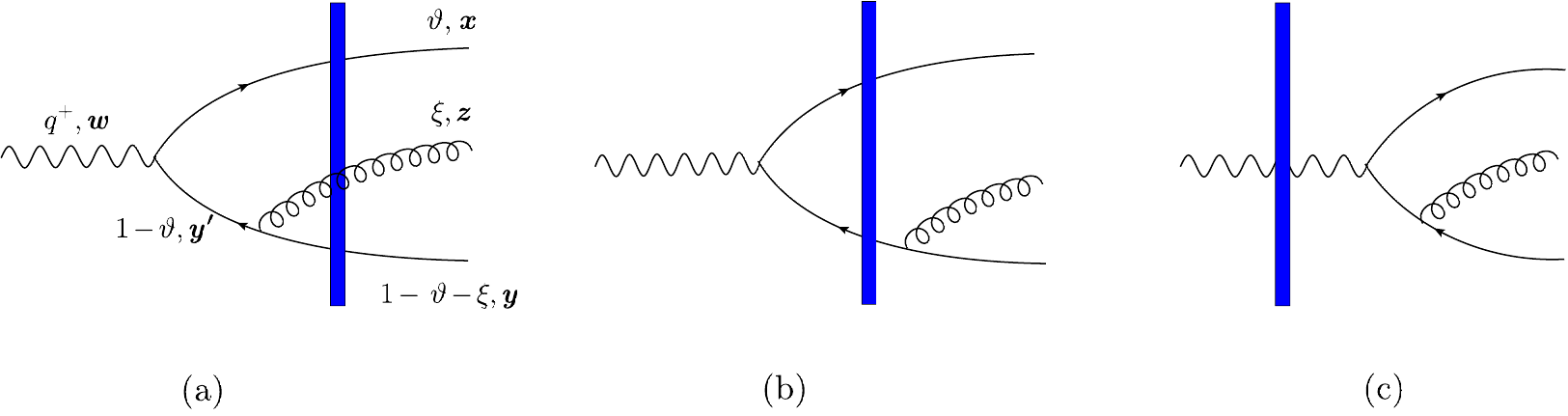}
  \caption{\small The three possible topologies for the quark-antiquark-gluon outgoing state in the case
  where the gluon is emitted by the antiquark. We only represent regular graphs, where the intermediate antiquark
  line represents an on-shell, propagating, fermion. See Fig.~\ref{fig:qbar-inst} for the respective
  instantaneous graphs.}
    \label{fig:NLO_Real}
    \end{figure}

\subsection{Real gluon emissions}
\label{sec:WFreal}

A single gluon emission by either the quark or the antiquark produces a $|q\bar q g\ran $ Fock component in the
final state, which contributes to $\order{g}$ to the outgoing wavefunction and hence to $\order{g^2}$ to the dijet cross-section.
Such an emission is generated by the term linear in $H_{_{\rm QCD}}$ in the expansion \eqref{pertU}  of the QCD evolution operator. With
reference to \eqn{outgammaqq}, it is clear that there are two possible time-orderings: the gluon can be emitted either before, or after,
the scattering between the $q\bar q$ pair and the nuclear shockwave. When the emitter is the antiquark,
these two possibilities are represented by graphs (a) and (b) in Fig.~\ref{fig:NLO_Real}. As a matter
of facts, there is also a third graph, illustrated in Fig.~\ref{fig:NLO_Real}.c, where the photon decays after crossing the shockwave.
But as already explained, this case is implicitly included in  \eqn{outgammaqq}, via the subtraction of the no-scattering limit
from the $S$-matrix.

Consider the initial-state emission first, cf.  Fig.~\ref{fig:NLO_Real}.a. The corresponding contribution to \eqn{outgammaqq}
reads (cf. Eqs.~\eqref{pertU} and \eqref{HQED})
\begin{align}\label{real-in}
&\hspace*{-1.cm}
\,(-i)^2(\hat S -1)\int_{-\infty}^0\rmd x^+ \int_{x^+}^0\rmd y^+H_{_{\rm QCD}}(y^+)H_{_{\rm QED}}(x^+)\,|\gamma\ran
\nonumber \\*[0.2cm]
&\hspace*{-1.cm}
=-(\hat S -1)\int_{-\infty}^0\rmd y^+\! \int_{-\infty}^{y^+}\rmd x^+
\rme^{i(E_{q_1\overline{q}_1g_1}-E_{q\overline{q}})y^++\epsilon y^+} |q_1\bar q_1 g_1\ran \lan q_1\bar q_1 g_1|
H_{_{\rm QCD}}|q\,\bar q\ran  \rme^{i(E_{q\overline{q}}-E_{\gamma})x^++\epsilon x^+}
\lan q\bar q| H_{_{\rm QED}}|\gamma\ran
\nonumber \\*[0.2cm]
&\hspace*{-1.cm}
=  |q_2\bar q_2 g_2\ran\lan q_2\bar q_2 g_2|\hat S -1 |q_1\bar q_1 g_1\ran \,
\frac{\lan q_1\bar q_1 g_1|H_{_{\rm QCD}}|q\,\bar q\ran}{E_{q_1\overline{q}_1g_1}-E_{\gamma}-2i\epsilon}
\,\frac{\lan q\bar q| H_{_{\rm QED}}\,|\gamma\ran}{E_{q\overline{q}}-E_{\gamma}-i\epsilon}\,.
\end{align}
When inserting intermediate Fock states in the second line, we used the fact that the action of $H_{_{\rm QCD}}$ on the
$|q\bar q\ran $ state consists in a gluon emission, leading to a state of the form $ |q_1\bar q_1 g_1\ran $,
whereas the subsequent action of the $S$-matrix cannot change the partonic content of that state, but only modify
the momenta and the discrete quantum numbers (polarisation and colour)  of the 3 partons.
By energy-momentum conservation, the energy denominators in \eqn{real-in} cannot vanish, hence 
one can safely let $\epsilon\to 0$ in the final result.

Consider similarly the final-state emission, cf.  Fig.~\ref{fig:NLO_Real}.b. By expanding the final evolution operator 
in \eqn{outgammaqq} to linear order, one finds the respective contribution as
\begin{align}\label{real-fin}
& (-i)^2 \int_0^{\infty}\rmd y^+ H_{_{\rm QCD}}(y^+)
(\hat S -1)\int_{-\infty}^0\rmd x^+H_{_{\rm QED}}(x^+)\,|\gamma\ran
\nonumber \\*[0.2cm]
&= -\, |q_2\bar q_2 g_2\ran \,
\frac{\lan q_2\bar q_2 g_2|H_{_{\rm QCD}}|q_1\bar q_1\ran}{E_{q_2\overline{q}_2g_2}-E_{q_1\overline{q}_1}}\,
\lan q_1\bar q_1|\hat S -1 |q\bar q\ran 
\,\frac{\lan q\bar q| H_{_{\rm QED}}\,|\gamma\ran}{E_{q\overline{q}}-E_{\gamma}}\,,
\end{align}
where we have omitted the intermediate steps as well as the $i\epsilon$ prescription (since unnecessary).

To summarise, the $q\bar qg$ Fock space component of the outgoing LCWF of the virtual photon reads
\begin{align}\label{outgammaRe}
\left| \gamma\right\rangle_{out}^{q\bar q g}\,=\,
\left|q_{2}\overline{q}_{2}g_2\right\rangle \Bigg\{ &
\lan q_2\bar q_2 g_2|\hat S -1 |q_1\bar q_1 g_1\ran \,
\frac{\lan q_1\bar q_1 g_1|H_{_{\rm QCD}}|q\bar q\ran}{E_{q_1\overline{q}_1g_1}-E_{\gamma}}
\nonumber \\*[0.2cm]
& +\frac{\langle q_{2}\overline{q}_{2}g_2
|H_{_{\rm QCD}}|
{q_{1}}{\overline{q}_{1}}\rangle }{E_{{q_{1}}{\overline{q}_{1}}}-E_{q_{2}\overline{q}_{2}g_2}}
\,\langle q_{1}{\overline{q} _{1}}  |\hat S-1| q\overline{q}\rangle \Bigg\}
\frac{ \lan q \bar q | H_{_{\rm QED}}\,|\gamma\ran}{E_{q\overline{q}}-E_{\gamma}},
\end{align}
where the first (second) line describes one initial-state (final-state) gluon emission. 

It is interesting to notice the structure of the energy denominators in \eqn{outgammaRe}. 
For the initial-state decays, they involve the difference between the
LC energies of the intermediate state and of the initial state, respectively; e.g. $E_{q_1\overline{q}_1g_1}-E_{\gamma}$.
For the final-state emissions, on the other hand, the energy of the initial state is replaced by that of the final state,
e.g. $E_{{q_{1}}{\overline{q}_{1}}}-E_{q_{2}\overline{q}_{2}g_2}$. These rules are in fact general and we shall see
other examples in what follows.

\subsection{Virtual corrections}
\label{sec:WFvirt}

\begin{figure}
  \centering
    \includegraphics[width=0.8\textwidth]{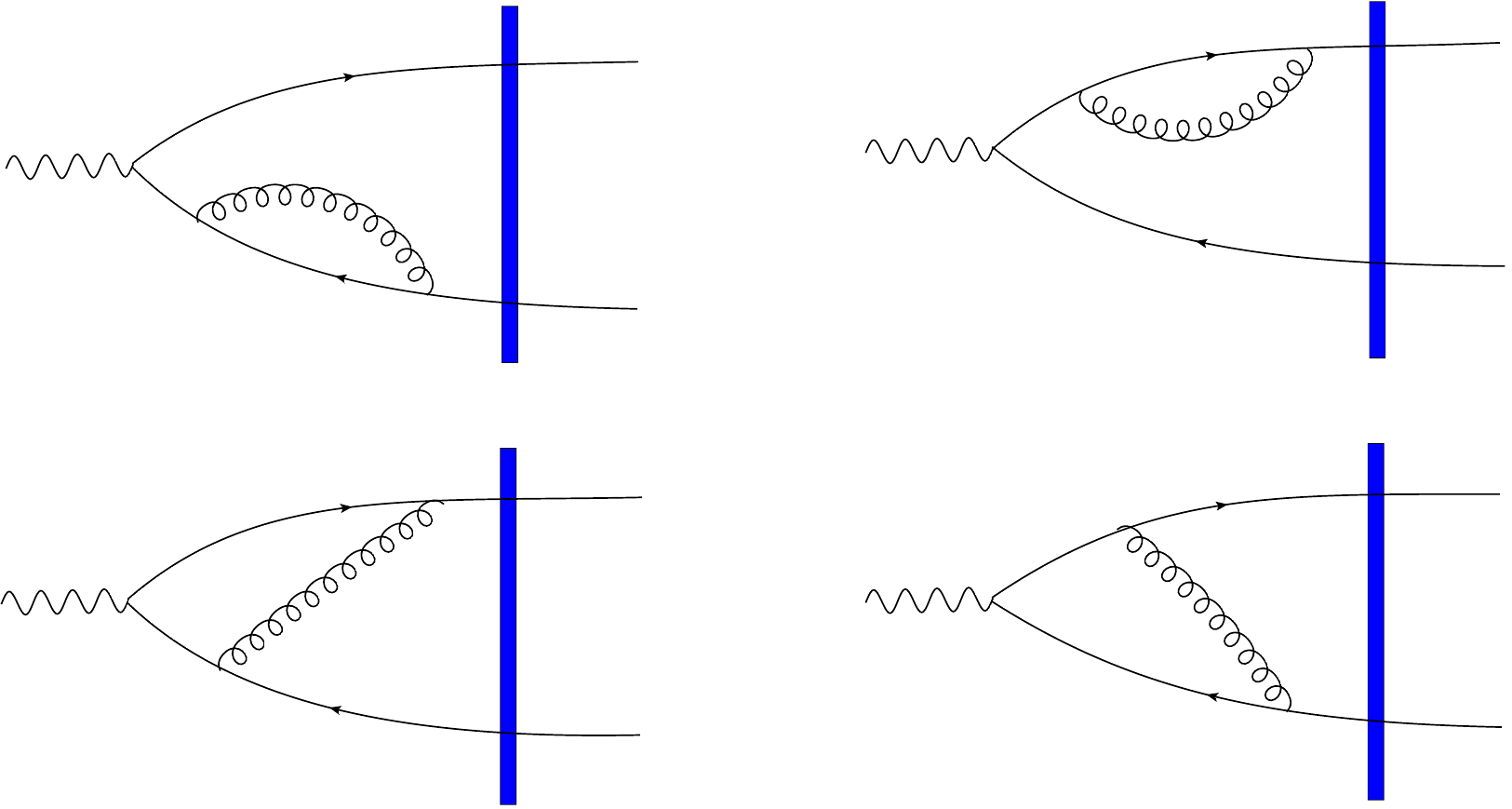}
  \caption{\small The 4 possible topologies for virtual corrections occurring in the initial state. }
    \label{fig:VirtIS}
    \end{figure}

\begin{figure}
  \centering
    \includegraphics[width=0.8\textwidth]{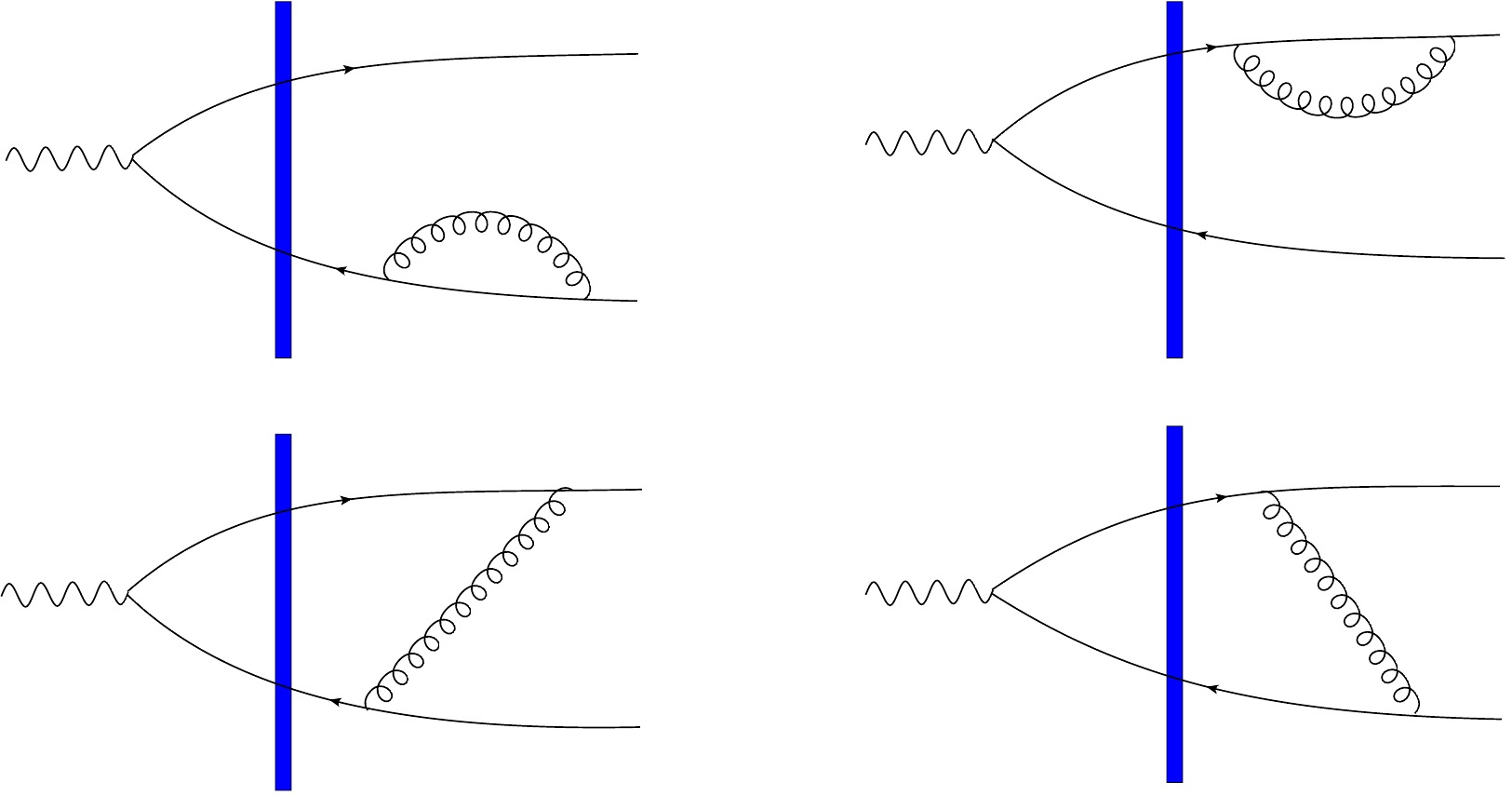}
  \caption{\small The 4 possible topologies for virtual corrections occurring in the final state. }
    \label{fig:VirtFS}
    \end{figure}

\begin{figure}
  \centering
    \includegraphics[width=0.8\textwidth]{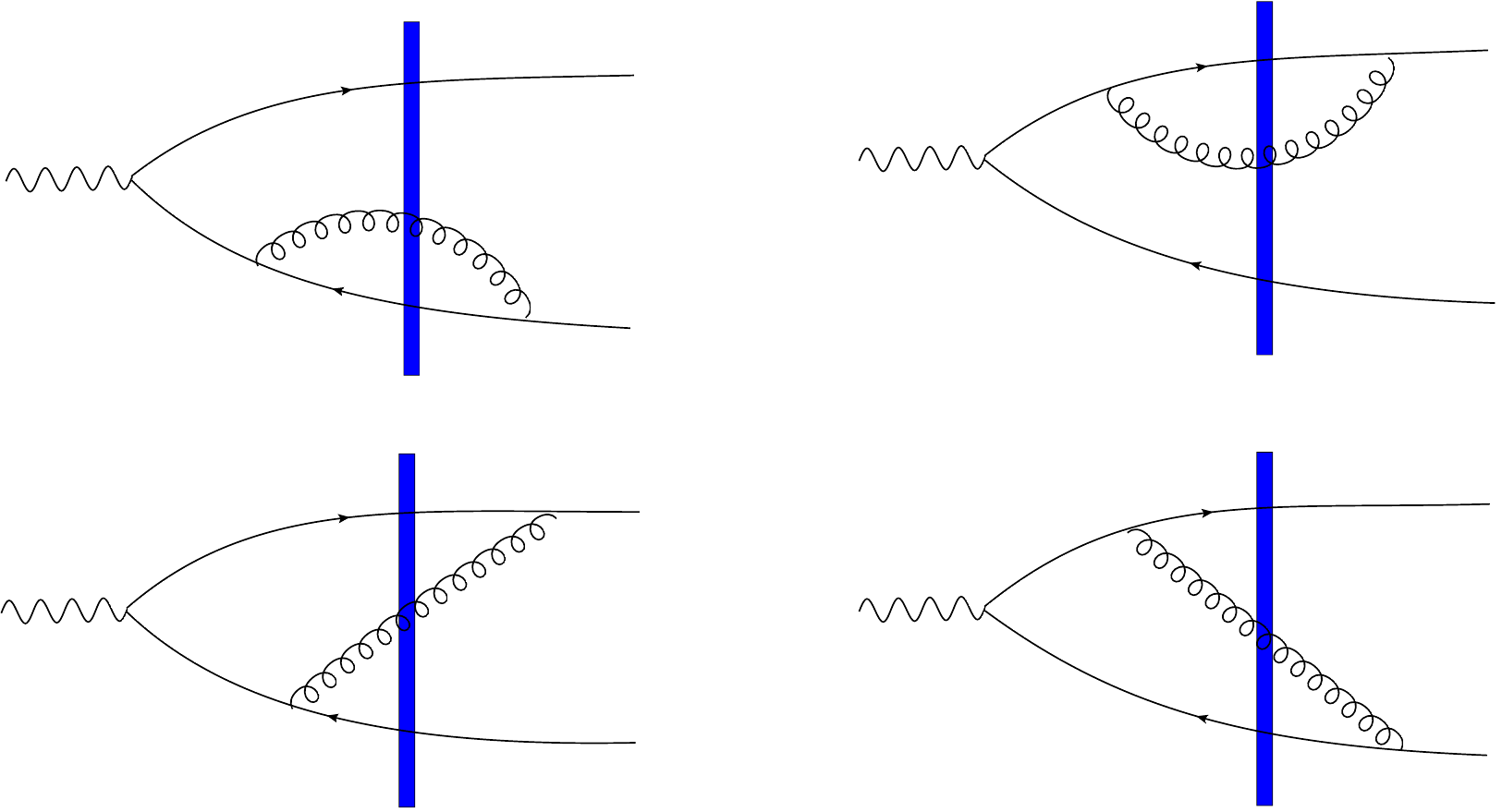}
  \caption{\small The 4 possible topologies for virtual corrections  in which the gluon crosses the shockwave.}
    \label{fig:VirtCR}
    \end{figure}

We now turn to the virtual corrections, that is, the one-loop contributions to the amplitude which are generated by the
second-order terms in the expansion of the evolution operators in \eqn{outgammaqq}.  These can be either self-energy graphs,
where the gluon is emitted and reabsorbed by the same fermion, or gluon exchange graphs,  where the gluon is emitted 
by the quark and reabsorbed by the anti-quark, or vice-versa. For both topologies, we encounter three types of time-orderings: 
initial-state evolution (cf. Fig.~\ref{fig:VirtIS}), final-state evolution (cf. Fig.~\ref{fig:VirtFS}), and mixed evolution, 
where the gluon is emitted before the scattering with the shockwave and is reabsorbed after that scattering
(cf. Fig.~\ref{fig:VirtCR}).

We start with the last case, where the gluon crosses the shockwave and thus can interact with it. 
We have already shown in \eqn{real-in} the
$q\bar q g$ state generated by first emitting a gluon and then scattering with the shockwave. To obtain the ``crossing'' virtual corrections, 
one must reabsorb this gluon after the scattering, which can be done by acting 
with the QCD Hamiltonian (the linear term in the expansion of $U_{_{\rm QCD}}(\infty, \,0)$) on the $q\bar q g$ state in  \eqn{real-in}.
Using 
\begin{align}
-i \int_0^{\infty}\rmd y^+ \lan q_3\bar q_3 | H_{_{\rm QCD}}(y^+)  |q_2\bar q_2 g_2\ran &\,=
-i \int_0^{\infty}\rmd y^+ \rme^{i(E_{q_3\overline{q}_3}-E_{q_2\overline{q}_2 g_2})y^+-\epsilon y^+} 
\lan q_3\bar q_3 | H_{_{\rm QCD}}|q_2\bar q_2 g_2\ran
\nonumber \\*[0.2cm]
&\,=
\frac{\left\langle q_{3}\overline{q}_{3}\right|H_{_{\rm QCD}}\left|q_{2}\overline{q}_{2} g_2\right\rangle }{E_{q_{3}\overline{q}_{3}}-E_{q_{2}\overline{q}_{2}g_2}+i\epsilon}\,,
\end{align}
one deduces (the upper label $C$ stays for ``crossing'')
\begin{align}\label{gammaCvirt}
\left| \gamma\right\rangle_{out}^{C,\,q\bar q}\,=-|q_3\bar q_3\ran
  \frac{\left\langle q_{3}\overline{q}_{3}\right|H_{_{\rm QCD}}\left|q_{2}\overline{q}_{2} g_2\right\rangle }{E_{q_{2}\overline{q}_{2}g_2}-
  E_{q_{3}\overline{q}_{3}}}\,\lan q_2\bar q_2 g_2|\hat S -1 |q_1\bar q_1 g_1\ran \,
\frac{\lan q_1\bar q_1 g_1|H_{_{\rm QCD}}|q\bar q\ran}{E_{q_1\overline{q}_1g_1}-E_{\gamma}}
\,\frac{\lan q\bar q| H_{_{\rm QED}}\,|\gamma\ran}{E_{q\overline{q}}-E_{\gamma}}\,,
\end{align}
where the energy denominators cannot vanish.

Consider now the one-loop graphs associated with initial-state evolution (cf. Fig.~\ref{fig:VirtIS}). To that aim, one must use the second-order term
in the expansion of $U_{_{\rm QCD}}(0, \,-\infty)$, cf. \eqn{pertU}, to first emit a gluon by the $q\bar q$ pair and then reabsorb it.
The relevant matrix element reads (compare to \eqn{real-in})
\begin{align}
&(-i)^3\int_{-\infty}^0\rmd y^+_2\! \int_{-\infty}^{y^+_2}\rmd y_1^+ \,\int_{-\infty}^{y^+_1}\rmd x^+
 \lan q_2\bar q_2| H_{_{\rm QCD}}(y^+_2) H_{_{\rm QCD}}(y^+_1) H_{_{\rm QED}}(x^+) |\gamma\ran
  \nonumber \\*[0.2cm]
&\qquad\,=-\frac{\left\langle q_{2}\overline{q}_{2}\right|H_{_{\rm QCD}}\left|q_{1}\overline{q}_{1} g_1\right\rangle }
{E_{q_2\overline{q}_2}-E_{\gamma}-3i\epsilon}
\frac{\lan q_1\bar q_1 g_1|H_{_{\rm QCD}}|q\,\bar q\ran}{E_{q_1\overline{q}_1g_1}-E_{\gamma}-2i\epsilon}
\,\frac{\lan q\bar q| H_{_{\rm QED}}\,|\gamma\ran}{E_{q\overline{q}}-E_{\gamma}-i\epsilon}\,.
 \end{align}
 After also adding the scattering with the nuclear target, one finds (the upper label $B$ stays for ``before'')
 \begin{align}\label{gammaBvirt}
\left| \gamma\right\rangle_{out}^{B,\,q\bar q}\,=-|q_3\bar q_3\ran
\lan q_3\bar q_3|\hat S -1 |q_2\bar q_2\ran \,
\frac{\left\langle q_{2}\overline{q}_{2}\right|H_{_{\rm QCD}}\left|q_{1}\overline{q}_{1} g_1\right\rangle }
{E_{q_2\overline{q}_2}-E_{\gamma}}
\frac{\lan q_1\bar q_1 g_1|H_{_{\rm QCD}}|q\,\bar q\ran}{E_{q_1\overline{q}_1g_1}-E_{\gamma}}
\,\frac{\lan q\bar q| H_{_{\rm QED}}\,|\gamma\ran}{E_{q\overline{q}}-E_{\gamma}}\,,
 \end{align}
 where the $i\epsilon$ prescription was again ignored, since unimportant.

The remaining case, that of the virtual corrections generated via final-state evolution (cf. Fig.~\ref{fig:VirtFS}), turns out to be more subtle.
To understand the difficulty, let us first present the formal respective result, as obtained after expanding the evolution operator 
 $U_{_{\rm QCD}}(\infty, \,0)$ to second order; this reads (the upper label $A$ stays for ``after'')
\begin{align}\label{gammaAvirt0}
\left| \gamma\right\rangle_{out}^{A,\,q\bar q}\,=-|q_3\bar q_3\ran
  \frac{\left\langle q_{3}\overline{q}_{3}\right|H_{_{\rm QCD}}\left|q_{2}\overline{q}_{2} g_2\right\rangle }{E_{q_{2}\overline{q}_{2}g_2}-
  E_{q_{3}\overline{q}_{3}}-i\epsilon}\,\frac{\lan q_2\bar q_2 g_2|H_{_{\rm QCD}}|q_1\bar q_1\ran}{E_{q_1\overline{q}_1}- 
  E_{q_{3}\overline{q}_{3}}-2i\epsilon} \,\lan q_1\bar q_1|\hat S -1 |q\bar q\ran \,
\,\frac{\lan q\bar q| H_{_{\rm QED}}\,|\gamma\ran}{E_{q\overline{q}}-E_{\gamma}}\,.
\end{align}
At a first sight, this contribution has the expected structure; e.g., the energy denominators associated with the QCD
transitions are built as differences between the energy of the state prior to the parton branching and that
of the final state $|q_3\bar q_3\ran$. Consider however the case of a self-energy correction,
where the gluon is emitted and reabsorbed by the same fermion --- quark or antiquark. 
In that case, the kinematics of the emitter is clearly the same before and after the loop, hence the intermediate state
$|q_1\bar q_1\ran$ is degenerate with the final state, $E_{q_1\overline{q}_1}=  E_{q_{3}\overline{q}_{3}}$,
and the respective denominator vanishes. This explains why we have carefully kept the $i\epsilon$ pieces in the QCD
energy denominators in \eqn{gammaAvirt0}. 

This difficulty is in fact well known: it is related to the proper definition of the final state (the ``wavefunction renormalisation''). A prescription to circumvent this difficulty has been proposed in \cite{Chen:1995pa}. 
For the physical problem at hand, it will be further discussed in the subsequent paper \cite{Yair:virtual},
which will be fully devoted to the virtual corrections.
From now on, in this paper we shall restrict ourselves to real gluon emissions alone.

\section{Dihadron production in DIS at leading order}
\label{sec:LO}

\begin{figure}
  \centering
    \includegraphics[width=0.9\textwidth]{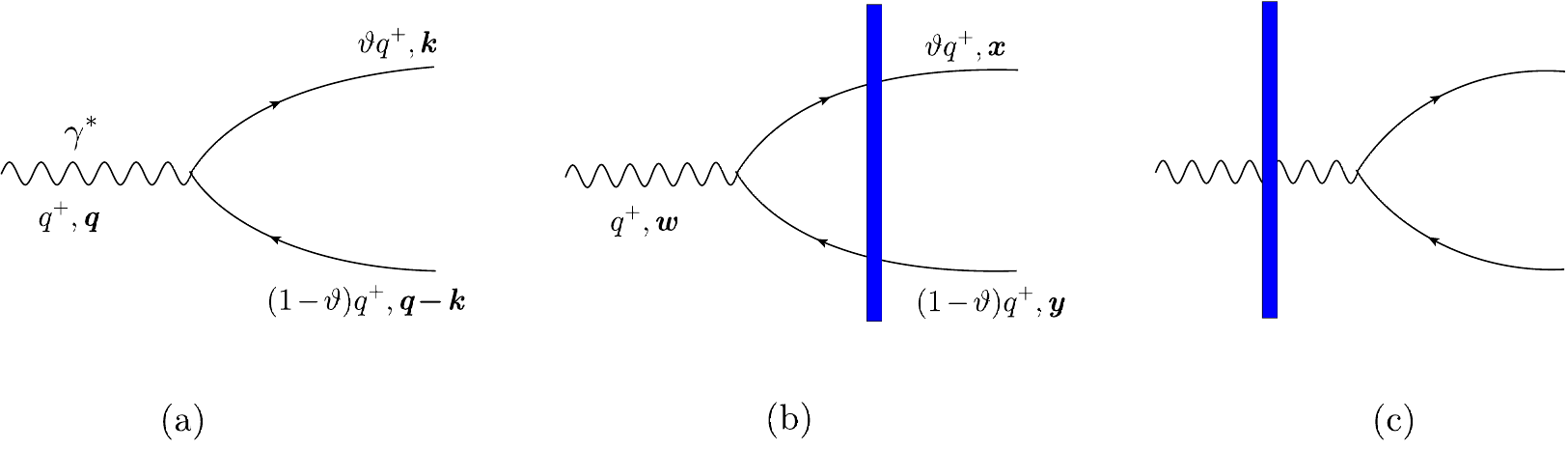}
  \caption{\small (a) The quark-antiquark fluctuation of the virtual photon. (b,c) The 2 graphs contributing
  to the leading-order scattering amplitude in coordinate space. }
    \label{fig:LO}
    \end{figure}

As a warm-up, in this section we shall derive the well-known result for the dijet (or dihadron) production 
in deep inelastic scattering at leading order \cite{Dominguez:2011wm}, 
by following the LCWF formalism outlined in Sect.~\ref{sec:LCWF}. The corresponding Feynman graphs 
are shown in Fig.~\ref{fig:LO}.

\subsection{The leading order photon outgoing state}

The outgoing state at leading order (LO) is shown in compact but formal notations in \eqn{gamma0qq}. In this
section, we shall explicitly compute this state, via the following strategy: First, we shall construct the 
quark-antiquark Fock state generated by the decay of the virtual photon, by working in
momentum space, where the Feynman rules of LCPT are most conveniently formulated; 
that is, we shall evaluate the graph in Fig.~\ref{fig:LO}.a.
Then, we shall perform a Fourier transform to the transverse coordinate representation, 
which is more convenient for computing the action of the $S$-matrix in the eikonal approximation;
we shall thus deduce the contributions of the graphs in Fig.~\ref{fig:LO}.b and c.

To LO, the  $q\bar q $ component of  the virtual photon LCWF reads
\begin{equation}\label{loexp}
\left|\gamma_{\lambda}(Q, q)\right\rangle_{q\overline{q}}^{(0)}
=\int_{0}^{\infty}\frac{dp^{+}}{2\pi}\,\frac{dk^{+}}{2\pi}\int\frac{d^{2}\bm{p}}{(2\pi)^{2}}\,\frac{d^{2}\bm{k}}{(2\pi)^{2}}\,\left|\overline{q}_{\lambda_{2}}^{\beta}(p)
q_{\lambda_{1}}^{\alpha}(k)\right\rangle \frac{\left\langle q_{\lambda_{1}}^{\alpha}(k)\overline{q}_{\lambda_{2}}^{\beta}(p)\left|\,{H}_{\gamma\rightarrow q\overline{q}}
\,\right|\gamma_{\lambda}(q)\right\rangle }{E_{q}(p) + E_{\overline{q}}(k)-E_{\gamma}(q, Q)}\,,
\end{equation}
with the following notations: $q=(q^{+}, \bm{q})$  is the 3-momentum of the incoming photon with virtuality $Q$, whereas  $k\equiv (k^+,\bk)$  and $p\equiv (p^+,\bm{p})$  similarly refer to the final quark and anti-quark.  Furthermore,
$\lambda=T$ or $L$  denotes the photon polarisation state, $\alpha,\beta$ are the fermion colour indices and $\lambda_1, \lambda_2$ are the respective helicity states. The matrix element in \eqn{loexp} involves
delta-functions for  3-momentum conservation, which imply $\bp=\bq-\bk$ and $p^+=q^+-k^+$.
The energy denominator is computed as
\begin{equation}\label{deno}
E_{q}(p) + E_{\overline{q}}(k)-E_{\gamma}(q, Q)\,=\,\frac{\bm{p}^{2}}{2p^{+}}
\,+\,\frac{\bm{k}^{2}}{2k^{+}}\,-\,\frac{\bm{q}^{2}-Q^{2}}{2q^{+}}
\,=\,\frac{\bm{\widetilde{k}}^{2}+\bar Q^{2}}{2\vartheta(1-\vartheta)q^{+}}\,,
\end{equation}
where $\vartheta$ is the quark longitudinal momentum fraction, $\bm{\widetilde{k}}$
is the  transverse momentum of the quark relative to its parent photon, and $\bar Q^2$
is a measure of the virtuality of the $q\bar q$ pair (see below):
\begin{equation}\label{newvar}
\vartheta\,\equiv\,\frac{k^{+}}{q^{+}}\,,\qquad \bm{\widetilde{k}}\equiv \bm{k}-\vartheta\bm{q}\,,
\qquad\bar Q^2\,\equiv\,\vartheta(1-\vartheta) Q^2\,.
\end{equation}

We shall consider in more detail the case of a virtual photon with transverse polarisation.
The relevant matrix element reads (for a quark flavour with electric charge $e_f$)
\begin{align}\label{gammaqq}
\left\langle q_{\lambda_{1}}^{\alpha}(p)\overline{q}_{\lambda_{2}}^{\beta}(k)\left|
{H}_{\gamma\rightarrow q\overline{q}}\right|\gamma_{T}^{i}(q)\right\rangle&\, =
(2\pi)^{3}\delta^{(3)}(q-k-p)\,\frac{ee_{f}\delta_{\alpha\beta}}{2\sqrt{2q^+}}\chi_{\lambda_{1}}^{\dagger}\left[
\frac{2\bm{q}^{i}}{q^{+}}-\frac{\bm{\sigma}\cdot\bm{k}}{k^{+}}\sigma^{i}-\sigma^{i}\frac{\bm{\sigma}\cdot\bm{p}}{p^+}\right]\chi_{\lambda_{2}}\nonumber\\*[0.2cm]
&\, =(2\pi)^{3}\delta^{(3)}(q-k-p)\,
\frac{e\,e_{f}\,\delta_{\alpha\beta}\,\varphi_{\lambda_{1}\lambda_{2}}^{ij}(\vartheta)}{2\sqrt{2(q^{+})^{3}}\,\vartheta(1-\vartheta)}\,\bm{\widetilde{k}}^{j}\,,
\end{align}
where $\chi_{\lambda}$ with $\lambda=\pm 1/2$ are the  usual helicity states and in obtaining the second line
we have used the following identity ($\sigma^i$ with $i=1,2,3$ are the usual Pauli matrices)
\begin{equation}
\frac{2\bm{q}^{i}}{q^{+}}-\frac{\bm{\sigma}\cdot\bm{k}}{k^{+}}\sigma^{i}-\sigma^{i}\frac{\bm{\sigma}\cdot(\bm{q}-\bm{k})}{q^{+}-k^{+}}\,=\,\frac{(2\vartheta-1)\delta^{ij}+i\varepsilon^{ij}\sigma^{3}}{\vartheta(1-\vartheta)q^{+}}\,(\bm{k}^{j}-\vartheta\bm{q}^{j}),
\end{equation}
together with the following definition:
\begin{equation}\label{phidef}
\varphi_{\lambda_{1}\lambda_{2}}^{ij}(\vartheta)\,\equiv\,\chi_{\lambda_{1}}^{\dagger}\left[(2\vartheta-1)\delta^{ij}+i\varepsilon^{ij}\sigma^{3}\right]\chi_{\lambda_{2}}\,=\,\delta_{\lambda_{1}\lambda_{2}}\left[(2\vartheta-1)\delta^{ij}+2i\varepsilon^{ij}\lambda_{1}\right].
\end{equation}
The coefficient $2\vartheta-1$ multiplying $\delta^{ij}$ in \eqn{phidef} arises as the difference between the
longitudinal momentum fractions, $\vartheta$ and $1-\vartheta$, of the quark and
the antiquark produced by the photon decay.

After inserting (\ref{gammaqq}) and (\ref{deno}) into Eq.~(\ref{loexp}), 
using 3-momentum conservation and replacing $\bm{k}\to \bm{\widetilde{k}}$ as the integration variable,
 one finds
\begin{align}\label{lowf}\hspace*{-0.5cm}
\left|\gamma_{T}^{i}(Q, q^{+}, \bm{q})\right\rangle_{q\overline{q}}^{(0)}
=\int_{0}^{1}d\vartheta\int d^{2}\bm{\widetilde{k}}\:\frac{ee_{f}\varphi_{\lambda_{1}\lambda_{2}}^{ij}(\vartheta)\,\sqrt{q^{+}}\,\bm{\widetilde{k}}^{j}}{(2\pi)^{3}\sqrt{2}\,\big(\bm{\widetilde{k}}^{2}+\bar Q^{2}\big)}\left|\bar{q}_{\lambda_{2}}^{\alpha}\big((1-\vartheta)q^{+}, (1-\vartheta)\bm{q}-\bm{\widetilde{k}}\big)\,q_{\lambda_{1}}^{\alpha}\big(\vartheta q^{+}, \bm{\widetilde{k}}+\vartheta\bm{q}\big)\right\rangle .
\end{align}
Its Fourier transform  to the transverse coordinate representation is readily obtained as
\begin{align}\label{lowfx}
\left|\gamma_{T}^{i}(Q, q^{+}, \bm{w})\right\rangle_{q\overline{q}}^{(0)}
&\,\equiv\int\frac{d^{2}\bm{q}}{(2\pi)^{2}}\,\rme^{-i\bm{w}\cdot\bm{q}}\left|\gamma_{T}^{i}(Q, q^{+}, \bm{q})\right\rangle_{q\overline{q}}^{(0)}\nonumber\\*[0.2cm]
&\, =\int_{\bm{x}, \bm{y}}\int_{0}^{1}d\vartheta\ \frac{iee_{f}\varphi_{\lambda_{1}\lambda_{2}}^{ij}(\vartheta)\sqrt{q^{+}}}{(2\pi)^{2}\sqrt{2}}\,\frac{\bm{R}^{j}}{R}\,\bar{Q}\,\rmK_{1}\left(\bar{Q}R\right)
\nonumber\\*[0.2cm]
&\qquad \times\delta^{(2)}\left(\bm{w}-\bm{c}\right)\left|\overline{q}_{\lambda_{2}}^{\alpha}\big((1-\vartheta)q^{+}, \bm{y}\big) q_{\lambda_{1}}^{\alpha}\big(\vartheta q^{+}, \bm{x}\big)\right\rangle.
\end{align}
Here $\bx$ and $\by$ are the transverse coordinates of the quark and the antiquark, respectively,
$\bm{R}\equiv\bm{x}-\bm{y}$ is their relative separation, $R=|\bm{R}|$, $\bm{c}=\vartheta\bm{x}+(1-\vartheta)\bm{y}$ is
 their center of energy; we have used the shorthand notation $\int_{\bm{x}}\equiv\int d^{2}\bm{x}$ for the transverse
 integrations.
 
At this level, it is straightforward to include the effect of the scattering with the nuclear target
(the shockwave) in the eikonal approximation:
the transverse coordinates, the longitudinal momenta, and the helicity states 
of the quark and the antiquark remain unchanged, but their colour states
get rotated by the interaction with the colour field generated by the left-moving partons from the target:
\beq
\hat S\left|\overline{q}^{\alpha}_{\lambda_{2}}
\big(p^+, \bm{y}\big) q_{\lambda_{1}}^{\alpha}\big(k^{+}, \bm{x}\big)\right\rangle\,=\,
\left(V(\bm{x})\,V^{\dagger}(\bm{y})\right)_{\alpha\beta}\,
\left|\overline{q}^{\beta}_{\lambda_{2}}
\big(p^+, \bm{y}\big) q_{\lambda_{1}}^{\alpha}\big(k^{+}, \bm{x}\big)\right\rangle
\eeq
where $V(\bm{x})$ and $V^{\dagger}(\bm{y})$ are Wilson lines in the fundamental representation of
the colour group:
\beq\label{SA}
 V(\bm{x})=\,{\rm T}\exp\left\{ ig\int dx^{+}\, t^{a}A^{-}_a(x^{+},\bm{x})\right\}, \qquad   
U(\bm{x})=\,{\rm T}\exp\left\{ ig\int dx^{+}\,T^{a}A_{a}^{-}(x^{+},\bm{x})\right\}.
    \eeq
  We have also introduced here the Wilson line $U(\bm{x})$ in the adjoint representation, for
   later convenience (this describe the colour precession of a gluon). In these equations,  $A^{-}_a$ is
   the colour field  representing Coulomb exchanges between the quark or the antiquark from
   the dipole and colour sources (quark and gluons) from the target.
   In the CGC effective theory, this field is random and must be averaged out at the level of the cross-section
   (see below).

Thus, finally, the $q\bar q$ outgoing component of the LCWF of a transverse virtual photon reads\footnote{As
compared to  \eqn{gamma0qq}, we renounce to the upper label {\it out}, to simplify the notation.}
\begin{align}\label{out_lo}
\left|\gamma_{T}^{i}(Q, q^{+}, \bm{w})\right\rangle_{q\overline{q}}^{(0)}
&\,=\,\int_{\bm{x}, \bm{y}}\,\int_{0}^{1}d\vartheta\ 
\frac{ie\,e_{f}\,\varphi_{\lambda_{1}\lambda_{2}}^{ij}(\vartheta)\,\sqrt{q^{+}}}{(2\pi)^{2}\sqrt{2}}\,
\frac{\bm{R}^{j}}{R}\,\bar{Q}\,\rmK_{1}(\bar{Q}R)\nonumber\\
&\,\times\left(V(\bm{x})\,V^{\dagger}(\bm{y})-1\right)_{\alpha\beta}\,\delta^{(2)}\left(\bm{w}-\bm{c}\right)\left|\overline{q}_{\lambda_{2}}^{\beta}((1-\vartheta)q^{+}, \bm{y})\,q_{\lambda_{1}}^{\alpha}(\vartheta q^{+}, \bm{x})\right\rangle .
\end{align}
This is a fully explicit version of the outgoing state \eqref{gamma0qq}, as written in a mixed Fourier representation
(longitudinal momentum and transverse coordinates). We recall that the subtraction of unity from the product
of Wilson lines inside the square bracket accounts for the process where the photon decays into a $q\bar q$
pair {\it after} crossing the shockwave, in which case there is no scattering  (see Fig.~\ref{fig:LO}.c).


Similarly, for a virtual photon with longitudinal polarisation we find the following result:
\begin{align}\label{outLO}
\left|\gamma_{L}(Q, q^{+}, \bm{w})\right\rangle_{q\overline{q}}^{(0)}
&\,=\,\int_{\bm{x}, \bm{y}}\,\int_{0}^{1}d\vartheta\,\frac{i\,e\,e_{f}\,\sqrt{q^{+}}}{(2\pi)^{2}\,\sqrt{2}}\,\bar{Q}\,\rmK_{0}(\bar{Q}R)\nonumber\\
&\times\left(V(\bm{x})\,V^{\dagger}(\bm{y})-1\right)_{\alpha\beta}\,\delta^{(2)}\left(\bm{w}-\bm{c}\right)\left|\overline{q}_{\lambda}^{\beta}((1-\vartheta)q^{+}, \bm{y})\,q_{\lambda}^{\alpha}(\vartheta q^{+}, \bm{x})\right\rangle .
\end{align}

\subsection{The dihadron cross-section at leading order \label{LO.res}}

Given the outgoing LCWF as computed in the previous section, we are now in a position to compute the
cross-section for dijet (or dihadron) production at leading order. In this approximation, the final ``jets'' are simply
the quark and the antiquark produced by the decay of the virtual photon and which are put on-shell by
their scattering off the hadronic target. The corresponding cross-section is obtained by simply
counting the number of $q\bar q$ pairs with a given kinematics in the outgoing state. For a transverse photon,
we can write
\beq
\frac{d\sigma_{T}^{\gamma A\rightarrow q\bar q+X}}
{dk_1^+d^{2}\bm{k}_{1}dk_2^+d^{2}\bm{k}_{2}}\,(2\pi)\delta(q^{+}-k_1^+-k_2^+)\,=\,
\frac{1}{2}\ \,
{}_{q\bar q}^{(0)}\!\left\langle \gamma_{T}^{i}(Q, q)\right|\,\mathcal{N}_{q}(k_{1})\,
\mathcal{N}_{\overline{q}}(k_{2})\,\left|\gamma_{T}^{i}(Q, q)\right\rangle_{q\bar q}^{(0)},
\eeq
where $\mathcal{N}_{q}(k_{1})$ and $\mathcal{N}_{\overline{q}}(k_{2})$ are particle number density operators
for bare quarks and antiquarks, and 
 the overall factor 1/2 comes from the average over the 2 transverse polarisations. The outgoing photon state
 in momentum space can be obtained by inverting the Fourier transform in \eqn{lowfx}. This gives
 \begin{equation}\begin{split}\label{defcross}
&\frac{d\sigma_{T}^{\gamma A\rightarrow q\bar q+X}}
{dk_1^+\,d^{2}\bm{k}_{1}\,dk_2^+\,d^{2}\bm{k}_{2}}\,(2\pi)\,\delta(q^{+}-k_1^+-k_2^+)\\
&\qquad\qquad =\,\frac{1}{2}\int d^{2}\bm{\overline{w}}\,d^{2}\bm{w}\ 
{}_{q\bar q}^{(0)}\!\left\langle \gamma_{T}^{i}(Q, q^{+}, \bm{\overline{w}})\right|\,\mathcal{N}_{q}
(k_{1})\,\mathcal{N}_{\overline{q}}(k_{2})\,\left|\gamma_{T}^{i}(Q, q^{+}, \bm{w})\right\rangle _{q\bar q}^{(0)},
\end{split}\end{equation}
where $\bm{w}$ and $\bm{\overline{w}}$ are the transverse coordinates of the virtual photon in the direct
amplitude (DA) and the complex conjugate amplitude (CCA), respectively, and
we have also set $\bq=0$ (this entails no loss of generality). The integrations over  $\bm{w}$
and $\bm{\overline{w}}$ simply remove the delta-functions like $\delta^{(2)}\left(\bm{w}-\bm{c}\right)$ in
\eqn{out_lo}. To proceed, it is preferable to stick to the transverse coordinate representation, that is, to use \eqn{out_lo} for the outgoing state together with the Fourier transform of the number density operators. E.g. for the quarks,
 we write
 \begin{equation}
\label{Nq}
\hat{\mathcal{N}}_{q}(k^{+},\bm{k})\,=\,\frac{1}{(2\pi)^{3}}\,b_{\lambda}^{\alpha\dagger}(k^{+},\bm{k})\,b_{\lambda}^{\alpha}(k^{+},\bm{k})\,=\,\frac{1}{(2\pi)^{3}}\int_{\overline{\bm{x}},\bm{x}}e^{i\bm{k}\cdot(\overline{\bm{x}}-\bm{x})}\,b_{\lambda}^{\alpha\dagger}(k^{+},\overline{\bm{x}})\,b_{\lambda}^{\alpha}(k^{+},\,\bm{x}),
\end{equation}
where the (bare) quark creation and annihilation operators satisfy the anti-commutation relation
 \begin{equation}
\left\{ b_{\lambda_{1}}^{\alpha}(k^{+},\,\bm{x}),\, b_{\lambda_{2}}^{\beta\dagger}(p^{+},\,\bm{y})\right\} =\,2\pi\,\delta_{\lambda_{1}\lambda_{2}}\,\delta^{\alpha\beta}\,\delta^{(2)}(\bm{x}-\bm{y})\,\delta(k^{+}-p^{+}).
\end{equation}

By using these relations  together with the corresponding one for the antiquarks, one  finds
\begin{equation}\begin{split}\label{transcross}
&\frac{d\sigma_{T}^{\gamma A\rightarrow q\bar q+X}}{dk_1^+d^{2}\bm{k}_{1}dk_2^+d^{2}\bm{k}_{2}}\,=\,\frac{2\alpha_{em}\,N_{c}}{(2\pi)^{6}q^{+}}\left(\vartheta^{2}+(1-\vartheta)^{2}\right)\,\left(\sum e_{f}^{2}\right)\,\delta(q^{+}-k_1^+-k_2^+)\\
&\times\,\int_{\overline{\bm{x}}, \bm{\overline{y}}, \bm{x}, \bm{y}}\,e^{-i\bm{k}_{1}\cdot(\bm{x}-\overline{\bm{x}})-i\bm{k}_{2}\cdot(\bm{y}-\overline{\bm{y}})}\,\frac{\bm{R}\cdot\bm{\overline{R}}}{R\,\overline{R}}\,\bar{Q}^{2}\,\rmK_{1}\left(\bar{Q}R\right)\,\rmK_{1}\left(\bar{Q}{\overline{R}}\right)\mathscr{W}\left(\bm{x}, \bm{y}, \bm{\overline{y}}, \bm{\overline{x}}\right),
\end{split}\end{equation}
with $\bm{\overline{R}}\equiv\bm{\overline{x}}-\bm{\overline{y}}$ and $\vartheta\equiv\frac{k_1^+}{q^{+}}$.
The sum over helicities has been performed as (cf. \eqn{phidef})
\beq
\varphi_{\lambda_{1}\lambda_{2}}^{ij}(\vartheta)\,
\varphi_{\lambda_{1}\lambda_{2}}^{il\,*}(\vartheta)\,
=\,2\delta^{jl}\left[1+(1-2\vartheta)^2\right]\,
=\,4\delta^{jl}\left[\vartheta^2+(1-\vartheta)^2\right].
\eeq
In writing \eqn{transcross}, we have also performed the average over the random colour fields $A^-_a$
in the target. This has generated 
the function $\mathscr{W}\left(\bm{x}, \bm{y}, \bm{\overline{y}}, \bm{\overline{x}}\right)$, which encodes the
effects of the scattering:
\begin{equation}\label{LOcolor}
\mathscr{W}\left(\bm{x}, \bm{y}, \bm{\overline{y}}, \bm{\overline{x}}\right)\,\equiv\,\mathcal{Q}\left(\bm{x}, \bm{y}, \bm{\overline{y}}, \bm{\overline{x}}\right)-\mathcal{S}\left(\bm{x}, \bm{y}\right)-\mathcal{S}\left(\bm{\overline{y}}, \bm{\overline{x}}\right)+1.
\end{equation}
The 3  non-trivial terms in the r.h.s. are (average) $S$-matrices describing the forward scattering of 
colourless systems made with up to four partons: a $q\bar q$ dipole in the DA,
\begin{equation}\label{lowils2}
\mathcal{S}\left(\bm{x}, {\bm{y}}\right)\,\equiv\,\frac{1}{N_{c}}\,\left\langle \mathrm{tr}\big(V(\bm{x})\,
V^{\dagger}({\bm{y}})\big)\right\rangle ,
\end{equation}
a similar $q\bar q$ dipole, $\mathcal{S}\left(\bm{\overline{y}}, \bm{\overline{x}}\right)$, in the CCA, and the
$q\bar q q\bar q$ quadrupole,
\begin{equation}
\label{quadrupole}
\mathcal{Q}\,(\bm{x}, \bm{y}, \overline{\bm{y}}, \overline{\bm{x}})\,\equiv\,\frac{1}{N_{c}}\,\left\langle
\mathrm{tr}\big(V(\bm{x})\,V^{\dagger}(\bm{y})\,V(\overline{\bm{y}})\,
V^{\dagger}(\overline{\bm{x}})\big)\right\rangle,
\end{equation}
where the colour flows connects the $q\bar q$ pair in the DA to that in the CCA.

%
For a virtual photon with longitudinal polarisation one similarly finds (cf. \eqn{outLO})
\begin{equation}\begin{split}\label{difflocross}
&\frac{d\sigma_{L}^{\gamma A\rightarrow q\bar q+X}}{dk_1^+d^{2}\bm{k}_{1}dk_2^+d^{2}\bm{k}_{2}}\,=\,\frac{8\alpha_{em}\,N_{c}}{(2\pi)^{6}q^{+}}\vartheta(1-\vartheta)\,\left(\sum e_{f}^{2}\right)\,\delta(q^{+}-k_1^+-k_2^+)\\
&\qquad\times\,\int_{\overline{\bm{x}}, \bm{\overline{y}}, \bm{x}, \bm{y}}\,e^{-i\bm{k}_{1}\cdot(\bm{x}-\overline{\bm{x}})-i\bm{k}_{2}\cdot(\bm{y}-\overline{\bm{y}})}\,\bar{Q}^{2}\,\rmK_{0}\left(\bar{Q}R\right)\,\rmK_{0}\left(\bar{Q}\overline{R}\right)\,\mathscr{W}\left(\bm{x}, \bm{y}, \bm{\overline{y}}, \bm{\overline{x}}\right).
\end{split}\end{equation}
Eqs.~\eqref{transcross} and \eqref{difflocross} are in agreement with previous calculations in the literature
 \cite{Dominguez:2011wm}.

To obtain the corresponding cross-sections for di-hadron production, $\gamma A\rightarrow h_1h_2+X$, is suffices
to convolute the above results for $q\bar q$ production with the fragmentation functions describing the probability
to find a hadron within the wavefunction of a quark, or antiquark:
\begin{align}
\label{2qFF}
\frac{d\sigma^{\gamma A\rightarrow h_1h_2+X}_\lo}{dk_1^+d^{2}\bm{k}_{1}dk_2^+d^{2}\bm{k}_{2}}
= \int \frac{d\zeta_1}{\zeta_1^3} \int \frac{d\zeta_2}{\zeta_2^3}\,
\frac{d\sigma_{\lo}^{\gamma A\rightarrow q\bar q+X}}
{dp_1^+d^{2}\bm{p}_{1}dp^+_2d^{2}\bm{p}_2}\Bigg |_{p_i=\frac{k_i}{\zeta_i}}
D_{h_1/q}(\zeta_1,\mu^{2})\,D_{h_2/\bar q}(\zeta_2,\mu^{2})\,.
 \end{align}
The lower script $p_i=k_i/\zeta_i$ on the $q\bar q$ cross-section stands for $p^+_i=k_i^+/\zeta_i$ and 
$\bm{p}_i=\bm{k}_i/\zeta_i$, with $i=1,\,2$.

\section{The tri-parton component of the transverse photon outgoing state}
\label{sec:qqgLCWF}

In this section, we compute the tri-parton (quark, antiquark, and gluon) 
Fock-space component $\left|\gamma_T^i\right\rangle_{q\overline{q}g}$
of the outgoing state produced by the evolution and the scattering
of an incoming state representing a virtual photon with transverse polarisation.
At leading order, this component involves two parton branchings: the decay of the virtual photon 
into a quark--antiquark pair ($\gamma_T\to q\bar q$), followed by
the emission of a gluon from either the quark ($q\to qg$), or the antiquark ($\overline{q}\to \overline{q} g$).
In practice, it is enough to consider one of these 2 cases --- say, gluon emission by the antiquark.
The contribution of the other case can then be simply obtained via symmetry operations.
Also, as explained in Sect.~\ref{sec:LCWF},  there is no need to explicitly compute the graphs 
where the photon decays after crossing the shockwave (see Fig.~\ref{fig:NLO_Real}.c)
--- their effects can be accounted for by subtracting the no-scattering limit $\hat S\to 1$
from the other contributions.
So, in practice, it is enough to compute the two graphs in Fig.~\ref{fig:NLO_Real}.a and b, together with
the corresponding ``instantaneous'' graphs, where the intermediate antiquark line is replaced with the
instantaneous piece of the fermion propagator (see Fig.~\ref{fig:qbar-inst}).

%
%

\subsection{Gluon emission by the antiquark}

We shall describe in detail the case where the gluon is emitted by the antiquark. We start with the regular graphs
in Figs.~\ref{fig:NLO_Real}.a and b.  The respective contributions to the $q\bar q g$ Fock-space component 
of the outgoing state can be inferred from  \eqn{outgammaRe}:
\begin{align}\label{qqgoutab}
\left| \gamma\right\rangle^{(a+b)}_{q\overline{\bm{q}}g}\,=\,
\left|q_{2}\overline{q}_{2}g_2\right\rangle \Bigg\{ &
\lan q_2\bar q_2 g_2|\hat S |q_1\bar q_1 g_1\ran \,
\frac{\lan q_1\bar q_1 g_1|{H}_{\overline{q}\rightarrow\overline{q}g}
|q\bar q\ran}{E_{q_1\overline{q}_1g_1}-E_{\gamma}}
\nonumber \\*[0.2cm]
& +\frac{\langle q_{2}\overline{q}_{2}g_2
|H_{\overline{q}\rightarrow\overline{q}g}|
{q_{1}}{\overline{q}_{1}}\rangle }{E_{{q_{1}}{\overline{q}_{1}}}-E_{q_{2}\overline{q}_{2}g_2}}
\,\langle q_{1}{\overline{q} _{1}}  |\hat S| q\overline{q}\rangle \Bigg\}
\frac{ \lan q \bar q | {H}_{\gamma\rightarrow q\overline{q}}|\gamma\ran}{E_{q\overline{q}}-E_{\gamma}}.
\end{align}
where the first  (second) term within the accolades describes gluon emission prior (after) the shockwave.
The QCD matrix elements ensure that $q_1=q$ in the first term and, respectively, $q_2=q_1$ in the second term.
Notice the subscript ${q\overline{\bm{q}}g}$ on the LCWF in the l.h.s.: the antiquark component is shown
in boldface to emphasise that this is the fermion which emits the gluon.

As for the leading-order calculation in the previous section, we shall first compute the LCWF in
the absence of scattering ($\hat S\to 1$) and in transverse momentum space. Then we shall construct its Fourier transform
to the transverse coordinate representation. Then it will be easy to insert the effects of the collision,
in the form of Wilson lines. Notice that, even after replacing $\hat S\to 1$, the two terms in 
\eqn{qqgoutab} are not identical:
they differ in the structure of their energy denominators, which in turn reflects the different time-orderings
of the gluon emission w.r.t. $x^+=0$ (the time of scattering).

\begin{figure}
  \centering
    \includegraphics[width=0.8\textwidth]{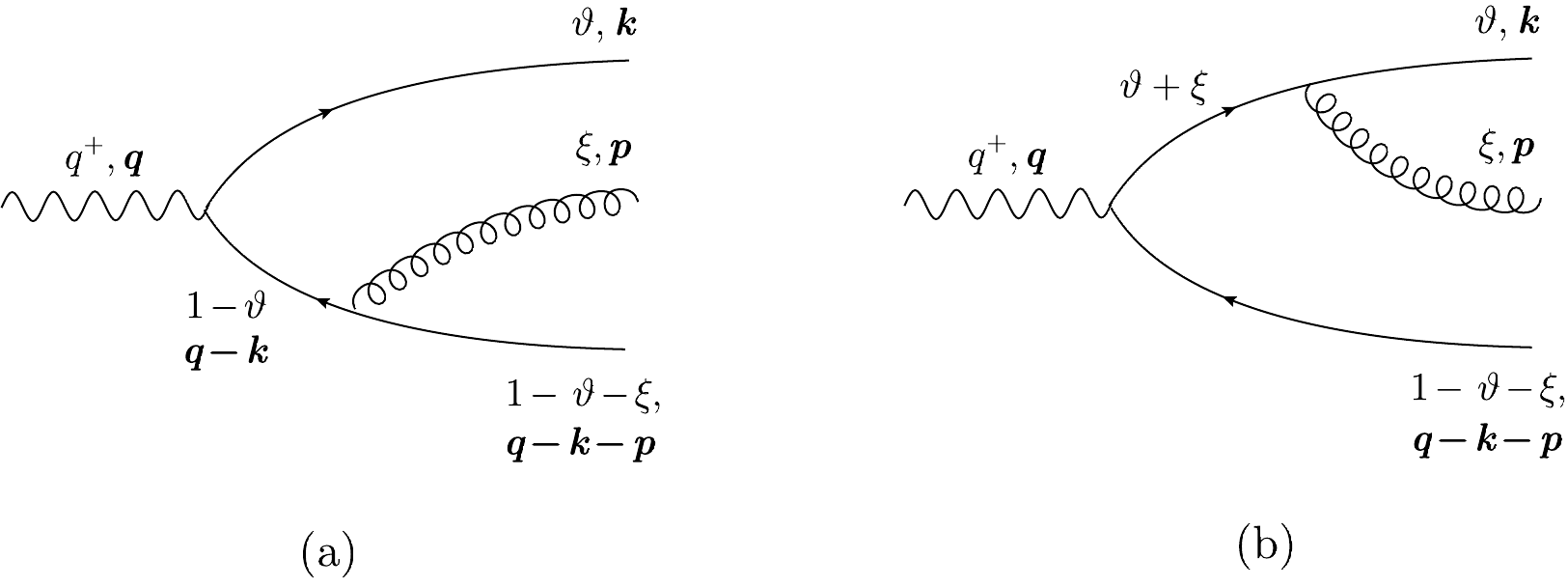}
  \caption{\small Gluon emission by the antiquark (a) or by the quark (b), in
  the transverse momentum representation.}
    \label{fig:Real_kT}
    \end{figure}

Our conventions for the momentum-space kinematics are summarised in Fig.~\ref{fig:Real_kT}.a.
Let us first work out the relevant energy denominators. They can all be constructed
form the LC energy differences at the emission vertices. For the photon decay
$\gamma\to q\bar q$, we have (we recall that $\bar Q^2=\vartheta(1-\vartheta) Q^2$)
\begin{align}\label{enden1}
E_{\overline{q}q}\,-\,E_{\gamma}&\,=\,\frac{1}{2q^+}\left[
\frac{\bm{k}^2}{\vartheta} + \frac{(\bm{q}-\bm{k})^2}{1-\vartheta}
 -\bm{q}^2 + Q^2\right]
\,=\,\frac{(\bm{k}-\vartheta\bm{q})^{2}+\bar{Q}^{2}}{2\vartheta(1-\vartheta)q^{+}},
\end{align}
whereas for the gluon emission $\overline{q}\to \overline{q} g$,
\begin{align}\label{enden2}
E_{q\overline{q}g}-E_{\overline{q}q}\,=\,E_{\overline{q}g}-E_{\overline{q}}&\,=\,\frac{1}{2q^+}\left[
\frac{\bm{p}^2}{\xi} + 
\frac{(\bm{q}-\bm{k}-\bm{p})^2}{1-\vartheta-\xi}  -
 \frac{(\bm{q}-\bm{k})^2}{1-\vartheta}
\right]
\nonumber\\*[0.2cm]
&\,=\,\frac{\left[(1-\vartheta)\bm{p}-\xi(\bm{q}-\bm{k})\right]^2
}{2\xi(1-\vartheta)(1-\vartheta-\xi)q^{+}}\,.
\end{align}
As expected, this energy difference would vanish for a collinear decay, i.e. in the case
where the gluon splitting fraction $z\equiv \xi/(1-\vartheta)$ controls not only its
longitudinal momentum ($p^+=z(q^+-k^+)$), but also its transverse momentum
($\bm{p}=z(\bm{q}-\bm{k})$).
The remaining energy denominator involves the sum of the two energy differences
written above:
\begin{align}\label{enden3}
E_{q\overline{q} g}-E_{\gamma}
\,=\,\frac{\vartheta(1-\vartheta)^2\bm{\widetilde{p}}^2\,+\,\xi(1-\vartheta-\xi) \big(\bm{\widetilde{k}}^{2}\,+\,
\bar{Q}^{2}\big)}{2\vartheta\xi(1-\vartheta)(1-\vartheta-\xi)q^{+}}\,,
\end{align}
where we introduced the ``shifted'' momenta
\beq\label{tildekp}
\bm{\widetilde{k}}\equiv\bm{k}-\vartheta\bm{q},\qquad\bm{\widetilde{p}}\equiv
\bm{p}-\frac{\xi}{1-\vartheta}(\bm{q}-\bm{k})\,,
\eeq
which physically express the deviations from collinearity
at the emission vertices and which are convenient to use
as integration variables when summing over the final states (see below).

%

The  matrix element for the virtual photon decay has already been shown in \eqn{gammaqq}.
That for the gluon emission from the antiquark reads \cite{Iancu:2020mos}
\begin{equation}\label{Hbarq}
\begin{split}
&\left\langle q_{\lambda_{3}}^{\gamma}(u)\,\overline{q}_{\lambda_{4}}^{\delta}(t)\,g_{i}^{a}(l)\left|{H}_{\overline{q}\rightarrow\overline{q}g}\right|q_{\lambda_{1}}^{\alpha}(s)\,\overline{q}_{\lambda_{2}}^{\beta}(k)\right\rangle \\
&=-(2\pi)^{6}\delta^{(3)}(s-u)\,\delta^{(3)}(k-t-l)\delta_{\gamma\alpha}
\delta_{\lambda_{3}\lambda_{1}}\,
\frac{gt_{\beta\delta}^{a}}{2\sqrt{2l^{+}}}\,
\chi_{\lambda_{2}}^{\dagger}\left[\frac{2\bm{l}^{i}}{l^{+}}-\sigma^{i}\frac{\bm{\sigma}\cdot\bm{t}}{t^{+}}-\frac{\bm{\sigma}\cdot\bm{k}}{k^{+}}\sigma^{i}\right]\chi_{\lambda_{4}}.
\end{split}\end{equation}
The structure of this matrix element is consistent with the fact that the antiquark should propagate 
backwards in time: the colour ($\delta$) and spin ($\lambda_{4}$) indices of the daughter antiquark formally
appear as the {\it initial} quantum numbers in the matrix element in the r.h.s. of \eqn{Hbarq}; similarly,
the respective indices $\beta$ and $\lambda_{2}$ of the emitter appear as {\it final} quantum numbers.

For the kinematics shown in Fig.~\ref{fig:Real_kT}, the above spinorial matrix element reads
\begin{equation}\label{Hbarq1}
\chi_{\lambda}^{\dagger}\left[\frac{2\bm{p}^{i}}{p^{+}}-\sigma^{i}\frac{\bm{\sigma}\cdot(\bm{q}-
\bm{k}-\bm{p})}{q^{+}-k^+-p^{+}}-\frac{\bm{\sigma}\cdot(\bm{q}-\bm{k})}{q^{+}-k^{+}}
\sigma^{i}
\right]\chi_{\lambda_{2}}\,=\,\frac{\tau_{\lambda\lambda_{2}}^{ij}(\xi,1-\vartheta-\xi)\,\bm{\widetilde{p}}^{j}}{\xi
(1-\vartheta-\xi)q^{+}}.
\end{equation}
with
\begin{equation}\label{taudef}
\tau_{\lambda\lambda_{2}}^{ij}(\xi,1-\vartheta-\xi)\,\equiv\,\chi_{\lambda}^{\dagger}\left[(2(1-\vartheta)-\xi)\delta^{ij}+i\xi
\varepsilon^{ij}\sigma^{3}\right]\chi_{\lambda_{2}}
\,=\,\delta_{\lambda\lambda_{2}}\left[(2(1-\vartheta)-\xi)\delta^{ij}+2i\xi\varepsilon^{ij}\lambda\right].
\end{equation}
The two arguments of this function are the longitudinal momentum fractions of the daughter partons:
the gluon ($\xi$) and the final antiquark ($1-\vartheta-\xi$).
The coefficient $2(1-\vartheta)-\xi$ multiplying $\delta^{ij}$ is recognised as the sum of the
momentum fractions, $1-\vartheta$ and $1-\vartheta-\xi$, of the antiquark prior and
respectively after the gluon emission.

Consider now the first term in \eqn{qqgoutab} with $\hat S=1$. This is of course the same as 
the $q\overline{q}g$ Fock state at the time of scattering, as shown in Fig.~\ref{fig:Real_kT}.a.
Using the energy denominators \eqref{enden1} and \eqref{enden3} together with the matrix elements in
Eqs.~\eqref{gammaqq} and \eqref{Hbarq}--\eqref{Hbarq1}, one finds
\begin{align}\label{qqgmom}
\left|\gamma^{i}_T
(Q,q^{+},\bm{q})\right\rangle^{(a)} _{q\overline{\bm{q}}g}&\,=-\int d^{2}\bm{\widetilde{k}}\,d^{2}\bm{\widetilde{p}}\,
\int_{0}^{1}d\vartheta\,\int_{0}^{1-\vartheta}d\xi\,
\nonumber\\*[0.2cm]
&\times\frac{ee_{f}g\,{q^{+}}\,\vartheta(1-\vartheta)\,
\varphi_{\lambda_{1}\lambda}^{ij}(\vartheta)\,\bm{\widetilde{k}}^{j}\,
\tau_{\lambda\lambda_{2}}^{mn}(\xi,1-\vartheta-\xi)\,
\bm{\widetilde{p}}^{n}\,t^a_{\alpha\beta}
}{2(2\pi)^{6}\sqrt{\xi}\left(\bm{\widetilde{k}}^{2}\,+\,\bar{Q}^{2}\right)\left[\vartheta(1-\vartheta)^{2}\bm{\widetilde{p}}^{2}\,+\,\xi(1-\vartheta-\xi) \big(\bm{\widetilde{k}}^{2}\,+\,
\bar{Q}^{2}\big)\right]}\nonumber\\*[0.2cm]
&\times\left|\overline{q}_{\lambda_{2}}^{\beta}\left((1-\vartheta-\xi)q^{+},  \bm{q}\!-\!\bm{k}\!-\!\bm{p}
\right)\,g_{m}^{a}\left(\xi q^{+}, \bm{p}\right)\,q_{\lambda_{1}}^{\alpha}\left(\vartheta q^{+}, \bm{k}
\right)\right\rangle,
\end{align}
where the transverse momenta of the final partons can be expressed in terms of 
 the integration variables $\bm{\widetilde{k}}$ and $\bm{\widetilde{p}}$ as follows (cf.  \eqn{tildekp})
\beq\label{finalpt}
\bm{k}=\bm{\widetilde{k}}+\vartheta\bm{q}\,,\quad
 \bm{p}= \bm{\widetilde{p}}+\xi \left(\bm{q}-\frac{\bm{\widetilde{k}}}{1-\vartheta}\right)
\,,\quad \bm{q}\!-\!\bm{k}\!-\!\bm{p}=(1-\vartheta-\xi)\left(\bm{q}-\frac{\bm{\widetilde{k}}}{1-\vartheta}\right)
- \bm{\widetilde{p}}\,.
\eeq

To prepare the Fourier transform to the transverse coordinate representation, let us
first introduce the corresponding representation for the 3-parton final state:
\begin{align}\label{qqgFT}
&\left|\overline{q}_{\lambda_{2}}^{\beta}\left((1-\vartheta-\xi)q^{+},  \bm{q}\!-\!\bm{k}\!-\!\bm{p}
\right)\,g_{m}^{a}\left(\xi q^{+}, \bm{p}\right)\,q_{\lambda_{1}}^{\alpha}\left(\vartheta q^{+}, \bm{k}
\right)\right\rangle=\nonumber\\
&\qquad =\int_{\bm{x}, \bm{y}, \bm{z}}\,  \rme^{i\bm{y}\cdot( \bm{q}-\bm{k}-\bm{p})
+i\bm{z}\cdot\bm{p} +i\bm{x}\cdot\bm{k}}
\left|\overline{q}_{\lambda_{2}}^{\beta}((1-\vartheta -\xi)q^{+}, \bm{y})\,g_{m}^{a}(\xi q^{+}, \bm{z})\,q_{\lambda_{1}}^{\alpha}(\vartheta q^{+}, \bm{x})\right\rangle ,
\end{align}
Also, we shall use 
the notations  $\bm{w}$ and $\bm{y}^\prime$ for the transverse coordinates of the incoming virtual photon
and of the intermediate antiquark, prior to the gluon emission,
respectively (see Fig.~\ref{fig:NLO_Real}.a). These are not independent coordinates: energy-momentum
conservation implies that $\bm{w}$ must coincide with the center-of-energy of the $q\bar q$ pair
produced by the decay of the virtual photon and also with that of the final 3-parton system;
similarly,  $\bm{y}^\prime$ must be the same as the center-of-energy of the $qg$ pair produced by
the decay of the intermediate anti-quark. These conditions imply
\begin{equation}
\label{defyw}
\bm{w}=\vartheta\bm{x}+(1-\vartheta)\bm{y}^\prime\,=\,
\vartheta\bm{x}+(1-\vartheta-\xi)\bm{y}+\xi\bm{z}\,,\qquad
\bm{y}^\prime=\frac{(1-\vartheta-\xi)\bm{y}+\xi\bm{z}}{1-\vartheta}\,.
\end{equation}
Using \eqn{finalpt} and the above relations for  $\bm{w}$ and $\bm{y}^\prime$, one can check that 
the exponent in \eqn{qqgFT} can be rewritten as
\begin{equation}\label{phase}
\rme^{i\left[\vartheta\bm{x}+(1-\vartheta-\xi)\bm{y}+\xi\bm{z}\right]\cdot\bm{q}-i(\bm{y}-\bm{z})\cdot\bm{\widetilde{p}}+i\left[(1-\vartheta)\bm{x}-(1-\vartheta-\xi)\bm{y}-\xi\bm{z}\right]\cdot\frac{\bm{\widetilde{k}}}{1-\vartheta}}=\rme^{i\bm{w}\cdot\bm{q}-i\bm{Y}\cdot\bm{\widetilde{p}}+i\bm{R}\cdot\bm{\widetilde{k}}},
\end{equation}
where we have introduced the notations $\bm{R}$ and $\bm{Y}$ for the transverse 
separations between the daughter partons after each of the emission vertices:
\begin{equation}\label{defRY}
\bm{R}\,\equiv\,\bm{x}-\bm{y}^\prime\,,\qquad \bm{Y}\,\equiv\,\bm{y}-\bm{z}\,.
\end{equation}
\eqn{phase} explains why it was more convenient to use the ``shifted'' transverse momenta 
$\bm{\widetilde{k}}$ and $\bm{\widetilde{p}}$ as integration variables, instead of the original
momenta $\bm{k}$ and $\bm{p}$: the shifted variables are conjugate to the transverse
separations between the daughter partons and thus facilitate the calculation of the Fourier
transform to the transverse coordinate representation (defined as in \eqn{lowfx}). 
Specifically, by using the following Fourier transform (see e.g. App. A in \cite{Beuf:2017bpd})
  \begin{equation}\begin{split} \label{fourier.4}
\int\frac{d^{2}\bm{k}}{2\pi}\frac{d^{2}\bm{p}}{2\pi}\,\frac{\bm{k}^{i}\,\bm{p}^{j}}{\left(\bm{k}^{2}+\bar Q^{2}\right)\left[a(\bm{k}^{2}+\bar Q^{2})+\bm{p}^{2}\right]}e^{i\bm{k}\cdot\bm{x}+i\bm{p}\cdot\bm{z}}\,=\,-\frac{\bm{x}^{i}\,\bm{z}^{j}}{\bm{z}^{2}\,\sqrt{\bm{x}^{2}+a\bm{z}^{2}}}\,\bar Q\,K_{1}\left(\bar Q\sqrt{\bm{x}^{2}+a\bm{z}^{2}}\right),
 \end{split}\end{equation}
one finds
\begin{align}\label{qqg0}
\left|\gamma^{i}_T(Q,q^{+},\bm{w})\right\rangle _{q\overline{\bm{q}}g}^{(a)}\,=\,&\,-\frac{ee_{f}g \,q^+ }
{2(2\pi)^4}
\int_{\bm{x}, \bm{y}, \bm{z}}
\int_{0}^{1}d\vartheta\int_{0}^{1-\vartheta}d\xi\
\frac{\vartheta}{\sqrt{\xi}}\,
\frac{\bm{R}^{j}\bm{Y}^{n}}{\bm{Y}^{2}}
\,\delta^{(2)}(\bm{w}-\bm{c})\nonumber\\*[0.2cm]
&\times
\varphi_{\lambda_{1}\lambda}^{ij}(\vartheta)\,\tau_{\lambda\lambda_{2}}^{mn}(\xi,1-\vartheta-\xi)
\,t^{a}_{\alpha\beta}\, \frac{Q\,\rmK_{1}(Q D)}{D}
\nonumber\\*[0.2cm]
&\times\left|\overline{q}_{\lambda_{2}}^{\beta}((1-\vartheta-\xi)q^{+}, \bm{y})\,g_{m}^{a}(\xi q^{+}, \bm{z})\,q_{\lambda_{1}}^{\alpha}(\vartheta q^{+}, \bm{x})\right\rangle ,
\end{align}
where $\bm{c}$ is the center-of-energy of the $q\bar q g$ system,
\beq
\bm{c}\,\equiv\,
\vartheta\bm{x}+(1-\vartheta-\xi)\bm{y}+\xi\bm{z}\,,
\eeq
while $D\equiv D(\bm{x},\bm{y},\bm{z},\vartheta,\xi)$ denotes the following transverse distance
(below, $\vartheta^\prime  \equiv 1-\vartheta-\xi$)
\begin{align}\label{defD}
 D^2(\bm{x},\bm{y},\bm{z},\vartheta,\xi)&\,\equiv\, \vartheta\vartheta^\prime \,(\bx-\by)^2 + \vartheta\xi \,(\bx-\bz)^2 + 
\vartheta^\prime \xi\, (\by-\bz)^2\,,\end{align}
which can be interpreted as the overall transverse size of the $q\bar q g$ partonic fluctuation at the time
of scattering ($x^+=0$). This size is limited by the virtuality of the space-like photon: $QD\lesssim 1$. (Indeed,
the modified Bessel function exponentially vanishes for large values $QD\gg 1$ of its argument.) It is interesting to notice
that $D^2$ can be equivalently rewritten as
\begin{align}\label{Dbarq}
 D^2\,=\,\vartheta(1-\vartheta) \bm{R}^2+
\frac{\xi(1-\vartheta-\xi)}{1-\vartheta}
\bm{Y}^{2}\,.
\end{align}
This rewriting shows that the virtuality $Q^2$ limits both the size $R=|\bm{x}-\bm{y}^\prime|$ of the $q\bar q$ pair produced
by the decay of the virtual photon, namely $R\lesssim 1/\bar Q$, and the size $Y=|\bm{y}-\bm{z}|$ of
the $g\bar q$ pair produced by the decay of the antiquark ($\bar q\to\bar q g$). 
In particular, when the gluon is soft, $\xi \ll 1$, the virtuality constraint on the gluon emission becomes quite loose:
 $Y^2\lesssim 1/(\xi Q^2)$; hence, despite the space-like
 virtuality, a soft gluon can propagate at relatively large distances from its emitter.
 
 Of course, this whole discussion of the virtuality limits on the transverse size of the partonic fluctuations applies only prior to
 the scattering ($x^+\le 0$). The collision can put the partons on their mass-shell and then they can move from each other
 arbitrarily far away.
 
At this point, it is straightforward to add the effects of the scattering in the eikonal
approximation and thus deduce our final result for the first term in the r.h.s. of \eqn{qqgoutab}:
it suffices to insert Wilson lines for the 3 partons which are crossing the shockwave. In practice,
this amounts to replacing
\beq
t^{a}_{\alpha\beta}\,\to\, U^{ab}(\bm{z})\Big[V(\bm{x})t^{b}V^{\dagger}(\bm{y})\Big]_{\alpha\beta}
\eeq
in the integrand of \eqn{qqg0}. At this step, it is also convenient to subtract the no-scattering limit ($\hat S\to 1$) 
from the final result\footnote{As explained in Sect.~\ref{sec:LCWF}, this subtraction is a convenient way
to include the contribution of the graph in  Fig.~\ref{fig:NLO_Real}.c, where the photon decays after crossing the
shockwave. In writing \eqn{qqg1}, we  keep the upper label $(a)$ although, strictly speaking,
this result involves contributions from both graphs (a) and (c) in Fig.~\ref{fig:NLO_Real}.}
and thus directly obtain the first term in the r.h.s. of  \eqn{outgammaRe}. Our final result 
for this term therefore reads 
\begin{align}\label{qqg1}
\left|\gamma^{i}_T(Q,q^{+},\bm{w})\right\rangle _{q\overline{\bm{q}}g}^{(a)}\,=\,&\,-\frac{ee_{f}g \,q^+}
{2(2\pi)^4}
\int_{\bm{x}, \bm{y}, \bm{z}}
\int_{0}^{1}d\vartheta\int_{0}^{1-\vartheta}d\xi\
\frac{\vartheta}{\sqrt{\xi}}\,
\frac{\bm{R}^{j}\bm{Y}^{n}}{\bm{Y}^{2}}
\,\delta^{(2)}(\bm{w}-\bm{c})\nonumber\\*[0.2cm]
&\times{\varphi_{\lambda_{1}\lambda}^{ij}(\vartheta)\,\tau_{\lambda\lambda_{2}}^{mn}(\xi,1-\vartheta-\xi)}
\,\frac{Q\,\rmK_{1}(Q D) }
{D}\Big[U^{ab}(\bm{z})V(\bm{x})t^{b}V^{\dagger}(\bm{y})-t^a\Big]_{\alpha\beta}
\nonumber\\*[0.2cm]
&\times\left|\overline{q}_{\lambda_{2}}^{\beta}((1-\vartheta-\xi)q^{+}, \bm{y})\,g_{m}^{a}(\xi q^{+}, \bm{z})\,q_{\lambda_{1}}^{\alpha}(\vartheta q^{+}, \bm{x})\right\rangle.
\end{align}

Consider now the second term in  \eqn{qqgoutab} which, we recall, corresponds to a gluon emission
in the final state (cf. Fig.~\ref{fig:NLO_Real}.b). In the absence of scattering ($\hat S=1$), this term
differs from the first term there only in one of the energy denominators: the energy difference $
E_{q\overline{q} g}-E_{\gamma}$ shown in \eqn{enden3} gets replaced by $
E_{q\overline{q}}- E_{q\overline{q} g}$, which is shown (up to a sign) in \eqn{enden2}.
So, the corresponding contribution (for $\hat S=1$, once again) can be obtained from \eqn{qqgmom}
by replacing 
\beq
\vartheta(1-\vartheta)^{2}\bm{\widetilde{p}}^{2}\,+\,\xi(1-\vartheta-\xi) \big(\bm{\widetilde{k}}^{2}\,+\,
\bar{Q}^{2}\big)\,\longrightarrow\,\vartheta(1-\vartheta)^{2}\bm{\widetilde{p}}^{2}\,,\eeq
in the denominator and changing the overall sign. When computing the Fourier transform to transverse
coordinates, the integrals over $\bm{\widetilde{p}}$ and $\bm{\widetilde{k}}$ factorise from each other.
The Wilson lines can then be easily added: in this case, they describe the scattering of the intermediate $q\bar q $ pair, with transverse coordinates $\bx$ and $\by'$ (see Fig.~\ref{fig:NLO_Real}.b). After also subtracting the no-scattering limit, one
eventually finds\footnote{Since the antiquark formally propagates backwards in time, the colour matrix $t^a$ associated
with the gluon emission vertex is located to the right of the Wilson lines describing the scattering of the
$q\bar q $ pair.}
\begin{align}\label{qqg2}
\left|\gamma^{i}_T(Q,q^{+},\bm{w})\right\rangle _{q\overline{\bm{q}}g}^{(b)}\,=\,&\,
\frac{ee_{f}g \,q^+}
{2(2\pi)^4}
\int_{\bm{x}, \bm{y}, \bm{z}}
\int_{0}^{1}d\vartheta\int_{0}^{1-\vartheta}d\xi\
\frac{\vartheta}{\sqrt{\xi}}\,
\frac{\bm{R}^{j}\bm{Y}^{n}}{\bm{Y}^{2}}
\,\delta^{(2)}(\bm{w}-\bm{c})\nonumber\\*[0.2cm]
&\times{\varphi_{\lambda_{1}\lambda}^{ij}(\vartheta)\,\tau_{\lambda\lambda_{2}}^{mn}(\xi,1-\vartheta-\xi)}
\,\frac{Q^2\rmK_{1}(\bar Q R) }
{\bar Q R}\Big[V(\bm{x})V^{\dagger}(\bm{y}^\prime)\,t^{a}-t^a\Big]_{\alpha\beta}
\nonumber\\*[0.2cm]
&\times\left|\overline{q}_{\lambda_{2}}^{\beta}((1-\vartheta-\xi)q^{+}, \bm{y})\,g_{m}^{a}(\xi q^{+}, \bm{z})\,q_{\lambda_{1}}^{\alpha}(\vartheta q^{+}, \bm{x})\right\rangle.
\end{align}

The above contributions in  Eqs.~\eqref{qqg1}  and \eqref{qqg2} can be combined with each other 
by introducing a convenient notation. Specifically, by using \eqref{defyw}, \eqref{defD}, and 
the following matrix identity relating the adjoint and the fundamental representations,
\begin{equation}\label{VUid}
V^{\dagger}(\bm{y}^\prime)t^{a}V(\bm{y}^\prime)=
U^{ab}(\bm{y}^\prime)t^{b}\ \Longrightarrow\ U^{ab}(\bm{y}^\prime)V(\bm{x})t^{b}V^{\dagger}(\bm{y}^\prime)=
V(\bm{x})V^{\dagger}(\bm{y}^\prime)t^{a}\,,
\end{equation}
one sees that the middle line of \eqref{qqg2} can be obtained from the corresponding 
line of \eqref{qqg1} by replacing both $\bm{y}$ and $\bm{z}$ by $\bm{y}^\prime$;
indeed, after this replacement, one has $QD\to \bar Q R$.
This replacement is merely formal (it should not be applied
to the ensemble of \eqn{qqg1}, but only to its middle line), but it is still useful in that it allows us to 
introduce a compact notation for the sum of Eqs.~\eqref{qqg1}  and \eqref{qqg2}:
\begin{align}\label{qqgout}
\left|\gamma^{i}_T(Q,q^{+},\bm{w})\right\rangle _{q\overline{\bm{q}}g}^{\rm reg}\,=\,&\,-\frac{ee_{f}g \,q^+}
{2(2\pi)^4}
\int_{\bm{x}, \bm{y}, \bm{z}}
\int_{0}^{1}d\vartheta\int_{0}^{1-\vartheta}d\xi\
\frac{\vartheta}{\sqrt{\xi}}\,
\frac{\bm{R}^{j}\bm{Y}^{n}}{\bm{Y}^{2}}
\,\delta^{(2)}(\bm{w}-\bm{c})\nonumber\\*[0.2cm]
&\times \left\{ \Phi_{\lambda_{1}\lambda_{2}}^{ijmn}(\vartheta,\xi)
\frac{Q\,\rmK_{1}(Q D) }
{D}\Big[U^{ab}(\bm{z})V(\bm{x})t^{b}V^{\dagger}(\bm{y})-t^a\Big]_{\alpha\beta}
-\left(\bm{y},\bm{z}\rightarrow\bm{y}^\prime\right)\right\}
\nonumber\\*[0.2cm]
&\times\left|\overline{q}_{\lambda_{2}}^{\beta}((1-\vartheta-\xi)q^{+}, \bm{y})\,g_{m}^{a}(\xi q^{+}, \bm{z})\,q_{\lambda_{1}}^{\alpha}(\vartheta q^{+}, \bm{x})\right\rangle ,
\end{align}
with an ``effective vertex'' which encompasses the spinor and helicity structure of the whole graph:
\beq\label{Phidef}
 \Phi_{\lambda_{1}\lambda_{2}}^{ijmn}(\vartheta,\xi)\equiv 
 \varphi_{\lambda_{1}\lambda}^{ij}(\vartheta)\,\tau_{\lambda\lambda_{2}}^{mn}(\xi,1-\vartheta-\xi)\,.
 \eeq
 
 \begin{figure}
  \centering
    \includegraphics[width=0.7\textwidth]{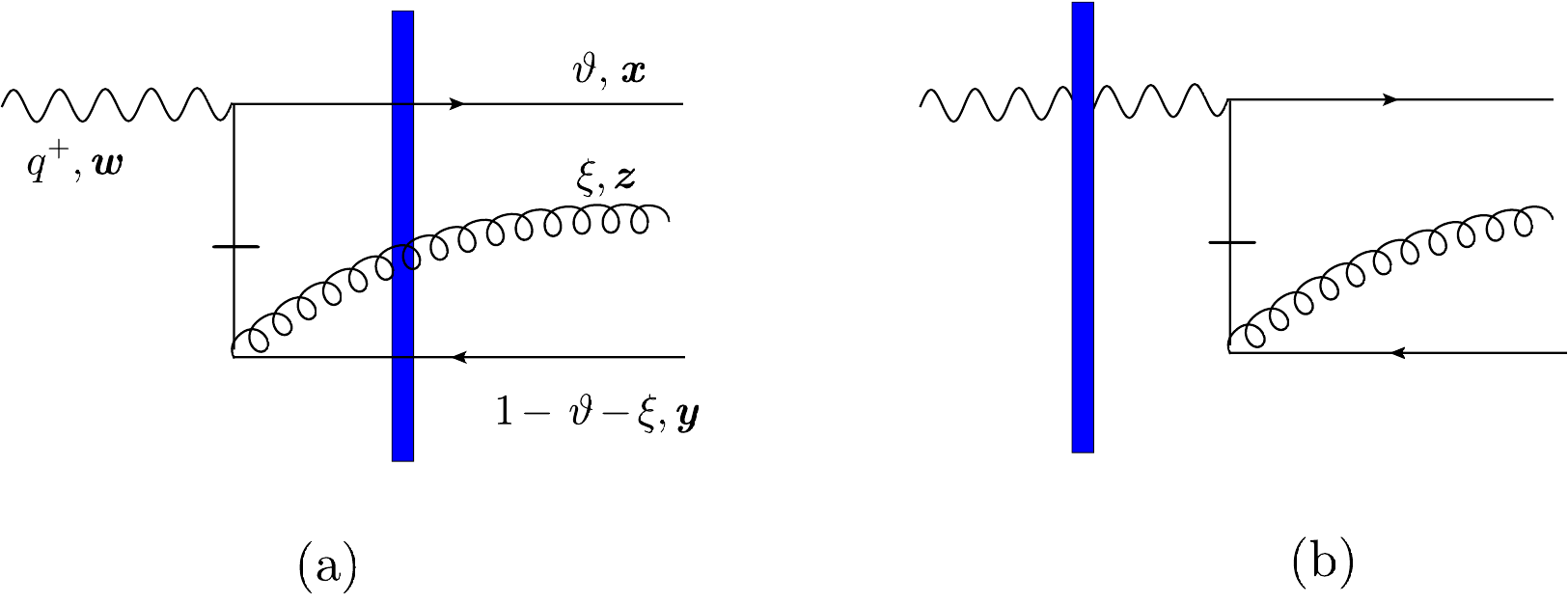}
  \caption{\small The two instantaneous graphs contributing to the quark-antiquark-gluon outgoing state in the case
  where the gluon is emitted by the antiquark.}
    \label{fig:qbar-inst}
    \end{figure}

In our notation in \eqn{qqgout}, we reintroduced the upper script  ``reg'' to recall that this is the contribution
of the ``regular'' graphs which involve the standard, non-local, piece of the intermediate antiquark
propagator.  On top on that, we also have the instantaneous contribution. This involves only 2 topologies,
since the photon decay and the gluon emission now occur at the same time (see Fig.~\ref{fig:qbar-inst}). 
The respective matrix element reads (see e.g. \cite{Beuf:2017bpd})
\begin{align}
\label{mat.inst1}
\left\langle q_{\lambda_{1}}^{\alpha}(k)\,\overline{q}_{\lambda_{2}}^{\beta}(s)\,g_{a}^{j}(p)\left|\,\mathsf{H}_{\gamma\rightarrow q
\overline{\bm{q}}g}\,\right|\gamma_{T}^{i}(q)\right\rangle 
=\,-(2\pi)^{3}\delta^{(3)}(q-k-p-s)
\frac{gee_{f}\delta_{\lambda_{1}\lambda_{2}}\,t_{\beta\alpha}^{a}}{4\sqrt{q^{+}p^{+}}}\,\frac{\delta^{ij}+2i\epsilon^{ij}\lambda_{1}}{s^{+}+p^{+}}.
\end{align}
A straightforward calculation, similar to that leading to \eqn{qqg1}, yields
\begin{align}\label{qqginst}
\left|\gamma^{i}_T(Q,q^{+},\bm{w})\right\rangle _{q\overline{\bm{q}}g}^{{\rm inst}}\,=\,&
\,\frac{ee_{f}g\,q^{+}}{2(2\pi)^{4}}\int_{\bm{x},\bm{y},\bm{z}}\int_{0}^{1}d\vartheta\,\int_{0}^{1-\vartheta}d\xi\ \sqrt{\xi}\,\frac{\vartheta(1-\vartheta-\xi)}{1-\vartheta}
\,\delta^{(2)}(\bm{w}-\bm{c})\nonumber\\*[0.2cm]
&\times\left(\delta^{imi}+2i\epsilon^{im}\lambda_{1}\right)\delta_{\lambda_{2}\lambda_{1}}\,\frac{Q\,\rmK_{1}(QD)}{D}\Big[U^{ab}(\bm{z})V(\bm{x})t^{b}V^{\dagger}(\bm{y})-t^{a}\Big]_{\alpha\beta}
\nonumber\\*[0.2cm]
&\times\left|\overline{q}_{\lambda_{2}}^{\beta}((1-\vartheta-\xi)q^{+},\bm{y})\,g_{m}^{a}(\xi q^{+},\bm{z})\,q_{\lambda_{1}}^{\alpha}(\vartheta q^{+},\bm{x})\right\rangle.
\end{align}
This has the same Wilson line structure as \eqn{qqg1}, as expected given the topology of the contributing diagrams in 
Fig.~\ref{fig:qbar-inst}. For what follows, it is convenient to observe that one can combine Eqs.~\eqref{qqgout} and \eqref{qqginst} in a unique expression by using a generalised version of the effective vertex \eqref{Phidef}, which reads
\beq\label{vertexPhi}
\Phi_{\lambda_{1}\lambda_{2}}^{ijmn}(\bm{x},\bm{y},\bm{z},\vartheta,\xi)\,\equiv\,\varphi_{\lambda_{1}\lambda}^{ij}(\vartheta)\,\tau_{\lambda\lambda_{2}}^{mn}(\xi,1-\vartheta-\xi)
-\delta_{\lambda_{1}\lambda_{2}}\delta^{nj}\,\frac{\xi(1-\vartheta-\xi)}{1-\vartheta}
\left(\delta^{im}+2i\epsilon^{im}\lambda_{1}\right)\frac{\bm{Y}^{2}}{\bm{R}\cdot\bm{Y}}\,. \eeq
It is understood that the second piece in \eqn{vertexPhi} vanishes after substituting $\bm{y}\to \bm{y}^\prime$ and $\bm{z}\to \bm{y}^\prime$, so the instantaneous contribution vanishes in this limit, as it should.

\begin{figure}[t]
  \centering
    \includegraphics[width=\textwidth]{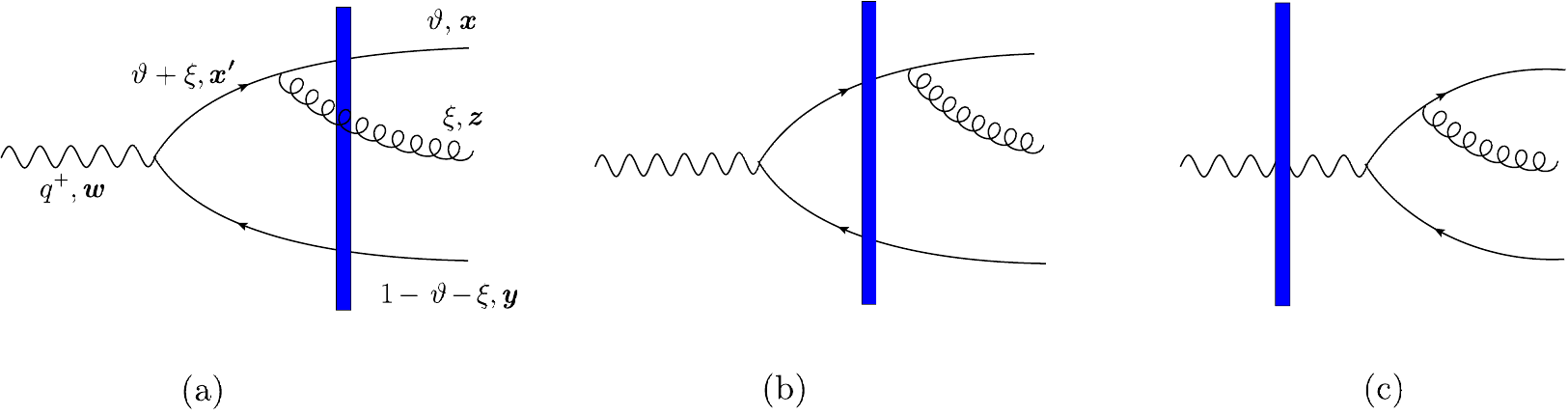}
  \caption{\small The three possible topologies in the case
  where the gluon is emitted by the quark. }
    \label{fig:NLO_Realq}
    \end{figure}

\subsection{Gluon emission by the quark}
\label{sec:gluonQ}

Our conventions for the case where the gluon is emitted by the quark are summarised in 
Fig.~\ref{fig:Real_kT}.b in transverse momentum space and in Fig.~\ref{fig:NLO_Realq}
in transverse coordinate space. (This last figure shows all the possible topologies for the
collision with the shock wave and for the case of the ``regular'' contributions.) 
There is in fact no need to explicitly compute these graphs: the respective result can be 
directly inferred from that in the previous section via suitable changes of variables which
amount to exchanging the quark and the antiquark, but such that
the final state remains unchanged. Specifically, the contribution of the
3 graphs in Fig.~\ref{fig:NLO_Realq} to the $q\bar q g$ Fock space component of the
LCWF of the transverse photon can be obtained from \eqn{qqgout} by
\texttt{(i)} exchanging the quark with the antiquark in the final state,
\texttt{(ii)} exchanging $\vartheta \leftrightarrow 1-\vartheta-\xi$, $\bx \leftrightarrow \by$,
$\lambda_1\leftrightarrow \lambda_2$, and $\alpha\leftrightarrow \beta$, 
\texttt{(iii)} taking the complex conjugate of the integrand, and \texttt{(iv)} changing the 
overall sign (to account for the fact that the colour charge of the quark is minus that
of the antiquark)\footnote{Notice that the leading-order LCWF in \eqn{out_lo} is invariant
under this combined set of symmetry operations, as it should: exchanging the quark and the antiquark
has no physical relevance in that case.}.

These transformations have the following consequences on the other 
transverse coordinates and distances in the problem:
\begin{equation}\label{RYX}
\bm{y}^\prime\,\to\,\frac{\vartheta\bm{x}+\xi\bm{z}}{\vartheta+\xi}
\equiv \bm{x}^\prime\,,\qquad
\bm{R}\equiv \bm{x}-\bm{y}^\prime\,\to\,  \bm{y}-\bm{x}^\prime\equiv -
\bm{\widetilde R}\,,\qquad  \bm{Y}\equiv\bm{y}-\bm{z}\,\to\,  
\bm{x}-\bm{z} \equiv \bm{X}\,,
\end{equation}
(Notice that $\bm{x}^\prime$ is the transverse coordinate 
of the intermediate quark, prior to the gluon emission; see Fig.~\ref{fig:NLO_Realq}.a).
Furthermore, the spinorial structures change as follows (recall \eqn{phidef})
\begin{align}\label{phiq}
\varphi_{\lambda_{1}\lambda}^{ij}(\vartheta)&\,\to\,\varphi_{\lambda_{2}\lambda}^{ij\,*}(1-\vartheta-\xi)
\,=\,\chi_{\lambda}^{\dagger}\left[(1-2(\vartheta+\xi))\delta^{ij}-i\varepsilon^{ij}\sigma^{3}\right]
\chi_{\lambda_{2}}\nonumber \\*[0.2cm]
&\,=\,- \,\varphi_{\lambda\lambda_{2}}^{ij}(\vartheta+\xi)\,=\,-\,
\delta_{\lambda\lambda_{2}}\left[(2(\vartheta+\xi)-1)\delta^{ij}+2i\varepsilon^{ij}\lambda\right]\,,
\end{align}
for the photon decay vertex  and, respectively (recall \eqn{taudef})
\begin{align}\label{tauq}
\tau_{\lambda\lambda_{2}}^{mn}(\xi,1-\vartheta-\xi)&\,\to\,
\tau_{\lambda\lambda_{1}}^{mn\,*}(\xi,\vartheta)
\,=\, \chi_{\lambda_{1}}^{\dagger}\left[(2\vartheta+\xi)\delta^{mn}-i\xi
\varepsilon^{mn}\sigma^{3}\right]\chi_{\lambda}\nonumber \\*[0.2cm]
&\,=\,
\tau_{\lambda_{1}\lambda}^{nm}(\xi,\vartheta)
\,=\,\delta_{\lambda\lambda_{1}}\left[(2\vartheta+\xi)\delta^{nm}+2i\xi\varepsilon^{nm}\lambda\right]\,,
\end{align}
for the vertex describing the gluon emission.

Remarkably, these transformations do not change the colour structure (including the Wilson lines) 
for the diagrams where the parton branchings occur either fully before or fully
after the collision. That is, the diagram in Fig.~\ref{fig:NLO_Realq}.a has exactly the same colour structure
as that in Fig.~\ref{fig:NLO_Real}.a and the same holds for the diagrams in Fig.~\ref{fig:NLO_Realq}.c
and Fig.~\ref{fig:NLO_Real}.c, respectively. This is so because the vertex for the photon decay has no 
incidence on the flow of colour, so the matrix $t^a$ describing the gluon emission 
can be ``commuted'' through the photon vertex without altering the colour structure.
On the other hand, the diagrams where the scattering occurs in the intermediate $q\bar q$ state,
that is those in Fig.~\ref{fig:NLO_Realq}.b and Fig.~\ref{fig:NLO_Real}.b respectively, do have 
different colour structures, since the matrix $t^a$ does not commute with 
the Wilson lines describing the collision. Yet, as in \eqn{qqgout}, the $S$-matrix structure of the
graph in Fig.~\ref{fig:NLO_Realq}.b can be obtained from that in Fig.~\ref{fig:NLO_Realq}.a
via an appropriate replacement of variables, which follows as well from the symmetry operations
above.

After making all these transformations and also changing the overall sign, one finds
(note the subscript ${\bm q}\overline{q}g$ on the outgoing state: the quark label is shown 
in boldface to emphasise that this is the source of the gluon emission):
\begin{align}\label{qqgoutq}
\left|\gamma^{i}_T(Q,q^{+},\bm{w})\right\rangle _{{\bm q}\overline{q}g}\,=\,&\,
\frac{ee_{f}g \,q^+}
{2(2\pi)^4}
\int_{\bm{x}, \bm{y}, \bm{z}}
\int_{0}^{1}d\vartheta\int_{0}^{1-\vartheta}d\xi\
\frac{1-\vartheta-\xi}{\sqrt{\xi}}\,
\frac{ \bm{\widetilde R}^{j}\bm{X}^{n}}{\bm{X}^{2}}
\,\delta^{(2)}(\bm{w}-\bm{c})\nonumber\\*[0.2cm]
&\times \left\{\widetilde\Phi_{\lambda_{1}\lambda_{2}}^{ijmn}(\bm{x},\bm{y},\bm{z},\vartheta,\xi)
\frac{Q\,\rmK_{1}(Q D) }
{D}\Big[U^{ab}(\bm{z})V(\bm{x})t^{b}V^{\dagger}(\bm{y})-t^a\Big]_{\alpha\beta}
-\left(\bm{x},\bm{z}\rightarrow\bm{x}^\prime\right)\right\}
\nonumber\\*[0.2cm]
&\times\left|\overline{q}_{\lambda_{2}}^{\beta}((1-\vartheta-\xi)q^{+}, \bm{y})\,g_{m}^{a}(\xi q^{+}, \bm{z})\,q_{\lambda_{1}}^{\alpha}(\vartheta q^{+}, \bm{x})\right\rangle ,
\end{align}
where the effective transverse size $D$ is the same as for an
emission by the antiquark, cf. \eqn{defD}. (Indeed, this distance
is invariant under the symmetry operations exchanging the quark and the antiquark.)
When replacing
$\bm{x},\bm{z}\rightarrow\bm{x}^\prime$, the argument of the Bessel function changes as
$QD \to \widetilde Q \widetilde R$, where $\widetilde R= |\bm{\widetilde R}|=|\bm{x}^\prime-\bm{y}|$ and
\beq\label{tildeQD}
\widetilde Q^2\equiv (\vartheta+\xi)(1-\vartheta-\xi) Q^2\,.
\eeq
The effective vertex $\widetilde \Phi$ is obtained via the appropriate transformations from \eqn{vertexPhi}
and reads
\beq\label{vertexPhi2}
 \widetilde \Phi_{\lambda_{1}\lambda_{2}}^{ijmn}(\bm{x},\bm{y},\bm{z},\vartheta,\xi)
 \,\equiv\,
 \tau_{\lambda_{1}\lambda}^{nm}(\xi,\vartheta)\,\varphi_{\lambda\lambda_{2}}^{ij}(\vartheta+\xi)
  -\delta_{\lambda_{1}\lambda_{2}}\delta^{nj}\,\frac{\xi\vartheta}{\vartheta+\xi}
\left(\delta^{im}-2i\epsilon^{im}\lambda_{1}\right)\frac
  {\bm{X}^2}{\bm{\widetilde R}\cdot\bm{X}}\,.
 \eeq
 Our final result for the $q\bar q g$  Fock-space component of the LCWF of the transverse photon
 is the sum of the contributions computed above, corresponding to emissions by the antiquark
 and by the quark, respectively:
 \beq\label{qqgWV}
 \left|\gamma_{T}^{i}(Q, q^{+}, \bm{w})\right\rangle _{q\overline{q}g}\,=\,
 \left|\gamma^{i}_{T}(Q,q^{+},\bm{w})\right\rangle _{q\overline{\bm{q}}g}\,+\,
 \left|\gamma^{i}_{T}(Q,q^{+},\bm{w})\right\rangle _{{\bm q}\overline{q}g}.
 \eeq
 
\begin{figure}[t]
  \centering
  \begin{subfigure}[t]{0.7\textwidth}
    \includegraphics[width=\textwidth]{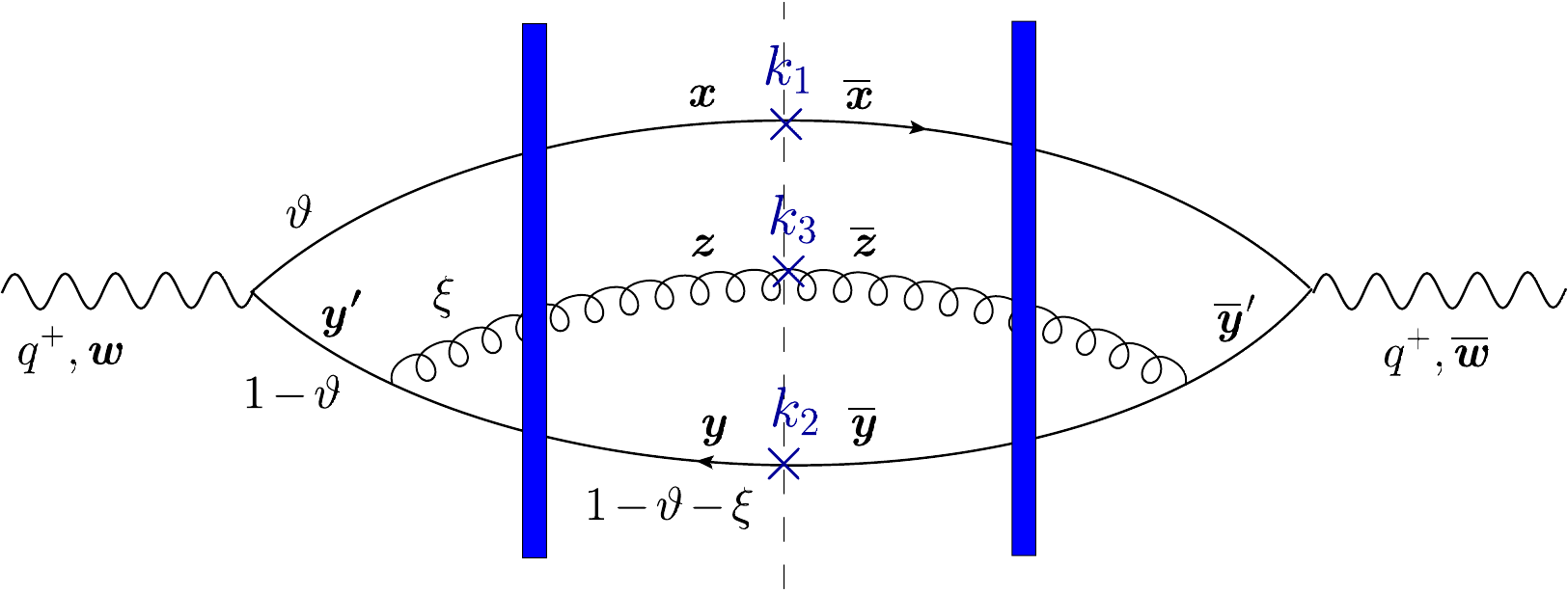}
   \caption{}\label{fig:3jets_DA}
  \end{subfigure}\\*[0.5cm]
  \begin{subfigure}[t]{0.7\textwidth}
    \includegraphics[width=\textwidth]{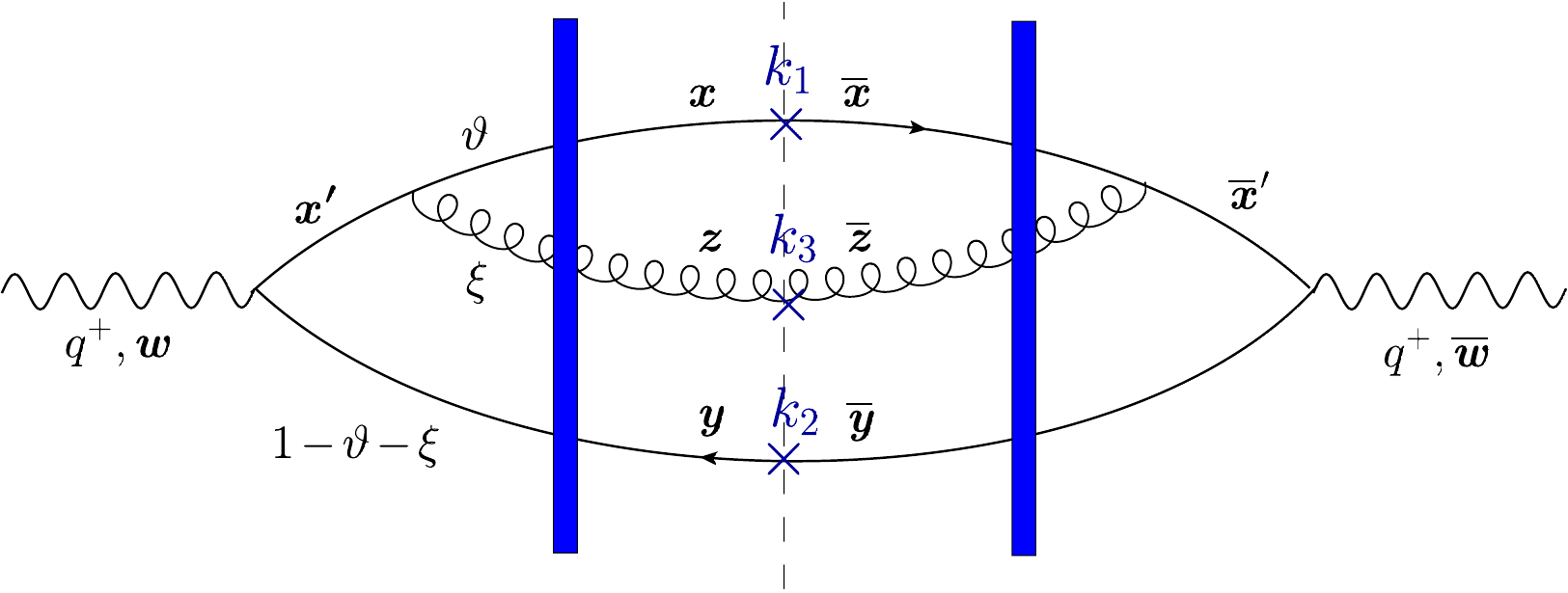}
   \caption{}\label{fig:3jets_DB}
  \end{subfigure}
  \caption{\small Two direct contributions to the leading-order cross-section for $q\bar q g$  production.
  The gluon is emitted and reabsorbed by the same fermion: the antiquark in figure (a) and the
   quark in figure (b). }
    \label{fig:3jets_D}
    \end{figure}

\section{Next-to-leading order corrections: the real terms}
\label{sec:NLOreal}

Given our previous results for the tri-parton Fock-space component of the outgoing wavefunction
of the transverse virtual photon, it is rather straightforward to deduce the ``real'' next-to-leading order
(NLO) corrections to the cross-section for producing a pair of quark-antiquark jets. This involves two
steps: first, one computes the (tree-level) cross-section for the production of three partons (quark, antiquark,
gluon); then, one integrates out the kinematics of the gluon which is not measured in the final state.

\subsection{The leading-order trijet cross-section}
The leading-order cross-section for $q\bar q g$  production in DIS
is computed similarly to \eqn{defcross}, that is, as the expectation value of the product of three number-density
operators (themselves built with Fock space operators for bare partons) on the photon LCWF in \eqn{qqgWV}:
\begin{equation}\begin{split}\label{3jetXsec}
&\frac{d\sigma^{\gamma A\rightarrow q\bar q g}}{dk_1^+d^{2}\bm{k}_{1}dk_2^+d^{2}\bm{k}_{2}dk_3^+d^{2}\bm{k}_{3}}\,(2\pi)\,\delta(q^{+}-k_1^+-k_2^+-k_3^+)\\
&=\frac{1}{2}\,\int_{\bm{\overline{w}}, \bm{w}}\,e^{-i\bm{q}\cdot(\bm{w}-\overline{\bm{w}})}\:_{q\overline{q}g}\!\left\langle \gamma_{\rmT}^{i}(Q, q^{+}, \bm{\overline{w}})\right|\,\mathcal{N}_{q}(k_{1})\,\mathcal{N}_{\overline{q}}(k_{2})\,\mathcal{N}_{g}(k_{3})\,\left|\gamma_{\rmT}^{i}(Q, q^{+}, \bm{w})\right\rangle _{q\overline{q}g}.
\end{split}\end{equation}
It is natural to distinguish between two types of contributions: {\it direct} contributions, where the
gluon is emitted and reabsorbed by the same fermion (quark or antiquark), and {\it interference}
terms, where the gluon is absorbed by the quark in the directed amplitude (DA) and reabsorbed 
by the antiquark in the complex conjugate amplitude (CCA), or vice-versa. 

\subsubsection{Direct  contributions}
\label{sec:direct3jet}

Two particular diagrams describing direct emissions --- one by the quark, the other one
by the antiquark --- are illustrated  in Fig.~\ref{fig:3jets_D}. For each type of emitter,
there are nine distinct topologies, corresponding to the 
product of the three graphs shown in Fig.~\ref{fig:NLO_Real} (for gluon emission by the antiquark)
and their complex conjugates.
(For simplicity, we explicitly discuss only the case of the regular graphs; the instantaneous contributions
are eventually added by modifying the effective vertices as shown in Eqs.~\eqref{vertexPhi} and
\eqref{vertexPhi2}.) Recall however that in our calculation of the LCWF
we have grouped the three graphs in Fig.~\ref{fig:NLO_Real} (or in Fig.~\ref{fig:NLO_Realq})
 into two terms: the last graph has been
used as a subtraction term for the two previous ones, thus making clear that this particular LCWF component 
vanishes in the absence of the scattering. Hence, the direct contribution to the 3-jet cross-section
will in fact contain only 4 terms.

As in the previous section, we first present our result for the emission by the antiquark.
A direct calculation using \eqn{3jetXsec} together with the LCWF in \eqn{qqgout} yields
\begin{align}\label{R1}
\frac{d\sigma^{(a), \rm direct}}{dk_1^+d^{2}\bm{k}_{1}dk_2^+d^{2}\bm{k}_{2}dk_3^+d^{2}\bm{k}_{3}}&\,=\,\frac{\alpha_{s}\alpha_{em}\,C_{F} N_{c}\,\vartheta^2{Q}^{2}}{2(2\pi)^{10}(q^{+})^{2}\xi}\,\left(\sum e_{f}^{2}\right)\delta(q^{+}-k_1^+-k_2^+-k_3^+)\nonumber\\*[0.2cm]
&\times\int_{\bm{x}, \bm{y}, \bm{z}, \overline{\bm{x}}, \bm{\overline{y}}, \bm{\overline{z}}}\,e^{-i\bm{k}_{1}\cdot(\bm{x}-\overline{\bm{x}})-i\bm{k}_{2}\cdot(\bm{y}-\overline{\bm{y}})-i\bm{k}_{3}\cdot(\bm{z}-\overline{\bm{z}})}\,\frac{\bm{R}^{i}\bm{Y}^{m}\,\bm{\overline{R}}^{\,j}\overline{\bm{Y}}^{\,n}}{\bm{Y}^{2}\,\bm{\overline{Y}}^{2}}\nonumber\\*[0.2cm]
&\times\Big[\mathcal{K}_{1}^{imjn}(\bm{x},\bm{y},\bm{z},\overline{\bm{x}},\bm{\overline{y}},\bm{\overline{z}},\vartheta,\xi,\bar{Q})\,\mathcal{W}(\bm{x},\bm{y},\bm{z},\overline{\bm{x}},\bm{\overline{y}},\bm{\overline{z}})\nonumber\\*[0.2cm]
&\left.-\,\big(\bm{y}, \bm{z}\rightarrow\bm{y}^{\prime}\big)\,-\,\big(\bm{\overline{y}}, \bm{\overline{z}}\rightarrow\bm{\overline{y}}^{\prime}\big)
\,+\,\big(\bm{y},\,\bm{z}\rightarrow\bm{y}^{\prime}\,\ \&\ \,\bm{\overline{y}},\,\bm{\overline{z}}\rightarrow\bm{\overline{y}}^{\prime}\big)
\right.\Big].
\end{align}
The notations here are similar to those in \eqn{qqgout} (see also Fig.~\ref{fig:3jets_D}). 
Transverse coordinates with a bar refer to the CCA.  
The longitudinal momentum fractions  $\theta$ and $\xi$ are the same
in the DA and in the CCA, since they are fully fixed by the kinematics of the final state,
as follows:
 \begin{equation}\label{longit1}
 \vartheta=\frac{k_1^+}{q^{+}},\qquad\xi=\frac{k_3^+}{q^{+}}\,.
 \end{equation}
 The $\delta$-function enforcing longitudinal momentum conservation implies
 $k^+_2=(1-\vartheta-\xi)q^+$, which in particular requires $\vartheta+ \xi\le 1$.

The tensorial kernel $\mathcal{K}_{1}^{imjn}$ is defined as
 \begin{align}\label{defK1}
\mathcal{K}_{1}^{imjn}(\bm{x},\bm{y},\bm{z},\overline{\bm{x}},\bm{\overline{y}},\bm{\overline{z}},\vartheta,\xi, {Q})&\,\equiv\,\Phi_{\lambda_{1}\lambda}^{lirm}(\bm{x},\bm{y},\bm{z}, \vartheta,\xi)\,\Phi_{\lambda_{1}\lambda}^{ljrn\,*}(\bm{\overline{x}},\overline{\bm{y}},\bm{\overline{z}},\vartheta, \xi)\nonumber\\*[0.2cm]
&\,\times\,\frac{\rmK_{1}\left({Q} D(\bm{x}, \bm{y}, \bm{z}, \vartheta, \xi)\right)\,\rmK_{1}\left({Q} D(\bm{\overline{x}}, \overline{\bm{y}}, \bm{\overline{z}}, \vartheta, \xi)\right)}{D(\bm{x}, \bm{y}, \bm{z}, \vartheta, \xi)\,D(\bm{\overline{x}}, \overline{\bm{y}}, \bm{\overline{z}}, \vartheta, \xi)}\,.
\end{align}
As already mentioned, the effective vertices in the numerator also include the contributions from the
 instantaneous graphs. However, these contributions become irrelevant in some important limits to be later discussed, such as the soft gluon limit $\xi\to 0$.  So, it will be useful to have an explicit expression for the
 product of effective vertices in \eqn{defK1} without these instantaneous terms (cf. \eqn{Phidef}):
 \begin{align}\label{phiphi}\hspace*{-0.6cm}
&\Phi_{\lambda_{1}\lambda}^{lirm}(\vartheta,\xi)\,\Phi_{\lambda_{1}\lambda}^{ljrn\,*}(\vartheta,\xi)\Big|_{{\rm non-inst.}}=\nn
&=\chi_{\lambda}^{\dagger}\left(\left[(2-2\vartheta-\xi)^{2}+\xi^{2}\right]\delta^{mn}+2i\xi(2-2\vartheta-\xi)\varepsilon^{mn}\sigma^{3}\right)\left(\left[1+(1-2\vartheta)^{2}\right]\delta^{ij}+2i(2\vartheta-1)\varepsilon^{ij}\sigma^{3}\right)\chi_{\lambda}\nn
&=8\left[(1-\vartheta-\xi)^{2}+(1-\vartheta)^{2}\right]\left[\vartheta^{2}+(1-\vartheta)^{2}\right]\delta^{ij}\delta^{mn}-8\xi(2\vartheta-1)(2-2\vartheta-\xi)\varepsilon^{ij}\varepsilon^{mn},
\end{align}
where we have also used $\chi_{\lambda}^{\dagger}\chi_{\lambda}=2$ and
$\chi_{\lambda_{1}}^{\dagger}\sigma^{3}\chi_{\lambda_{2}}=2\lambda_{1}\delta_{\lambda_{1}\lambda_{2}}$.

The first piece in the last equation, proportional to $\delta^{ij}\delta^{mn}$, generates the {\it scalar} products of the transverse
separations in the DA and, respectively, the CCA, whereas the second piece, proportional to $\varepsilon^{ij}\varepsilon^{mn}$,
generates their {\it vectorial} products, followed by a final scalar product between
the 2 vectors previously generated\footnote{Notice that $\varepsilon^{ij}=\varepsilon^{ij3}$ for any $i,j=1,2$ and that a vector
product between 2 transverse vectors, like $\bm{R}\times\bm{\overline{R}}$, is a 3-dimensional vector which is
oriented along the third axis, i.e. the collision axis.}:
\beq
\bm{R}^{i}\bm{Y}^{m}\,\bm{\overline{R}}^{\,j}\overline{\bm{Y}}^{\,n}\delta^{ij}\delta^{mn}
=\big(\bm{R}\cdot \bm{\overline{R}}\big)\big(\bm{Y}\cdot\overline{\bm{Y}}\big),\qquad
\bm{R}^{i}\bm{Y}^{m}\,\bm{\overline{R}}^{\,j}\overline{\bm{Y}}^{\,n}
\varepsilon^{ij}\varepsilon^{mn}=\big(\bm{R}\times\bm{\overline{R}}\big)\cdot\big(\bm{Y}\times\overline{\bm{Y}}\big).
\eeq
The structure of the squared vertex in \eqn{phiphi} is consistent with that
reported in Refs.~\cite{Altinoluk:2020qet} (see Eq.~(2.27) there) and \cite{Taels:2022tza} (see Eqs.~(9.31-32)).
It furthermore agrees with Eq.~(B.5) in  \cite{Caucal:2021ent}, which refers to the case where the gluon is 
emitted the final state (i.e. after the collision with the shockwave) in both the DA and the CCA\footnote{As we 
shall demonstrate in Sect.~\ref{sec:realDGLAP}, the piece proportional to $\varepsilon^{ij}\varepsilon^{mn}$
does not contribute to the cross-section in this particular case (gluon emission  in the final state), in agreement
with the results in  \cite{Caucal:2021ent,Taels:2022tza}.}.
(For the other cases, the sums over spins and helicities were not explicitly computed in 
 \cite{Caucal:2021ent}, so it is difficult to directly compare the results.)

The effects of the collision are encoded in the function $\mathcal{W}$,  defined 
as the following linear combination of partonic $S$-matrices:
\begin{align}\label{defW}
\mathcal{W}(\bm{x}, \bm{y}, \bm{z}, \bm{\overline{x}}, \overline{\bm{y}}, \bm{\overline{z}})\,\equiv\,S_{q\overline{q}g q \overline{q} g}(\bm{x}, \bm{y}, \bm{z}, \bm{\overline{x}}, \overline{\bm{y}}, \bm{\overline{z}})\,-\,S_{q\overline{q}g}(\bm{x}, \bm{y}, \bm{z})\,-\,S_{q\overline{q}g}^{\dagger}(\bm{\overline{x}}, \overline{\bm{y}}, \bm{\overline{z}})\,+\,\mathbf{1}.
\end{align}
The first term in the r.h.s., with the most complex structure, corresponds to the topology depicted
in Fig.~\ref{fig:3jets_DA}: all three partons scatter in both the DA and the CCA. So, this is built with a total
of six Wilson lines --- two in the adjoint representation and four in the fundamental one. Using  Fierz identities like \eqref{VUid},
it is possible to express all the adjoint Wilson lines in terms of fundamental ones. The final result
looks particularly simple and suggestive in the large $N_c$ limit and will be also exhibited below.
Specifically, one has
\begin{align}\label{S6}
S_{q\overline{q}gq\overline{q}g}(\bm{x}, \bm{y}, \bm{z}, \bm{\overline{x}}, \bm{\overline{y}}, \bm{\overline{z}})&\,=\,\frac{1}{C_{F}\,N_{c}}\left\langle \left[U^{\dagger}(\bm{\overline{z}})\,U(\bm{z})\right]^{ab}\,\mathrm{tr}\left[V(\bm{\overline{y}})\,t^{a}\,V^{\dagger}(\bm{\overline{x}})\,V(\bm{x})\,t^{b}\,V^{\dagger}(\bm{y})\right]\right\rangle \nonumber\\*[0.2cm]
&\simeq\,\mathcal{Q}(\bm{x}, \bm{z}, \bm{\overline{z}}, \bm{\overline{x}})\,\mathcal{Q}(\bm{z}, \bm{y}, \bm{\overline{y}}, \bm{\overline{z}})\,,
\end{align}
where the approximate equality in the second line holds at large $N_c$: in this limit, the 6-parton $S$-matrix
of interest reduces to the product of two fundamental quadrupoles.

Furthermore, the second term $S_{q\overline{q}g}(\bm{x}, \bm{y}, \bm{z})$ in \eqn{defW} corresponds
to a diagram where the DA describes initial-state evolution (both parton branchings occur prior to the
scattering), whereas the CCA describes final-state evolution (the two parton branchings occur after the
scattering). Hence, Wilson lines must be attached only to the three partons from the DA, yielding
\begin{equation}\begin{split}\label{S3}
S_{q\overline{q}g}(\bm{x}, \bm{y}, \bm{z})\,=\,\frac{1}{C_{F}\,N_{c}}\left\langle U^{ba}(\bm{z})\,\mathrm{tr}\left[V(\bm{x})t^{a}V^{\dagger}(\bm{y})t^{b}\right]\right\rangle \,\simeq\,\mathcal{S}(\bm{x}, \bm{z})\,\mathcal{S}(\bm{z}, \bm{y}).
\end{split}\end{equation}
A similar interpretation holds for the third term, with however the DA and the CCA being interchanged with 
each other. Finally the unit term in \eqn{defW} represents the case where the evolution is restricted to
the final state for both the DA and the CCA.

The above considerations also show that at large $N_c$ the overall $S$-matrix structure 
can be fully expressed in terms of dipoles and quadrupoles in the fundamental representation of the colour
group:
\begin{align}\label{colorstr}
\mathcal{W}(\bm{x}, \bm{y}, \bm{z}, \bm{\overline{x}}, \bm{\overline{y}}, \bm{\overline{z}})\simeq
\mathcal{Q}(\bm{x}, \bm{z}, \bm{\overline{z}}, \bm{\overline{x}})
\mathcal{Q}(\bm{z}, \bm{y}, \bm{\overline{y}}, \bm{\overline{z}})-\mathcal{S}(\bm{x}, \bm{z})\,\mathcal{S}(\bm{z}, \bm{y})-\mathcal{S}(\bm{\overline{z}}, \bm{\overline{x}})\,\mathcal{S}(\bm{\overline{y}}, \bm{\overline{z}})+\mathbf{1}.
\end{align}

The remaining three terms within the square brackets in \eqn{R1}, for which we use the concise notation
introduced in \eqn{qqgout}, refer to situations where the scattering in the 3-parton final state gets replaced by scattering in the $q\bar q$ intermediate state (in either the DA, or the CCA, or both). 
As suggested by the notations, the respective
expressions for the kernel and for the $S$-matrix are obtained from Eqs.~\eqref{defK1}
and \eqref{defW} by simultaneously replacing $\bm{y}\rightarrow\bm{y}^{\prime}$ and 
$\bm{z}\rightarrow\bm{y}^{\prime}$, and similarly for the CCA. 
For later purposes, it will be useful to have explicit expressions for the respective
colour structures (at large $N_c$, for simplicity). For the second term, denoted as 
$\big(\bm{y}, \bm{z}\rightarrow\bm{y}^{\prime})$, one has
  \begin{align}\label{colorstr2}
\mathcal{W}(\bm{x}, \bm{y}^{\prime}, \bm{y}^{\prime},\bm{\overline{x}}, \bm{\overline{y}}, \bm{\overline{z}})\simeq
\mathcal{Q}(\bm{x},  \bm{y}^{\prime}, \bm{\overline{z}}, \bm{\overline{x}})
\mathcal{S}(\bm{\overline{y}}, \bm{\overline{z}})-\mathcal{S}(\bm{x},  \bm{y}^{\prime})-\mathcal{S}(\bm{\overline{z}}, \bm{\overline{x}})\,\mathcal{S}(\bm{\overline{y}}, \bm{\overline{z}})+\mathbf{1},
\end{align}
whereas for the third term $(\bm{\overline{y}}, \bm{\overline{z}}\rightarrow\bm{\overline{y}}^{\prime}\big)$, it reads
\begin{align}\label{colorstrbis}
\mathcal{W}(\bm{x}, \bm{y}, \bm{z}, \bm{\overline{x}}, \bm{\overline{y}}^{\prime}, \bm{\overline{y}}^{\prime})\simeq
\mathcal{Q}(\bm{x}, \bm{z}, \bm{\overline{y}}^{\prime}, \bm{\overline{x}})
\mathcal{S}(\bm{z}, \bm{y})-\mathcal{S}(\bm{x}, \bm{z})\,\mathcal{S}(\bm{z}, \bm{y})-\mathcal{S}(\bm{\overline{y}}^{\prime}, \bm{\overline{x}})+\mathbf{1}.
\end{align}
Finally, for the fourth term, as obtained by simultaneously replacing
$\big(\bm{y}, \bm{z}\rightarrow\bm{y}^{\prime})$ and $(\bm{\overline{y}}, \bm{\overline{z}}\rightarrow\bm{\overline{y}}^{\prime}\big)$, one finds
\begin{align}\label{colorfact}
\mathcal{W}(\bm{x}, \bm{y}^{\prime}, \bm{y}^{\prime}, \bm{\overline{x}}, \bm{\overline{y}}^{\prime}, \bm{\overline{y}}^{\prime})\simeq
\mathcal{Q}(\bm{x}, \bm{y}^{\prime}, \bm{\overline{y}}^{\prime}, \bm{\overline{x}})-\mathcal{S}(\bm{x}, \bm{y}^{\prime})-\mathcal{S}(\bm{\overline{y}}^\prime, \bm{\overline{x}})+\mathbf{1}.
\end{align}
Incidentally, we recognise here the same colour structure as at leading order, cf. \eqn{LOcolor}. This should not
be a surprise: this last term corresponds to Feynman graphs where the gluon is emitted in the final state,
i.e. after the scattering with the target. 

At this level, it is easy to write down the other direct contribution, where the gluon is emitted and
reabsorbed by the quark (see Fig.~\ref{fig:3jets_DB}). 
This is obtained from \eqn{R1} by replacing $\vartheta \rightarrow 1-\vartheta-\xi$,
$\bm{R}\to \bm{\widetilde R}$, $\bm{Y}\to \bm{X}$, $\bm{y}^\prime
\to \bm{{x}}^\prime$, and similarly for the transverse coordinates with a bar. 
The corresponding kernel  $\widetilde{\mathcal{K}}_{1}^{imjn}$ is defined as in \eqn{defK1}, but with
the effective vertex $ \widetilde \Phi_{\lambda_{1}\lambda_{2}}^{ijmn}$ from \eqn{vertexPhi2}. The identity  \eqref{phiphi}
remains true, but with $\vartheta \rightarrow 1-\vartheta-\xi$,  of course. The ``subtracted'' terms
representing the topologies where the shockwave is inserted in the intermediate $q\bar q$ state, 
are now obtained via the replacements 
$\big(\bm{x}, \bm{z}\rightarrow\bm{x}^{\prime})$ or/and $(\bm{\overline{x}}, \bm{\overline{z}}\rightarrow\bm{\overline{x}}^{\prime}\big)$. 

The $S$-matrix structure in \eqn{defW} remains unchanged, for the reasons explained
 in Sect.~\ref{sec:gluonQ}. This means that the
colour structure for the particular graph depicted in Fig.~\ref{fig:3jets_DB} is exactly the same
as for that in  Fig.~\ref{fig:3jets_DA}. But this is not true anymore for the other graphs, where
 the shockwave is inserted in the intermediate $q\bar q$ state. For instance, for
the piece  denoted as $\big(\bm{x}, \bm{z}\rightarrow\bm{x}^{\prime})$, one finds
\beq\label{colorstr3}
\mathcal{W}(\bm{x}^{\prime}, \bm{y}, \bm{x}^{\prime}, \bm{\overline{x}}, \bm{\overline{y}}, \bm{\overline{z}})\simeq
\mathcal{S}(\bm{\overline{z}}, \bm{\overline{x}})
\mathcal{Q}(\bm{x}^{\prime}, \bm{y}, \bm{\overline{y}}, \bm{\overline{z}})-\mathcal{S}(\bm{x}^{\prime}, \bm{y})-\mathcal{S}(\bm{\overline{z}}, \bm{\overline{x}})\,\mathcal{S}(\bm{\overline{y}}, \bm{\overline{z}})+\mathbf{1},
\eeq
which is indeed different from the corresponding result for an emission from the antiquark, that is, the term
$\big(\bm{y}, \bm{z}\rightarrow\bm{y}^{\prime})$, as shown  in  \eqn{colorstr2}.

\begin{figure}[t]
  \centering
    \includegraphics[width=0.6\textwidth]{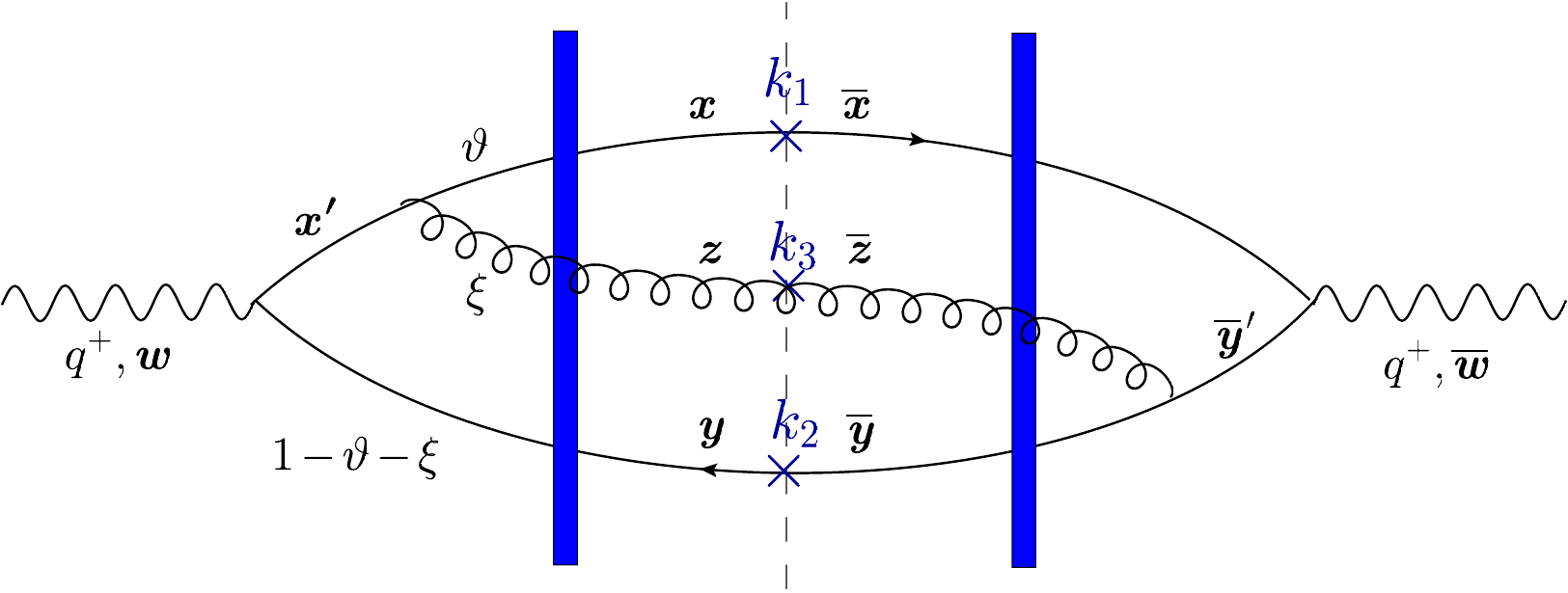}
     \caption{\small A particular interference graph contributing to the 
  leading-order cross-section for $q\bar q g$  production. }
    \label{fig:3jets_inter}
    \end{figure}

\subsubsection{Interference terms}
\label{sec:interf}

Fig.~\ref{fig:3jets_inter} shows a particular interference term, in which the gluon is emitted by
the quark in the DA and reabsorbed by the antiquark in the CCA. As for the direct terms, there are
9 possible topologies, but only 4 terms due to our peculiar way of regrouping terms. The other
type of diagrams, where the gluon is emitted by the antiquark and absorbed by the quark, are
related to those of the first type via complex conjugation. Hence the overall contribution of the
interference graphs reads
\begin{align}\label{R2}
\frac{d\sigma^{\rm inter}}{dk_1^+d^{2}\bm{k}_{1}dk_2^+d^{2}\bm{k}_{2}dk_3^+d^{2}\bm{k}_{3}}&\,=-\,\frac{\alpha_{s}\alpha_{em}\,C_{F} N_{c}\,\vartheta(1-\vartheta-\xi)
Q^2}{2(2\pi)^{10}(q^{+})^{2}\xi}\,\left(\sum e_{f}^{2}\right)\delta(q^{+}-k_1^+-k_2^+-k_3^+)\nonumber\\*[0.2cm]
&\times2\,{\mathcal Re}
\int_{\bm{x}, \bm{y}, \bm{z}, \overline{\bm{x}}, \bm{\overline{y}}, \bm{\overline{z}}}\,e^{-i\bm{k}_{1}\cdot(\bm{x}-\overline{\bm{x}})-i\bm{k}_{2}\cdot(\bm{y}-\overline{\bm{y}})-i\bm{k}_{3}\cdot(\bm{z}-\overline{\bm{z}})}\,\frac{\bm{\widetilde R}^{i}\bm{X}^{m}\,\bm{\overline{R}}^{\,j}\overline{\bm{Y}}^{\,n}}{\bm{X}^{2}\,\bm{\overline{Y}}^{2}}\nonumber\\*[0.2cm]
&\times\Big[\mathcal{K}_{2}^{imjn}(\bm{x},\bm{y},\bm{z},\overline{\bm{x}},\bm{\overline{y}},\bm{\overline{z}},\vartheta,\xi,
{Q})\,\mathcal{W}(\bm{x},\bm{y},\bm{z},\overline{\bm{x}},\bm{\overline{y}},\bm{\overline{z}})\nonumber\\*[0.2cm]
&\left.-\,\big(\bm{x}, \bm{z}\rightarrow\bm{x}^{\prime}\big)\,-\,\big(\bm{\overline{y}}, \bm{\overline{z}}\rightarrow\bm{\overline{y}}^{\prime}\big)
\,+\,\big(\bm{x},\,\bm{z}\rightarrow\bm{x}^{\prime}\,\ \&\ \,\bm{\overline{y}},\,\bm{\overline{z}}\rightarrow\bm{\overline{y}}^{\prime}\big)
\right.\Big].
\end{align}
where the new kernel is defined as 
\begin{align}\label{defK2}
\mathcal{K}_{2}^{imjn}(\bm{x},\bm{y},\bm{z},\overline{\bm{x}},\bm{\overline{y}},\bm{\overline{z}},\vartheta,\xi,{Q})&\,\equiv\,\widetilde{\Phi}_{\lambda_{1}\lambda}^{lirm}(\bm{x},\bm{y},\bm{z}, \vartheta,\xi)\,\Phi_{\lambda_{1}\lambda}^{ljrn\,*}(\bm{\overline{x}},\overline{\bm{y}},\bm{\overline{z}},\vartheta, \xi)\nonumber\\*[0.2cm]
&\,\times\,\frac{\rmK_{1}\left({Q} D(\bm{x}, \bm{y}, \bm{z}, \vartheta, \xi)\right)\,\rmK_{1}\left({Q} D(\bm{\overline{x}}, \overline{\bm{y}}, \bm{\overline{z}}, \vartheta, \xi)\right)}{ D(\bm{x}, \bm{y}, \bm{z}, \vartheta, \xi)\,D(\bm{\overline{x}}, \overline{\bm{y}}, \bm{\overline{z}}, \vartheta, \xi)}\,.
\end{align}
The product of effective vertices with the instantaneous pieces excluded is evaluated similarly to \eqn{phiphi} and reads
\begin{align}\label{phiphi2}
&\widetilde{\Phi}_{\lambda_{1}\lambda}^{lirm}(\vartheta,\xi)\,\Phi_{\lambda_{1}\lambda}^{ljrn\,*}(\vartheta,\xi)\Big|_{{\rm non-inst.}}=
\nonumber\\
&=\,8\left[1-\xi-2\vartheta(1-\xi-\vartheta)\right]\left[\xi(1-\xi)+2\vartheta(1-\xi-\vartheta)\right]\delta^{ij}\delta^{mn}-8\xi(1-2\vartheta-\xi)^{2}\varepsilon^{ij}\varepsilon^{mn}.
\end{align}
Like in \eqn{phiphi}, the ``squared'' vertex is the sum of two contributions, one generating scalar products of transverse vectors,
the other one giving rise to vectorial products. The structure in \eqn{phiphi2} appears to be consistent with the previous results 
in the literature  \cite{Altinoluk:2020qet,Caucal:2021ent,Taels:2022tza}, whenever direct comparisons are 
possible (see e.g. Eq.~(2.32) in \cite{Altinoluk:2020qet}  and Eq.~(B.7) in \cite{Caucal:2021ent}).

The general $S$-matrix (colour) structure is the same as for the direct contributions, that is, it is given by
the function $\mathcal{W}$ shown in \eqn{defW}. When specialised to the ``subtraction term'' denoted as
$\big(\bm{x}, \bm{z}\rightarrow\bm{x}^{\prime}\big)$, this function takes the form in \eqn{colorstr3}, 
whereas for the
term $\big(\bm{\overline{y}}, \bm{\overline{z}}\rightarrow\bm{\overline{y}}^{\prime}\big)$, it is shown in 
 \eqn{colorstrbis}. Finally, for the ``double-subtracted term'' $(\bm{x},\,\bm{z}\rightarrow\bm{x}^{\prime}\,\ \&\ \,\bm{\overline{y}},\,\bm{\overline{z}}\rightarrow\bm{\overline{y}}^{\prime})$, one finds
 \beq\label{colorstr4}
\mathcal{W}(\bm{x}^{\prime}, \bm{y}, \bm{x}^{\prime}, \bm{\overline{x}}, \bm{\overline{y}}^{\prime}, 
\bm{\overline{y}}^{\prime})\simeq
\mathcal{S}(\bm{\overline{y}}^{\prime}, \bm{\overline{x}})
\mathcal{S}(\bm{x}^{\prime}, \bm{y})-\mathcal{S}(\bm{x}^{\prime}, \bm{y})-\mathcal{S}(\bm{\overline{y}}^{\prime}, \bm{\overline{x}})+\mathbf{1}.
\eeq

\subsection{The real NLO corrections to quark-antiquark production in DIS}

The real NLO corrections to (forward) dihadron (or dijet) production in DIS at small $x$ are obtained from the 
leading-order trijet results in the previous subsection, by integrating out the kinematics of the
parton that is not measured in the final state. In what follows we shall explicitly consider only the
case where the unmeasured parton is the gluon (so the ``dijet'' is a $q\bar q $ pair, as at leading order).  
This is indeed the most interesting case, in that it overlaps with both the JIMWLK and the DGLAP 
evolutions and it corresponds to the virtual
corrections to be discussed later on. The results for the two other cases, where one measures either
a quark-gluon pair, or an antiquark-gluon one, can be simply obtained via similar manipulations.

One therefore has
\begin{align}
\label{2q-v0}
\frac{d\sigma^{\gamma A\rightarrow q\overline{q}+X}_\rnlo}{dk_1^+d^{2}\bm{k}_{1}dk_2^+d^{2}\bm{k}_{2}}
=\int dk^+_3 d^2\bm{k}_{3}\,
\frac{d\sigma^{\gamma A\rightarrow q\bar q g}}{dk_1^+d^{2}\bm{k}_{1}dk_2^+d^{2}\bm{k}_{2}dk_3^+d^{2}\bm{k}_{3}}\,,
\end{align}
where the subscript ``rNLO'' stays for real next-to-leading order corrections. The trijet cross-section in the
r.h.s. is the sum of direct and interference contributions, as shown in  Eqs.~\eqref{R1} and \eqref{R1}.
By inspection of these expressions, it is clear that the integral over $k^+_3 $ can be trivially performed 
by using the $\delta$-function for longitudinal momentum conservation (which leaves
a step-function $\Theta(q^+-k^+_1-k^+_2)$ constraining the final momenta), 
whereas the integral over 
$\bm{k}_{3}$ yields a factor $(2\pi)^2\delta^{(2)}(\bm{z}-\overline{\bm{z}})$, 
which allows one to identify the coordinates of the unmeasured gluon in the DA and the CCA, respectively. 
The result of \eqn{2q-v0} can be succinctly written as
\begin{align}
\label{2quarks}
\frac{d\sigma_\rnlo^{\gamma A\rightarrow q\overline{q}+X}}{dk_1^+d^{2}\bm{k}_{1}dk_2^+d^{2}\bm{k}_{2}}=\,
(2\pi)^2\,\frac{d\sigma^{\gamma A\rightarrow q\bar q g}}{dk_1^+d^{2}\bm{k}_{1}dk_2^+d^{2}\bm{k}_{2}dk_3^+d^{2}\bm{k}_{3}}\,\Bigg |_{k^+_3=q^+-k^+_1-k^+_2\,,\ 
\bm{z}=\overline{\bm{z}}},
\end{align}
where it is understood that the trijet cross-section  in the r.h.s. is given by expressions like \eqref{R1}, but without
the $\delta$-function expressing the conservation of longitudinal momentum. It is understood that the
longitudinal momentum fractions of $\vartheta$ and $\xi $ must now be expressed in terms of the
respective momenta of the two measured particles, that is (cf. \eqn{longit1})
\begin{equation}\label{longit2}
 \vartheta=\frac{k_1^+}{q^{+}},\qquad\xi=1-\frac{k_1^+ + k_2^+}{q^{+}}\,,
 \end{equation}
where it is understood that $\vartheta+\xi\le 1$.

The fact that the gluon is not measured brings some simplifications in the structure of the
$S$-matrix in \eqn{defW}: when the gluon scatters in both the DA and the CCA, as is the case
for the first term  in the r.h.s. of \eqn{defW} (recall Fig.~\ref{fig:3jets_DA}), the associated Wilson
lines compensate each other by unitarity, $U^{\dagger}(\bm{z}) U(\bm{z}) =1$, and then
the product of quadrupoles appearing in the second line of  \eqn{S6} reduces to a product
of dipoles (we consider large $N_c$, for simplicity):
\begin{equation}\label{6q1}
S_{q\overline{q}gq\overline{q}g}(\bm{x}, \bm{y}, \bm{z}, \bm{\overline{x}}, \bm{\overline{y}}, \bm{\overline{z}}
=\bm{z})
\,\simeq\,\mathcal{S}(\bm{x},\overline{\bm{x}})\,\mathcal{S}(\bm{y},\overline{\bm{y}}).
 \end{equation}
But for the other terms  in  \eqn{defW}, the identification  $\bm{z}=\overline{\bm{z}}$ brings no
special simplification, because the gluon is not anymore interacting on {\it both} sides of the cut.
Altogether, \eqn{colorstr} gets replaced by
 \begin{align}\label{Wnlo}
\mathcal{W}(\bm{x}, \bm{y}, \bm{z}, \bm{\overline{x}}, \bm{\overline{y}}, \bm{\overline{z}}=\bm{z})\simeq
\mathcal{S}(\bm{x},\overline{\bm{x}})\,\mathcal{S}(\bm{y},\overline{\bm{y}})-
\mathcal{S}(\bm{x}, \bm{z})\,\mathcal{S}(\bm{z}, \bm{y})-\mathcal{S}(\bm{{z}}, \bm{\overline{x}})\,\mathcal{S}(\bm{\overline{y}}, \bm{{z}})+\mathbf{1}.
\end{align}

As for the remaining three terms within the square brackets in \eqn{R1}, where the gluon is emitted after
the scattering (at least, on one side of the cut), it is clear that there is no special simplification in the respective
colour structure. In evaluating these terms, the identification $\bm{z}=\overline{\bm{z}}$ must be performed
only at the very end,  {\it after} making the replacements $\big(\bm{y}, \bm{z}\rightarrow\bm{y}^{\prime})$ 
or/and $(\bm{\overline{y}}, \bm{\overline{z}}\rightarrow\bm{\overline{y}}^{\prime}\big)$.  
(The two types of identifications do not commute with each other,  as one can easily check.)
In particular, the only effect of the identification $\bm{z}=\overline{\bm{z}}$ on an expression like
that in \eqn{colorfact} refers to the ``primed'' variables  $\bm{y}^\prime$ and $\bm{\overline{y}}^\prime$,
which depend upon $\bm{z}$ and respectively $\overline{\bm{z}}$, as visible in \eqn{defyw}.

\section{The soft gluon limit: recovering the B-JIMWLK evolution}
\label{sec:soft}

In this section, we will show that in the limit where the gluon emission is soft ($\xi\to 0$), our previous
results for the NLO corrections reproduce the expected part of the B-JIMWLK evolution of the leading-order
cross-section \eqref{transcross}. By the ``expected part'' we mean those terms in the B-JIMWLK equation
for the quadrupole $\mathcal{Q}\,(\bm{x}, \bm{y}, \overline{\bm{y}}, \overline{\bm{x}})$ in which the soft
gluon is emitted by the leg at $\bm{x}$ or at $\bm{y}$ (i.e. in the DA) and is reabsorbed 
by the leg at $ \overline{\bm{x}}$ or at $ \overline{\bm{y}}$ (in the CCA). All the other pieces of
the evolution equation for the quadrupole, as well as the complete evolution equations for the
dipole $S$-matrices $\mathcal{S}\left(\bm{x}, \bm{y}\right)$ and 
$\mathcal{S}\left(\bm{\overline{y}}, \bm{\overline{x}}\right)$ which enter the LO colour structure in \eqn{LOcolor}
will be generated by the virtual NLO corrections to the dijet cross-section, as we shall later check.
 
 For more generality, we shall start with the case where the soft gluon is measured as well, that is,
 with the LO cross-section for producing a $q\bar q g$ triplet, but such that $\xi\to 0$. In this slightly more 
general case, we will recognise the action of the ``production'' version of
the JIMWLK Hamiltonian \cite{Kovner:2006ge,Kovner:2006wr,Iancu:2013uva}. This
is a generalised version of the JIMWLK Hamiltonian which, when acting on the cross-section for particle production in dilute-dense ($pA$ or $eA$) collisions, leads to the emission of an additional, soft,
gluon, which is measured in the final state. The (relevant part of the) standard  B-JIMWLK equation
for the quadrupole \cite{Dominguez:2011gc,Iancu:2011ns,Iancu:2011nj}
 will be eventually obtained after identifying $\bm{z}=\overline{\bm{z}}$.

The soft gluon limit is obtained from our previous result by keeping only the leading non-trivial terms
 in the limit $\xi\to 0$. One then encounters several types of simplifications. 
 
 First one can neglect the recoil of the emitter (quark or antiquark) at the respective emission vertex,
 meaning that its transverse coordinate is not modified by the soft emission. In practice this means that
 we can neglect the difference between ``primed'' and ``unprimed'' coordinates:
$\bm{y}^\prime\simeq \bm{y}$ for an emission by the antiquark  (recall Fig.~\ref{fig:NLO_Real})
and similarly $\bm{x}^\prime\simeq \bm{x}$ 
 for an emission by the quark (cf. Fig.~\ref{fig:NLO_Realq}). This also implies
 that the transverse size of the $q\bar q $ is not affected by a soft gluon emission; indeed, when $\xi\to 0$,
  Eqs.~\eqref{defRY},  \eqref{defD}  and \eqref{RYX} imply $R\simeq \widetilde R\simeq |\bm{x}-\bm{y}|$,
  whereas $D^2\simeq \vartheta(1-\vartheta) (\bm{x}-\bm{y})^2$.
  From now on, in this section we shall use the notation $R\equiv |\bm{x}-\bm{y}|$ (and similarly
   $ \overline R\equiv| \overline{\bm{x}}-\overline{\bm{y}}|$ in the complex conjugate amplitude).
 One also has $\widetilde Q^2\simeq \bar Q^2 = \vartheta(1-\vartheta) Q^2$.
 
Furthermore, one can take  the limit $\xi\to 0$ within the emission kernels \eqref{defK1}  and
\eqref{defK2}, which entails important simplifications --- both in the effective vertices and in 
the energy denominators. In particular, the contributions of the instantaneous graphs  to the
effective vertices vanish in this limit, cf. Eqs.~\eqref{vertexPhi} and \eqref{vertexPhi2},
and the same is true for the pieces proportional to $\varepsilon^{ij}\varepsilon^{mn}$ in the 
squared vertices \eqref{phiphi} and  \eqref{phiphi2} (indeed, all these contributions are suppressed
by additional factors of $\xi$). If one combines
the respective limits of the emission kernels with the explicit factors involving $\xi$ and $\vartheta$ in equations
like \eqref{R1} and \eqref{R2},
one finds that the dependence upon the longitudinal fractions  $\xi$ and $\vartheta$ becomes
the same for all the 3 contributions to the trijet cross-section --- the direct  contribution of the antiquark \eqref{R1}, 
the similar one due to the quark, and the interference terms \eqref{R2} ---,  and it is such that
the photon decay vertex factorises out from the gluon emission:
\begin{align}\label{Ksoft}\hspace*{-0.5cm}
\frac{\vartheta^{2}}{\xi}\mathcal{K}_{1}^{imjn}\simeq
\frac{(1-\vartheta)^{2}}{\xi}\widetilde
{\mathcal{K}}_{1}^{imjn}\simeq
\frac{\vartheta(1-\vartheta)}{\xi}\mathcal{K}_{2}^{imjn}\simeq
16\delta^{ij}\delta^{mn} \,\vartheta(1-\vartheta)\,\frac{\vartheta^2+\left(1-\vartheta\right)^{2}}{\xi}\,
\frac{K_1(\bar Q R) K_1(\bar Q \overline R)}{R \overline R}\,,\nn
\end{align}
So, the only difference between these three channels which subsists in the soft gluon limit refers
the gluon emission vertex, which involves the appropriate transverse separation, $\bm{X}=\bx-\bz$ or $\bm{Y}=\by-\bz$,
between the gluon and its emitter. 

 \begin{figure}[!t]\center
 \includegraphics[scale=0.95]{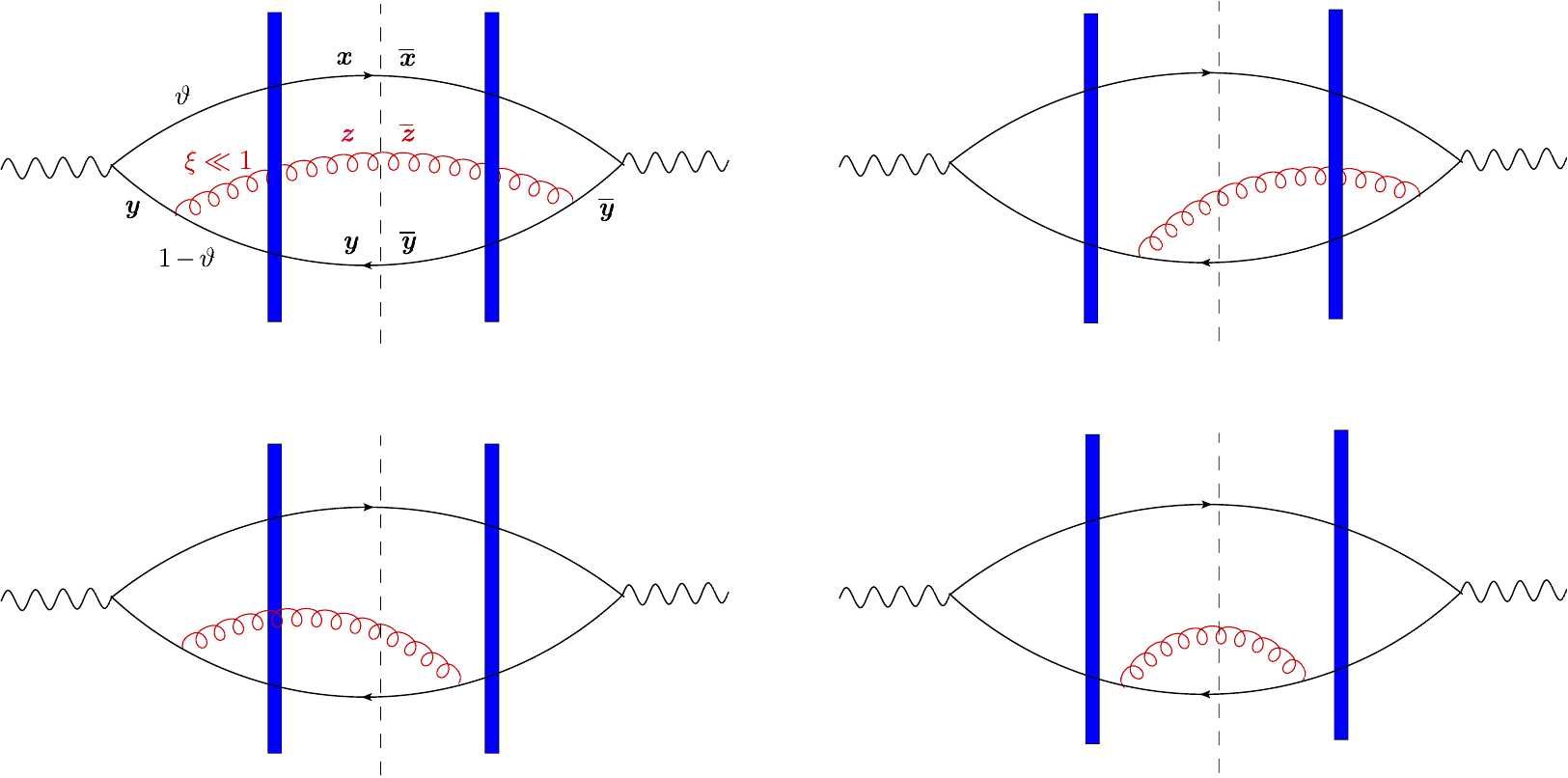}
\caption{The four diagrams which give the dominant contribution to the piece of the cross-section in \eqref{R1}
(direct emission by the antiquark) in
the limit where the emitted gluon is soft ($\xi\ll 1$).}
\label{3jets-soft}
\end{figure}

\begin{figure}[!t]\center
 \includegraphics[scale=0.95]{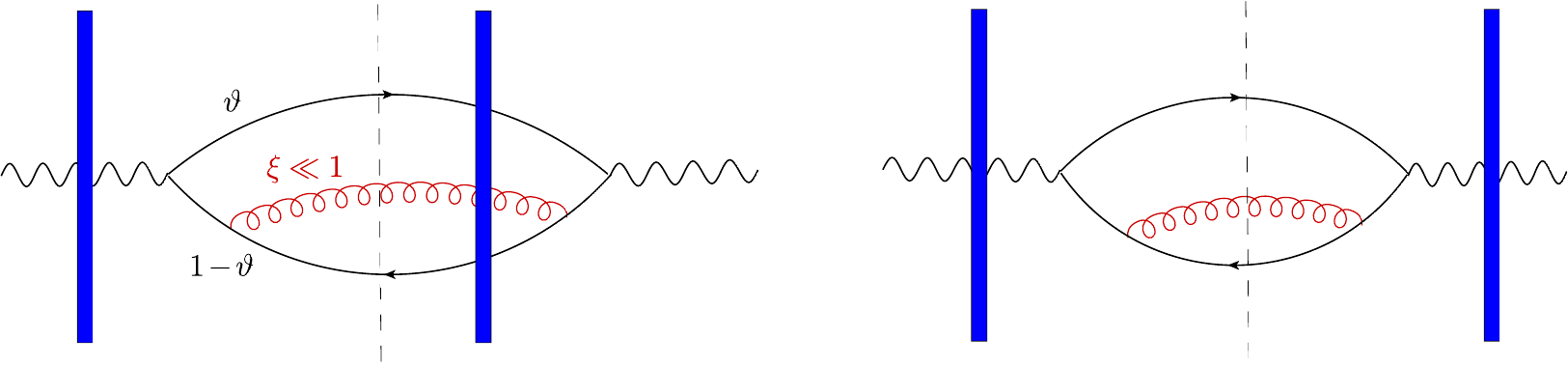}
\caption{Diagrams which contribute to the $q\bar q g$ cross-section \eqref{R1}
in general, but do not survive in the soft limit $\xi\ll 1$.}
\label{3jets-nonsoft}
\end{figure}

The ``degeneracy'' in \eqn{Ksoft} has yet another important consequence: it implies that
the four terms contributing to any of the three channels under consideration (i.e. the 
four terms within the square brackets in \eqn{R1}, or those within the square brackets in \eqn{R2})
are now multiplied by exactly the same emission kernel. Consider e.g.  \eqn{R1}, which we recall
represents direct emissions by the antiquark: after the simplifications
in  \eqn{Ksoft}, an operation like $\big(\bm{y}, \bm{z}\rightarrow\bm{y}^{\prime})$ --- which now
reduces to $\bm{z}\rightarrow\bm{y}$, since $\bm{y}=\bm{y}^{\prime}$ anyway --- has no effect on
the kernel, but only on the $S$-matrix structure exhibited in \eqn{defW}. Accordingly, the colour structures
associated with the four terms  in \eqn{R1} are simply added to each other and many of the individual 
$S$-matrices (12 over a total of 16) cancel in their sum. The 4 surviving terms correspond to the 4 graphs
shown in Fig.~\ref{3jets-soft} for the case of \eqn{R1}.
The 12 missing terms, which in fact correspond to only 5 distinct topologies
(due to our rearrangement of the three amplitude graphs in Fig.~\ref{fig:NLO_Real}),
are those where the photon has crossed the shockwave prior to its decay, in either the DA, or the CCA, or both.
That is, these are the contributions to the cross-section which involve the graph in Fig.~\ref{fig:NLO_Real}.c
on at least one side of the cut (two such diagrams are shown in Fig.~\ref{3jets-nonsoft}). 

Physically,
these simplifications can be understood as follows: the soft gluon with longitudinal momentum $p^+=\xi
q^+$ and transverse momentum $\bm{p}$ has a short formation time $\tau_p=2\xi q^+/\bm{p}^2$,
hence in order to affect the scattering (and the ensuing particle production), it must occur relatively
close to the shockwave, with a distance $\Delta x^+\sim \tau_p$ away from it. The diagrams shown in Fig.~\ref{3jets-nonsoft} are suppressed in this limit since the photon
decay must occur even closer to the shockwave (at least, on one side of the cut), hence the respective
phase-space is considerably reduced compared to that available to the diagrams in Fig.~\ref{3jets-soft},
where the position of the photon vertex is unconstrained.

This physical interpretation becomes transparent by inspection of the energy denominators 
corresponding to the three graphs in Fig.~\ref{fig:NLO_Real}. For the first 2 graphs,
in Fig.~\ref{fig:NLO_Real}.a and \ref{fig:NLO_Real}.b respectively,
there is only one energy denominator involving the gluon. In the soft limit $\xi\ll 1$, the respective
energy difference is dominated by the gluon light-cone energy $p^-=\bm{p}^2/2\xi q^+$
(recall Eqs.~\eqref{enden2} and \eqref{enden3}):
\beq
E_{q\overline{q}g}-E_{\overline{q}q}\,\simeq\,E_{q\overline{q} g}-E_{\gamma}\,\simeq\,
\frac{\bm{p}^2}{2q^+\xi} \,,\eeq
which shows that the longitudinal phase-space for the soft gluon emission is of order $\tau_p\sim\xi$,
as anticipated.
However, the denominator corresponding to the graph in Fig.~\ref{fig:NLO_Real}.c
involves the product $(E_{q\overline{q}}-E_{q\overline{q}g})(E_{\gamma}-
E_{q\overline{q}g})$ of two large energy differences, hence that terms scales like
$\tau_p^2\sim\xi^2$ and is strongly suppressed as $\xi\to 0$.

After exploiting all the aforementioned simplifications and adding together the direct and interference
contributions, the soft gluon limit of the trijet cross-section is obtained as 
 \begin{align}\label{qggeik}
 \hspace*{-0.6cm}
&\frac{d\sigma^{\gamma A\rightarrow q\bar q g}}{dk_1^+\,d^{2}\bm{k}_{1}\,dk_2^+\,d^{2}\bm{k}_{2}\,dk_3^+\,d^{2}\bm{k}_{3}}\,\bigg |_{\xi\to 0} =
\frac{2\alpha_{em}\,N_{c}}{(2\pi)^{6}q^{+}}\left(\vartheta^{2}+(1-\vartheta)^{2}\right)\,\left(\sum e_{f}^{2}\right)\,\delta(q^{+}-k_1^+-k_2^+)\nonumber \\*[0.2cm]
&\qquad\times\,\int_{\overline{\bm{x}}, \bm{\overline{y}}, \bm{x}, \bm{y}}\,e^{-i\bm{k}_{1}\cdot(\bm{x}-\overline{\bm{x}})-i\bm{k}_{2}\cdot(\bm{y}-\overline{\bm{y}})}\,\frac{\bm{R}\cdot\bm{\overline{R}}}{R \overline{R}}\,\bar{Q}^{2}\,\rmK_{1}\big(\bar{Q}R\big)\rmK_{1}\big(\bar{Q} \overline{R}\big)
\nonumber \\*[0.2cm]
&\qquad\times
\frac{4}{(2\pi)^{4}}\,
\frac{\alpha_{s} C_{F}}{\xi q^{+}}
\int_{\bm{\overline{z}},\bm{z}}\,e^{-i\bm{k}_{3}\cdot(\bm{z}-\overline{\bm{z}})}
\,\Delta_{\rm real} \mathcal{Q}(\bm{x}, \bm{y}, \bm{\overline{y}}, \bm{\overline{x}}\,;\, \bm{z},\overline{\bm{z}}) \end{align}
where we have also ignored the soft momentum $k_2^+$ in the
$\delta$--function for longitudinal momentum conservation. In the first two lines of this formula, 
we have factorised the would-be leading-order cross-section
\eqref{transcross}, but without the respective $S$-matrix structure \eqref{LOcolor}.
The third line encodes the information about the soft gluon emission: the emission kernels in
transverse coordinate space and the colour structures corresponding to all possible
topologies. Specifically, one finds
 \begin{align}\hspace*{-0.4cm}
 \label{deltaQreal}
 \Delta_{\rm real} \mathcal{Q}(\bm{x}, \bm{y}, \bm{\overline{y}}, \bm{\overline{x}}\,;\, \bm{z},\overline{\bm{z}}) 
 &\,\equiv \left[
\frac{\bm{X}\cdot\overline{\bm{X}}}{\bm{X}^{2}\overline{\bm{X}}^{2}} +
\frac{\bm{Y}\cdot\overline{\bm{Y}}}{\bm{Y}^{2}\overline{\bm{Y}}^{2}} 
- \frac{\bm{X}\cdot\overline{\bm{Y}}}{\bm{X}^{2}\overline{\bm{Y}}^{2}} 
- \frac{\bm{Y}\cdot\overline{\bm{X}}}{\bm{Y}^{2}\overline{\bm{X}}^{2}} \right]
\mathcal{Q}(\bm{x}, \bm{z}, \bm{\overline{z}}, \bm{\overline{x}})\,\mathcal{Q}(\bm{z}, \bm{y}, \bm{\overline{y}}, \bm{\overline{z}})
\nonumber \\*[0.2cm] 
&\, +\left[
\frac{\bm{X}\cdot\overline{\bm{X}}}{\bm{X}^{2}\overline{\bm{X}}^{2}} +
\frac{\bm{Y}\cdot\overline{\bm{Y}}}{\bm{Y}^{2}\overline{\bm{Y}}^{2}} \right]
\mathcal{Q}(\bm{x}, \bm{y}, \bm{\overline{y}}, \bm{\overline{x}})\nonumber \\*[0.2cm] 
&\,
 -\left[\frac{\bm{X}\cdot\overline{\bm{Y}}}{\bm{X}^{2}\overline{\bm{Y}}^{2}} 
+ \frac{\bm{Y}\cdot\overline{\bm{X}}}{\bm{Y}^{2}\overline{\bm{X}}^{2}} \right]
\mathcal{S}(\bm{x}, \bm{y})
\mathcal{S}(\bm{\overline{y}}, \bm{\overline{x}})
\nonumber \\*[0.2cm] 
&\,-\left[\frac{\bm{X}\cdot\overline{\bm{X}}}{\bm{X}^{2}\overline{\bm{X}}^{2}}- \frac{\bm{X}\cdot\overline{\bm{Y}}}{\bm{X}^{2}\overline{\bm{Y}}^{2}} \right]
\mathcal{Q}(\bm{x}, \bm{y}, \bm{\overline{y}}, \bm{\overline{z}})\,\mathcal{S}(\overline{\bm{z}},\overline{\bm{x}})
\nonumber \\*[0.2cm] 
&\,-\left[\frac{\bm{Y}\cdot\overline{\bm{Y}}}{\bm{Y}^{2}\overline{\bm{Y}}^{2}} 
- \frac{\bm{Y}\cdot\overline{\bm{X}}}{\bm{Y}^{2}\overline{\bm{X}}^{2}} \right]
\mathcal{Q}(\bm{x}, \bm{y}, \bm{\overline{z}}, \bm{\overline{x}})\,\mathcal{S}(\overline{\bm{y}},\overline{\bm{z}})
\nonumber \\*[0.2cm] 
&\,-\left[\frac{\bm{X}\cdot\overline{\bm{X}}}{\bm{X}^{2}\overline{\bm{X}}^{2}}- \frac{\bm{Y}\cdot\overline{\bm{X}}}{\bm{Y}^{2}\overline{\bm{X}}^{2}} \right]
\mathcal{S}({\bm{x}},{\bm{z}})\,
\mathcal{Q}(\bm{z}, \bm{y}, \bm{\overline{y}}, \bm{\overline{x}})
\nonumber \\*[0.2cm] 
&\,-\left[\frac{\bm{Y}\cdot\overline{\bm{Y}}}{\bm{Y}^{2}\overline{\bm{Y}}^{2}} 
- \frac{\bm{X}\cdot\overline{\bm{Y}}}{\bm{X}^{2}\overline{\bm{Y}}^{2}} \right]\mathcal{Q}(\bm{x}, \bm{z}, \bm{\overline{y}}, \bm{\overline{x}})\, \mathcal{S}(\bm{z},\bm{y}).
 \end{align}
The diagrammatic interpretation of the various terms in this lengthy expression can be easily traced
back by inspection of their respective emission kernel and $S$-matrix structure. For instance the 4 terms
proportional to $({\bm{Y}\cdot\overline{\bm{Y}}})/({\bm{Y}^{2}\overline{\bm{Y}}^{2}})$ correspond 
to the 4 graphs  in Fig.~\ref{3jets-soft}. The associated colour structures are recognised as
the first terms in Eqs.~\eqref{colorstr},  \eqref{colorstr2}, \eqref{colorstrbis}, and
 \eqref{colorfact}, respectively (with $\bm{y}=\bm{y}^{\prime}$, of course). 
 
The expression in the third line of \eqn{qggeik} represents indeed 
the action of the ``production'' JIMWLK Hamiltonian on the original quadrupole 
$\mathcal{Q}(\bm{x}, \bm{y}, \bm{\overline{y}}, \bm{\overline{x}})$ from \eqn{LOcolor}.
This can be verified e.g. by comparing \eqn{deltaQreal} with Eq.~(A.5) in 
Ref.~\cite{Iancu:2013uva}. In making such a comparison, it is useful to recall that
at large $N_c$ one can approximate  $2\alpha_s C_F \simeq \alpha_s N_c \equiv\pi\abar$.
Furthermore, the standard version of the B-JIMWLK equation
for the quadrupole \cite{Dominguez:2011gc,Iancu:2011ns,Iancu:2011nj}
is easily recovered after integrating out both the transverse momentum $\bm{k}_{3}$
(which leads to identifying $\bm{z}=\overline{\bm{z}}$) and the longitudinal momentum 
$k^+_3=\xi q^+$ (which introduces the measure factor $dk^+_3=q^+ d \xi$) of the
soft gluon. The result can be written as en evolution equation with $\ln(1/\xi)$, namely,
\begin{align}\label{evolQ}
\xi\frac{\partial \mathcal{Q}(\bm{x}, \bm{y}, \bm{\overline{y}}, \bm{\overline{x}})}
{\partial \xi}\Big |_{\rm real} \,=\,\frac{\abar}{2\pi}\int_{\bm{z}}
 \Delta_{\rm real} \mathcal{Q}(\bm{x}, \bm{y}, \bm{\overline{y}}, \bm{\overline{x}}\,;\, \bm{z}=\overline{\bm{z}}),
\end{align}
whose r.h.s. is easily checked to precisely coincide with all the relevant terms in e.g. Eq.~(2.17)
from Ref.~\cite{Iancu:2011ns} (the terms which are missing in this comparison will be generated
by the virtual NLO corrections to dijet production, as we shall see \cite{Yair:virtual}).

\section{The collinear limit: recovering the DGLAP evolution}
\label{sec:realDGLAP}

An another interesting limit of our previous results for the ``real'' NLO corrections to the
dihadron production is the limit where the gluon is emitted in the final state and it is nearly collinear
with its emitter. In this limit, the cross-section is expected to develop a logarithmic divergence
from the integral over the transverse momentum of the gluon relative to its emitter. Physically, 
this divergence signals the fact that the gluon is emitted longtime after the interaction and should
be viewed as a part of the wavefunction of the measured parton. For this interpretation
to be correct, the collinear divergence should factorise from the ``hard process'' 
--- here, the photon decay into a $q\bar q$ pair and the scattering between this pair and the
nuclear target --- and be recognised as the first step in the 
DGLAP evolution of the fragmentation function of the parent parton.
In this section, we shall verify that this is indeed the case for the gluon emission by the antiquark.

Since our results for the cross-sections are expressed in the transverse coordinate
representation, it is convenient to also formulate the collinear limit in this representation.
For definiteness, les us consider the case where the gluon is emitted by the antiquark.
In coordinate space, the collinear limit for this emission is the limit where the transverse separation 
$\bm{Y}=\bm{y}-\bm{z}$ between the daughter partons (the antiquark and the gluon)
is much larger than the separation $\bm{R}= \bm{x}- \bm{y}^\prime$ between the $q\bar q$ pair 
produced by the photon decay: $ |\bm{Y}|\gg |\bm{R}|$. 
To see this, it is convenient to change the transverse momentum
variables for the final $\bar q g$ pair from $\bm{k}_{2}$ and $\bm{k}_{3}$, to $\bm{K}$
and $\bm{P}$, defined as
\beq\label{coll-phase}
\bm{K}\equiv \bm{k}_{2}+\bm{k}_{3},\quad
\bm{P}\equiv \frac{(1-\vartheta-\xi)\bm{k}_{3}-\xi  \bm{k}_{2}}{1-\vartheta}
\quad\Longrightarrow\quad 
e^{-i\bm{k}_{2}\cdot \bm{y}-i\bm{k}_{3}\cdot \bm{z}}\,=\,
e^{-i\bm{K}\cdot \bm{y}^\prime+i\bm{P}\cdot \bm{Y}}\,.
\eeq
Clearly,  $\bm{K}$ is the total transverse momentum of the $\bar q g$ pair 
and is conjugated to the transverse position $\bm{y}^\prime$ of its center-of-energy (cf. \eqn{defyw}), 
whereas $\bm{P}$ is the  {\em relative}  transverse momentum of that pair
and is conjugated to the transverse separation $\bm{Y}$.  The collinear limit corresponds
to  $\bm{P}\to 0$. (Indeed, when $\bm{P}=0$, the gluon takes a transverse momentum fraction equal to its
longitudinal splitting fraction $\xi/(1-\vartheta)$, meaning it is collinear with the antiquark; recall the discussion after
\eqn{enden2}.) The structure of the phases in \eqn{coll-phase} confirms that the collinear limit $\bm{P}\to 0$ corresponds
to large values of $Y=|\bm{Y}|$, as anticipated. In turn, this implies that the gluon must be emitted after the scattering
with the nuclear shockwave, in order to escape the virtuality barrier.

Indeed, as discussed in relation with \eqn{Dbarq}, the amplitude for a gluon emission occurring prior to the collision 
--- as described by the LCFW \eqref{qqg1} (or, equivalently, by the first term in  \eqn{qqgout}) --- is
proportional to $\rmK_{1}(Q  D)$, which implies $Y^2\lesssim 1/(\xi Q^2)$. On the other hand, there is no 
such a restriction for  a gluon emitted after the collision,  cf. Fig.~\ref{fig:NLO_Real}.b  --- the respective
LCWF \eqn{qqg1} (or the second term in  \eqn{qqgout}) is instead proportional to $\rmK_{1}(\bar Q R)$,
which is independent of $Y$. This argument also shows that the
instantaneous piece of the effective vertex \eqref{vertexPhi} does not contribute to the collinear limit.
Clearly, a similar argument holds for gluon emissions by the quark.

From the experience with the DGLAP evolution, we expect that the terms expressing interferences 
between gluon
emissions by the quark and the antiquark (cf. Sect.~\ref{sec:interf}) should not contribute in the collinear limit. 
The respective argument is quite obvious 
in momentum space, since the gluon cannot be simultaneously collinear to both the quark and the antiquark
(except for special configurations of zero measure). It is interesting to see how this argument works in
the transverse coordinate representation. Consider a final-state gluon which is emitted by the antiquark
in the direct amplitude (DA) and by the quark in the complex-conjugate amplitude (CCA). 
Since the final momenta are the same on both sides of the cut, one can use the momentum variables  
$\bm{K}$ and $\bm{P}$ defined in \eqn{coll-phase} (instead of $\bm{k}_2$ and $\bm{k}_3$)
in both the DA and the CCA. In both cases, $\bm{P}$
is conjugated to the transverse separation between the gluon and the final antiquark
--- that is,  to $\bm{Y}=\bm{y}-\bm{z}$ in the DA and, respectively, to
 $\bm{\overline{Y}}=\bm{\overline{y}}-\bm{\overline{z}}$ in the CCA. The collinear limit for 
 an emission by the antiquark corresponds to $P_\perp=|\bm{P}|\to 0$, which implies that $Y$
 and $\overline{Y}$ are simultaneously large. In the DA, where the  transverse position $\bm{y}^\prime$
 of the intermediate antiquark is different from the final one $\bm{y}$, the limit  $Y\to \infty$ brings
 no constraint on the transverse size  $R= |\bm{x}- \bm{y}^\prime|$ of the $q\bar q$ pair
 prior to the gluon emission\footnote{The coordinate $\bm{y}^\prime$ of the intermediate antiquark is the
 same as the center of energy of the final quark-gluon pair, cf. \eqn{defyw}. So $\bm{y}^\prime$ can be close to  
 $\bm{x}$ despite the fact that the {\it final} antiquark and the gluon are very far away from each other: in general,
 they are also far away from both the quark and the intermediate antiquark.}.
  But in the CCA, where the gluon is emitted by the quark, one has 
  $\bm{\overline{y}}^\prime=\bm{\overline{y}}$, while the transverse coordinate  $\bm{\overline{x}}^\prime$
  of  the intermediate quark coincides with the center
 of energy of the final quark-gluon pair (cf. \eqn{RYX}). So, the distance
  $\widetilde{R}\equiv |\bm{\overline{x}}^\prime-\bm{\overline{y}}|$ is typically commensurable 
 with $\overline{Y}$ and can be arbitrarily large. Accordingly,  this interference term is strongly damped by the
 Bessel function $\rmK_{1}(\widetilde Q \widetilde R)$ with $\widetilde R \sim \overline{Y}\gg 1/\widetilde Q$
 (recall \eqn{tildeQD}).
  
\begin{figure}[!t]\center
 \includegraphics[scale=1.]{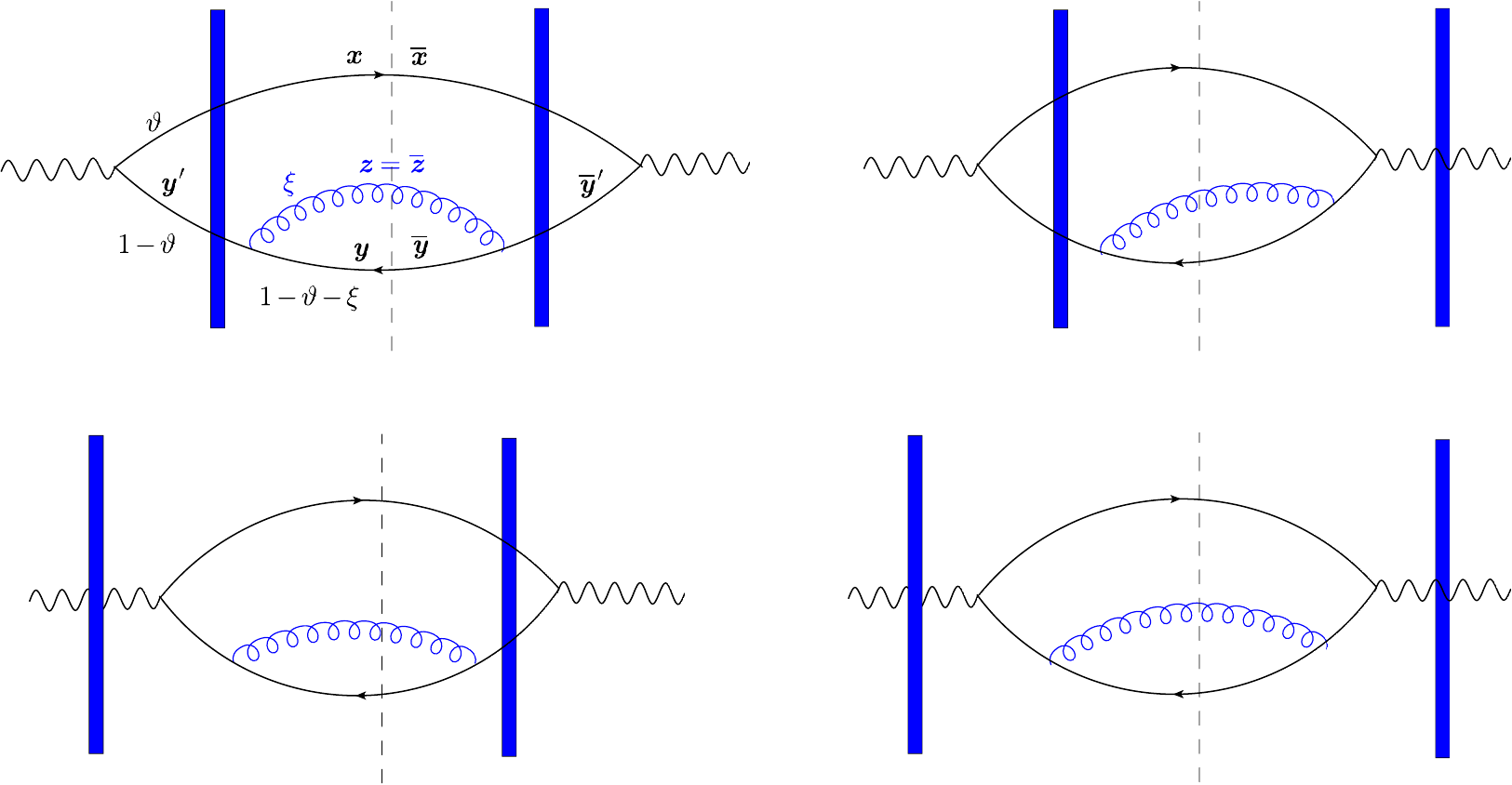}
\caption{The four graphs surviving in the collinear limit for the gluon emission by the antiquark. 
These graphs are void of final-state interactions.}
\label{fig:qDGLAP}
\end{figure}

  To summarise, the dominant contributions to the tri-parton cross-section in the collinear limit
  come from {\it direct} gluon emissions (by either the quark, or the antiquark) 
  in the {\it final state} (i.e. after the collision with the SW). E.g., when the emitter is the antiquark,
  these contributions  correspond to the 4 graphs shown in Fig.~\ref{fig:qDGLAP}.  Their sum is
  contained in the last term, denoted as $\big(\bm{y}, \bm{z}\rightarrow\bm{y}^{\prime})\,\&\,(\bm{\overline{y}}, \bm{\overline{z}}\rightarrow\bm{\overline{y}}^{\prime}\big)$, in \eqn{R1}.   The associated colour structure, 
 shown in \eqn{colorfact}, is the same as at leading-order, cf. \eqn{LOcolor}, 
 which is a necessary condition for factorisation. 
   
To complete the proof of factorisation in the collinear limit, one must demonstrate that 
the tensorial kernel ${\mathcal{K}}_{1}^{imjn}$ multiplying the fourth term in \eqn{R1}
factorises in this limit. To start with, let us check that the last piece, proportional to 
$\varepsilon^{ij}\varepsilon^{mn}$, in the product  \eqref{phiphi} of the effective vertices 
does not contribute to this limit. To that aim, it is convenient to change the integration variables in
 \eqn{R1} from $\bx,\by, \bz$ to $\bx$, $\by',\bm{Y}$, so that the Fourier phases take the form
in  \eqn{coll-phase} (and similarly in the CCA). This is convenient since, in these new variables and
for final-state gluon emissions, the Fourier transforms over $\bm{Y}$ and $\bm{\overline{Y}}$
factorise from the other integrations\footnote{This is perhaps most obvious by inspection of the respective
amplitude in \eqn{qqg2}: the variable $\bm{Y}$ enters this amplitude only via the gluon emission kernel
$\propto \bm{Y}^{n}/\bm{Y}^{2}$.} and are easily computed as
\beq
\int_{\bm{Y},\bm{\overline{Y}}}\,\rme^{i\bm{P}\cdot (\bm{Y}-\bm{\overline{Y}})}
\frac{\bm{Y}^{m}\,\overline{\bm{Y}}^{\,n}}{\bm{Y}^{2}\,\bm{\overline{Y}}^{2}}\,=\,(2\pi)^2\frac{\bm{P}^{m} \bm{P}^{n}}
{\bm{P}^4}\,.
\eeq
When this structure is multiplied with the squared vertex \eqref{phiphi}, the last term in the latter,
which is antisymmetric in $m$ and $n$, will clearly yield a vanishing contribution.
Using the remaining part of \eqn{phiphi}, which is symmetric in $m$ and $n$,
 together with the fact that $Q D\to \bar Q R$ for final state emissions, one finds
\begin{align}\label{Kcoll}
\frac{\vartheta^{2}}{\xi}\mathcal{K}_{1}^{imjn}\bigg |_{\rm 4th\, term}\,=\,32
\delta^{ij}\delta^{mn} \,{\vartheta}P_{\gamma\to q}(\vartheta)\,
P_{q\to g}\left(\frac{\xi}{1-\vartheta}\right)\frac{\rmK_1(\bar Q R) \rmK_1(\bar Q \overline R)}{R \overline R}\,,
\end{align}
where 
\beq
P_{\gamma\to q}(z)\equiv \frac{z^2+(1-z)^2}{2}\,,\qquad
P_{q\to g}(z)\equiv \frac{1+(1-z)^2}{2z}\,,
\eeq
are the DGLAP splitting functions for the processes $\gamma\to q\bar q$ and $q\to qg$, respectively.

The factorisation of the collinear gluon emission from the hard process is now manifest: first, the virtual
photon fluctuates into a $q\bar q$ pair with transverse coordinates $\bm{x}$ and $\bm{y}^\prime$,
respectively. Then this pair scatters off the nuclear SW, with the $S$-matrix shown in  \eqn{colorfact}.
Finally, the antiquark emits a gluon with splitting fraction $z={\xi}/{(1-\vartheta)}$.

Putting things together, one finds (after also integrating over the phase-space of the unmeasured gluon,
which removes the $\delta$-function for longitudinal momentum conservation and identifies  $\bm{z}=\overline{\bm{z}}$) 
\begin{align}\label{qqcoll1}
\frac{d\sigma^{\gamma A\rightarrow q\overline{q}+X}_\rnlo}{dk_1^+d^{2}\bm{k}_{1}dk_2^+d^{2}\bm{k}_{2}}
\bigg |_{(a),\, \rm coll}
&\simeq \frac{4\alpha_{em}\,N_{c}}{(2\pi)^{6}(q^+)^2(1-\vartheta)}\,P_{\gamma\to q}(\vartheta)\,\left(\sum e_{f}^{2}\right)\nonumber\\*[0.2cm]
 &\times\,\int_{\overline{\bm{x}}, \bm{\overline{y}}, \bm{x}, \bm{y}}\,e^{-i\bm{k}_{1}\cdot(\bm{x}-\overline{\bm{x}})-i\bm{k}_{2}\cdot(\bm{y}-\overline{\bm{y}})}\,\frac{\bm{R}\cdot\bm{\overline{R}}}{R \overline{R}}\,\bar{Q}^{2}\,\rmK_{1}\big(\bar{Q}R\big)\rmK_{1}\big(\bar{Q} \overline{R}\big)
\nonumber \\*[0.2cm]
&\times\Big[\mathcal{Q}(\bm{x}, \bm{y}^{\prime}, \bm{\overline{y}}^{\prime}, \bm{\overline{x}})-\mathcal{S}(\bm{x}, \bm{y}^{\prime})-\mathcal{S}(\bm{\overline{y}}^\prime, \bm{\overline{x}})+\mathbf{1}\Big]
 \nonumber\\*[0.2cm] &\times  \frac{4\alpha_s C_F}{(2\pi)^2}\,
P_{q\to g}\left(\frac{\xi}{1-\vartheta}\right)\int_{\bm{z}} \frac{\bm{Y}\cdot\overline{\bm{Y}}}{\bm{Y}^{2}\overline{\bm{Y}}^{2}} 
\,,
 \end{align}
where the longitudinal fractions $\vartheta$ and $\xi$ are given by \eqn{longit2}.
 
At this level, it is convenient to replace the integration variables $\bm{y}$ and $\bm{\overline{y}}$
by  $\bm{y}^\prime$ and $\bm{\overline{y}}^{\prime}$,  respectively  (since the latter are the actual
arguments of both the photon decay probability (e.g.  $\bm{R}=\bm{x}- \bm{y}^\prime$)
and the $S$-matrix structure), and also to use $\vartheta$ and $z_2$ as independent
longitudinal momentum fractions, instead of $\vartheta$ and $\xi$. Here, $z_2$ is
the splitting fraction of the daughter antiquark at that $\bar q\to\bar q g$ vertex:
 \beq\label{defz2}
 z_2\equiv 1- \frac{\xi}{1-\vartheta} = \frac{k_2^+}{q^+-k_1^+}\quad\Longrightarrow\quad
\bm{y}^\prime=z_2\bm{y}+(1-z_2)\bm{z}\,. \eeq
One thus finds
\begin{align}\label{qqcoll}
\frac{d\sigma^{\gamma A\rightarrow q\overline{q}+X}_\rnlo}{dk_1^+d^{2}\bm{k}_{1}dk_2^+d^{2}\bm{k}_{2}}
\bigg |_{(a),\, \rm coll}
&\simeq \frac{4\alpha_{em}\,N_{c}}{(2\pi)^{6}(z_2q^+)^2(1-\vartheta)}\,P_{\gamma\to q}(\vartheta)\,\left(\sum e_{f}^{2}\right)\nonumber\\*[0.2cm]
 &\times\,\int_{\overline{\bm{x}}, \bm{\overline{y}}^\prime, \bm{x}, \bm{y}^\prime}\,e^{-i\bm{k}_{1}\cdot(\bm{x}-\overline{\bm{x}})-i{\bm{k}_{2}}\cdot(\bm{y}^\prime-\overline{\bm{y}}^\prime)/{z_2}}\,\frac{\bm{R}\cdot\bm{\overline{R}}}{R \overline{R}}\,\bar{Q}^{2}\,\rmK_{1}\big(\bar{Q}R\big)\rmK_{1}\big(\bar{Q} \overline{R}\big)
\nonumber \\*[0.2cm]
&\times\Big[\mathcal{Q}(\bm{x}, \bm{y}^{\prime}, \bm{\overline{y}}^{\prime}, \bm{\overline{x}})-\mathcal{S}(\bm{x}, \bm{y}^{\prime})-\mathcal{S}(\bm{\overline{y}}^\prime, \bm{\overline{x}})+\mathbf{1}\Big]
 \nonumber\\*[0.2cm] &\times  \frac{4\alpha_s C_F}{(2\pi)^2}\,
P_{q\to q}\left(z_2\right)\int_{\bm{z}} \frac{(\bm{y}^\prime-\bm{z})\cdot(\overline{\bm{y}}^\prime-\bm{z})}{(\bm{y}^\prime-\bm{z})^{2}(\overline{\bm{y}}^\prime-\bm{z})^{2}} 
\,,
 \end{align}
where we have also used $P_{q\to g}(1-z_2) =P_{q\to q}(z_2) $. 
Via these manipulations, 
the collinear divergence has been isolated in the last term and also factorised from the leading-order
cross-section for the production of a pair of jets (cf. \eqref{transcross}): 
a quark with 3-momentum $k_1=(k_1^+,\bm{k}_1)$
and an antiquark with 3-momentum $p=(p^+,\bm{p})$, where $p^+=k_2^+/z_2$ and 
$\bm{p}=\bm{k}_2/z_2$. 

The collinear divergence refers to the  large--$|\bm{z}|$ limit of the last integral in \eqn{qqcoll}. 
To exhibit this singularity, it is convenient to make an excursion through momentum space and
 introduce a low-momentum cutoff $\Lambda$ on the transverse momentum of the
 unmeasured gluon, meaning an upper cutoff $\sim 1/\Lambda$ on the transverse separations
 $|\bm{y}^\prime-\bm{z}|$ and $|\overline{\bm{y}}^\prime-{\bm{z}}|$.
 One thus finds
 \beq\label{colldiv}
\int_{\bm{z}} \frac{(\bm{y}^\prime-\bm{z})\cdot(\overline{\bm{y}}^\prime-\bm{z})}{(\bm{y}^\prime-\bm{z})^{2}(\overline{\bm{y}}^\prime-\bm{z})^{2}} =
 \int\frac{d^2\bm{p}}{\bm{p}^2}\,e^{-i\bm{q}\cdot(\bm{y}^\prime-\overline{\bm{y}}^\prime)}\,
 \Theta(p-\Lambda)\,=\,\pi\ln\frac{1}{r^2\Lambda^2}
 \,=\,\pi\ln\frac{1}{r^2\mu^2} + \pi\ln\frac{\mu^2}{\Lambda^2}
 \,,\eeq
 where $r\equiv |\bm{y}^\prime-\overline{\bm{y}}^\prime|$ and it is understood that $\Lambda \ll 1/r$
 (since one is eventually interested in the limit $\Lambda \to 0$).
 In the last step, we have split the result between a would-be divergent piece which is independent
 of $r$ and which will be shortly absorbed into the DGLAP evolution and an $r$--dependent piece
 which is a part of the NLO correction to the hard factor. The factorisation scale $\mu$ is {\it a priori}
 arbitrary (within the range $\Lambda < \mu < 1/r$), but  ideally it should be chosen of order 
 $1/r \sim |\bm{k}_2|/z_2$, to minimise the NLO corrections. In practice though, it is more 
 convenient to use $\mu^2=
  {\rm max}(\bm{k}_1^2, \,\bm{k}_2^2)$, so that the same factorisation scale also applies 
  to a collinear emission by the quark.
  
The would-be singular contribution to the last line in \eqn{qqcoll}  describes the
evolution of the final antiquark state via the emission of a quasi-collinear gluon. This 
piece must be subtracted from the NLO correction and absorbed into the 
DGLAP evolution of the (anti)quark fragmentation function into hadrons:
  \beq\label{frag1}
D_{h/q}(\zeta,\mu^{2})\,=\,\int_{\zeta}^1\frac{dz}{z}\left[\delta(1-z)+
 \frac{\alpha_s C_F}{\pi}
P_{q\to q}(z)\,\ln\frac{\mu^2}{\Lambda^2}\right]D_{h/q}^{(0)}\left(\frac{\zeta}{z}\right),
\eeq
where $D_{h/q}^{(0)}(\zeta)$ is the (non-perturbative) fragmentation function of a bare quark.
Hence, after adding the collinearly-divergent piece of \eqn{qqcoll} 
to the LO dihadron cross-section, one finds an approximate version of \eqn{2qFF}
where the (first step in the) DGLAP evolution is included only on the final antiquark state:
\begin{align}
\label{2qFFnlo}
\frac{d\sigma^{\gamma A\rightarrow h_1h_2+X}_\lo}{dk_1^+d^{2}\bm{k}_{1}dk_2^+d^{2}\bm{k}_{2}}
\simeq \int \frac{d\zeta_1}{\zeta_1^3} \int \frac{d\zeta_2}{\zeta_2^3}\,
\frac{d\sigma_{\lo}^{\gamma A\rightarrow q\bar q+X}}
{dp_1^+d^{2}\bm{p}_{1}dp^+_2d^{2}\bm{p}_2}\Bigg |_{p_i=\frac{k_i}{\zeta_i}}
D_{h_1/q}^{(0)}(\zeta_1)\,D_{h_2/q}(\zeta_2,\mu^{2})\,.
 \end{align}
In particular, the $\delta$-function for longitudinal momentum
conservation inherent in the partonic cross-section, cf. \eqn{transcross}, 
reproduces the normalisation factor
in  \eqn{qqcoll}, via the following identity\footnote{To make contact with
the result in \eqn{qqcoll}, which is written at partonic level, one can use
$D_{h/q}^{(0)}(\zeta)\to \delta(1-\zeta)$.}:
\beq
\int \frac{dz}{z^3}\,\delta\left(q^+ - k_1^+ - \frac{k_2^+}{z}\right) = \frac{1}{z_2^2q^+(1-\vartheta)}\,,
  \eeq
with $\vartheta=k_1^+/q^+$ and $z_2$ as defined in \eqn{defz2}.

Clearly, the case of a collinear emission by the quark can be similarly treated and leads to the analog
of \eqn{2qFFnlo} in which one step in the DGLAP evolution is included on the final quark state.

\section*{Acknowledgements} 
We would like to thank Nestor Armesto, Guillaume Beuf, Paul Caucal, Adrian Dumitru, Yuri Kovchegov, Tuomas Lappi, 
Michael Lublinsky, Cyrille Marquet, Farid Salazar and Pieter Taels for insightful discussions during the gestation of this work.
The work of E.I. is supported in part by the Agence Nationale de la Recherche project  ANR-16-CE31-0019-01.   
Y.M acknowledges financial support from Xunta de Galicia (Centro singular de investigación de Galicia accreditation 2019-2022); the “María de Maeztu” Units of Excellence program MDM2016-0692 and the Spanish Research State Agency under project PID2020-119632GB-I00; European Union ERDF; MSCA RISE 823947 ”Heavy ion collisions: collectivity and precision in saturation physics” (HIEIC). This work was also supported under the European Union’s Horizon 2020 research and innovation programme by the European Research Council (ERC, grant agreement No. ERC-2018-ADG-835105 YoctoLHC) and by the STRONG-2020 project (grant agreement No. 824093). The content of this article does not reflect the official opinion of the European Union and responsibility for the information and views expressed therein lies entirely with the authors.

%
\providecommand{\href}[2]{#2}\begingroup\raggedright\endgroup

\end{document}